\documentclass[%
 preprint, 
 amsmath,amssymb,
 aps, physrev,
]{revtex4-2}

\usepackage{subcaption}
\usepackage{graphicx}
\usepackage{dcolumn}
\usepackage{bm}

\usepackage{refcount}
\newsavebox{\lsbox}
\newcounter{LScount}
\makeatletter
\newcommand*\LS@register[1]{\global\@namedef{LS@rot@#1}{}}
\AtBeginDocument{%
  \@tempcnta\z@
  \loop\advance\@tempcnta\@ne
    \@ifundefined{r@LS@page@\the\@tempcnta}{}%
      {\expandafter\LS@register\expandafter{\getpagerefnumber{LS@page@\the\@tempcnta}}}%
  \ifnum\@tempcnta<50 \repeat}
\AddToHook{shipout/before}{%
  \ifdefined\pdfpageattr
    \@ifundefined{LS@rot@\number\value{page}}%
      {\global\pdfpageattr{}}{\global\pdfpageattr{/Rotate 90}}%
  \fi}
\newenvironment{landscape}
  {\clearpage\begin{lrbox}{\lsbox}\begin{minipage}{\textheight}%
   \stepcounter{LScount}\label{LS@page@\arabic{LScount}}}
  {\end{minipage}\end{lrbox}%
   \noindent\makebox[\textwidth][c]{\rotatebox[origin=c]{90}{\usebox{\lsbox}}}%
   \clearpage}
\makeatother

\begin{document}

\preprint{APS/123-QED}

\title{\textbf{Mach-disk formation and shock-structure transitions in underexpanded coflowing jets} 
}%

\author{Ganesh Dhungana}
\author{Srijan Satyal}%
\author{Nek Sharan}%
 \email{Contact author: nsharan@auburn.edu}
\affiliation{%
 Department of Aerospace Engineering, Auburn University, Auburn, AL 36849
}%



\date{\today}

\begin{abstract}
The near-field shock structures of underexpanded sonic jets exiting into a subsonic coflow are investigated over a range of nozzle pressure ratio (NPR) and coflow-to-nozzle-exit velocity ratio ($U_c$), representative of a propulsive nozzle in subsonic flight. Time-averaged statistics from fully-resolved axisymmetric simulations and (inviscid) method-of-characteristics (MOC) analysis are utilized to understand how the coflow alters the shock-cell structures, in particular, the Mach-disk formation. It is well established that increasing NPR leads to a transition from regular reflection (characterized by oblique shocks) to Mach reflection (characterized by a near-normal shock segment called the Mach disk) at the jet centerline. This study systematically investigates the effects of coflow on Mach-disk formation and shock-structure transitions. It is found that the coflow has an opposite influence to that of NPR, \textit{i.e.}, a strong coflow shrinks the Mach disk size until it vanishes, reverting the centerline Mach reflection to regular reflection. Equivalently, the NPR for the transition from regular to Mach reflection increases with increasing $U_c$. This effect has previously been attributed to a reduction in the jet boundary inclination at the nozzle lip, which confines the lip Prandtl–Meyer fan to a smaller angle and weakens the embedded shock. We show that the jet inclination is determined by the non-uniform pressure the coflow imposes along the jet boundary, which is the primary driver of the shock-structure transitions in coflowing jets. 
The non-uniform pressure weakens the boundary-reflected compression waves and orients them at shallower angles so that the embedded shock reflects regularly or fails to form. MOC analysis with simulation-informed non-uniform pressure boundary condition reproduces the transition behavior with increasing coflow. Coflow also lengthens the first shock cell linearly, which is accurately estimated by a simple correction to the Prandtl's classical shock-cell length scaling.
\end{abstract}

\maketitle


\section{Introduction\label{sec:Introduction}}
An underexpanded jet involves a flow exiting a nozzle with a static pressure higher than the ambient pressure. This study focuses on underexpanded jets exiting from a convergent choked nozzle. The degree of underexpansion is typically characterized by the pressure ratio (PR = $p_e/p_\infty$) or the nozzle pressure ratio (NPR = $p_0/p_\infty$), where $p_0$ denotes the total pressure at the nozzle inlet, and $p_e$ and $p_\infty$ denote the nozzle-exit and ambient static pressure, respectively. For an isentropic flow (of a calorically perfect gas) through a convergent–divergent nozzle, \textit{i.e.}, a nozzle flow without any internal shock waves, the two ratios are related through the nozzle-exit Mach number ($M_e$), given by

\begin{equation}
    \frac{\mathrm{NPR}}{\mathrm{PR}}
    =
    \frac{p_0}{p_e}
    =
    \left[1+\frac{\gamma-1}{2}M_e^2\right]^{\frac{\gamma}{\gamma-1}}.
    \label{eq:nozzle-isentropic-relation}
\end{equation}
The near-field flow structures of an underexpanded jet are primarily governed by the static-pressure mismatch between the nozzle exit and the ambient, characterized by PR, and the nozzle-exit Mach number ($M_e$). For a choked convergent nozzle, $M_e = 1$ is fixed and NPR is directly proportional to PR \textit{i.e.} $\mathrm{NPR} = 1.893\,\mathrm{PR}$ for $\gamma = 1.4$ from Eq.~\eqref{eq:nozzle-isentropic-relation}. So, either of the two ratios can be equivalently used to classify the flow regimes. 
Following the classification discussed by \citet{franquet2015free}, the near-field flow structures can be classified into four regimes . 
At low pressure ratios ($1 \lesssim \mathrm{NPR} \lesssim 1.9$), below the choking threshold and outside the range considered in this study, the nozzle is unchoked and the jet issues subsonically with its exit pressure matched to the ambient. Moderately underexpanded jets ($2 \lesssim \mathrm{NPR} \lesssim 4$) develop the typical diamond-shaped shock-cell pattern, while highly underexpanded jets ($4$--$5 \lesssim \mathrm{NPR} \lesssim 7$) form series of Mach-cells each containing a barrel shock terminated by a Mach disk. Finally, in extremely underexpanded jets ($\mathrm{NPR} \gtrsim 7$), the potential core is dominated by a single barrel, with no further shock structures formed downstream. The present work focuses on the transition from shock-cell-dominated to Mach-disk-dominated flows, so the moderately and highly underexpanded regimes are of interest here and are described in detail below.

\begin{figure}[hbtp]
        \centering
        \includegraphics[width=0.8\textwidth]{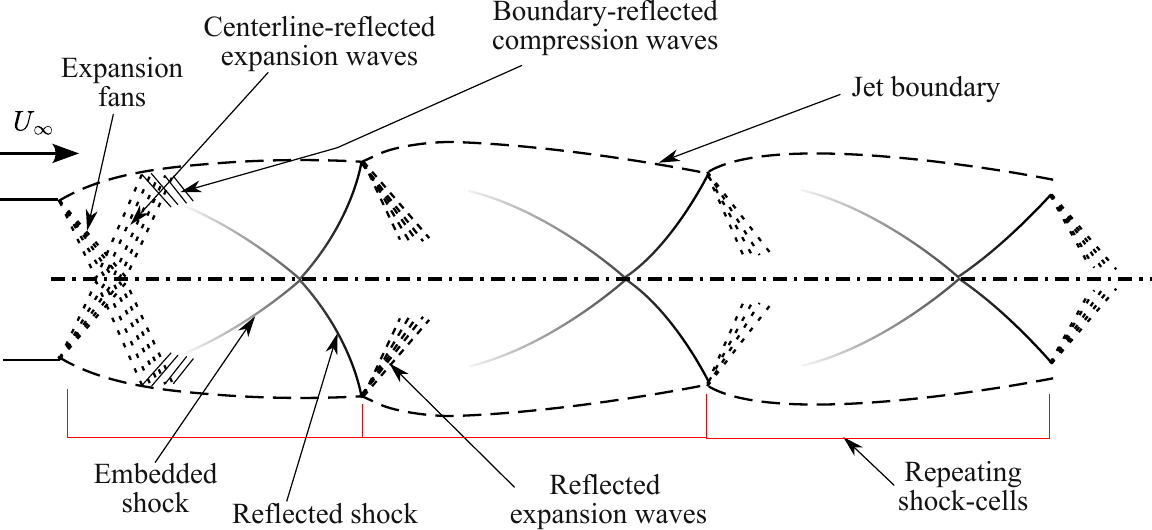}
    \caption{Moderately underexpanded jet (adapted from \cite{franquet2015free}).}
    \label{fig:regular-reflection-schematic}
\end{figure}

In moderately underexpanded jets ($2 \lesssim \textrm{NPR} \lesssim 4$), the formation of the shock-cell structure (Fig.~\ref{fig:regular-reflection-schematic}) can be understood through a simple inviscid argument. As the flow exits the nozzle, an expansion fan originates at the nozzle lip, accelerating the jet to reduce the pressure along the jet boundary to the ambient value $p_\infty$. The expansion waves then traverse downstream, and 
reflect from the centerline as expansion waves of a similar kind to turn the flow parallel to the centerline. The core flow in the jet, having passed through both the incident and reflected fans, over-expands to a pressure below the ambient level. When the reflected expansion waves subsequently reach the jet boundary, the constant-pressure condition there does not permit the boundary pressure to fall below the ambient value, and the expansion waves must therefore reflect as compression waves. These compression waves recompress the core to a pressure above $p_\infty$ and, over most of this moderately underexpanded regime, coalesce into a curved oblique shock, called the embedded shock (or the intercepting, barrel, or incident shock), which reflects regularly at the centerline as an oblique shock (Fig.~\ref{fig:regular-reflection-schematic}). 
The waves thus reflect successively between the constant-pressure jet boundary (a streamline in the inviscid description) and the centerline, producing the characteristic diamond-shaped cell pattern: reflections from the centerline preserve the wave type, \textit{i.e.}, a compression (expansion) wave reflects as a compression (expansion) wave, while reflections from the boundary convert expansion waves into compression waves and vice versa. In the ideal inviscid limit this cell pattern would repeat indefinitely; in a real (viscous) flow, the shock cells weaken downstream as the growing turbulent shear layer thickens and progressively absorbs the wave system.

\begin{figure}[hbtp]
        \centering
        \includegraphics[width=0.8\textwidth]{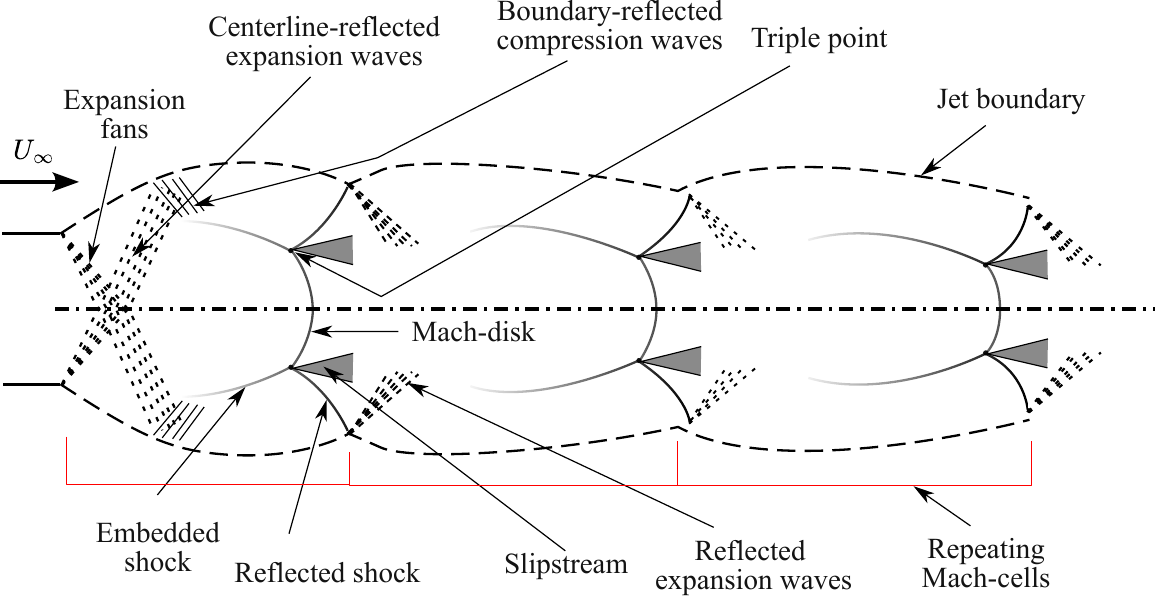}
    \caption{Highly underexpanded jet (adapted from \cite{franquet2015free})}
    \label{fig:mach-reflection-schematic}
\end{figure}

Highly underexpanded jets ($4$--$5 \lesssim \textrm{NPR} \lesssim 7$) retain the same underlying wave system as moderately underexpanded jets, but the stronger lip expansion due to larger pressure mismatch produces boundary-reflected compression waves that coalesce closer to the nozzle into a stronger, more steeply inclined embedded shock, as illustrated in Fig.~\ref{fig:mach-reflection-schematic}. \citet{chang1973mach} performed an inviscid method-of-characteristics analysis to predict the formation of such embedded shocks and showed that their strength, curvature, and alignment determine the type of reflection at the centerline. For highly underexpanded jets, regular reflection of the embedded shock becomes infeasible. The reflected shock must turn the flow behind the embedded shock back parallel to the centerline. If the turning required exceeds the maximum turning angle attainable at the local Mach number, no attached oblique-shock solution exists for the reflected shock. Consequently, a Mach reflection occurs, and the embedded shock terminates in a Mach disk (the axisymmetric counterpart of the Mach stem in planar flow), as shown in Fig.~\ref{fig:mach-reflection-schematic}. A reflected shock forms at the intersection of the embedded shock and the Mach disk. The pressure and flow direction downstream of the reflected shock match those downstream of the Mach disk, while other flow properties, such as velocity magnitude, density, and entropy, remain discontinuous between these two post-shock regions. These two regions are separated by a contact discontinuity, the slipstream (slip line in planar flow), manifesting as a thin annular shear layer in the viscous jet. The slipstream originates at the common intersection of the embedded shock, the reflected shock, and the Mach disk, known as the triple point (Fig.~\ref{fig:mach-reflection-schematic}).

As discussed earlier, NPR determines the strength of the lip-centered expansion fan and hence of the boundary-reflected compression waves, and thereby governs the transition of the embedded-shock reflection at the centerline from regular to Mach reflection. Few studies explicitly identify the NPR at which this transition occurs; substantial work instead examines the variation of the Mach-disk diameter with NPR (or PR) experimentally \cite{crist1966study, ashkenas1966structure, gibbings1972flow, addy1981effects, antsupov1974properties, carlson1964normal, dattorre1965parameters, werle1970freejet, driftmyer1972correlation, ewan1986structure, love1955some}, theoretically \cite{murzinov1971similarity, jiang2022theoretical, chang1973mach}, and numerically, often in conjunction with experiments \cite{dambrosio1999physical, sommerfeld1994structure, lee2004supersonic, muraoka2022onset}. While most of these studies consider jets from convergent nozzles, several also address jets from convergent-divergent nozzles \cite{love1955some, carlson1964normal, dattorre1965parameters, werle1970freejet, chang1973mach}. If the transition is defined as the NPR at which the Mach-disk diameter vanishes, then, extrapolating the measured diameters of Refs.~\cite{antsupov1974properties, crist1966study, addy1981effects, gibbings1972flow} to zero, \citet{muraoka2022onset} obtained a transition NPR of approximately 3.8. From their own direct numerical simulations of underexpanded axisymmetric jets, however, they found the transition to occur at a lower value, in the range $\mathrm{NPR} \approx 3.08$--$3.12$. The transition NPR for a jet in a quiescent ambient thus remains unsettled, with reported values ranging from approximately 3.1 to 3.9.

The literature discussed in the previous paragraph represents ground-test conditions, in which the underexpanded jet issues into a quiescent ambient (\textit{i.e.}, $U_\infty = 0$ in Figs.~\ref{fig:regular-reflection-schematic} and~\ref{fig:mach-reflection-schematic}). In flight, by contrast, the ambient fluid moves relative to the nozzle, so that a nozzle operating at off-design conditions exhausts its underexpanded jet into a co-flowing ambient stream (\textit{i.e.} $U_\infty \neq 0$ in Figs.~\ref{fig:regular-reflection-schematic} and~\ref{fig:mach-reflection-schematic}), hereafter referred to as coflow. In this study, we restrict our analysis to subsonic coflow velocities, which are relevant to propulsive nozzles in flight at low to high subsonic speeds. Since the near-field shock structures of jet, and in particular the presence of a Mach disk, marks a substantial change in the state of the jet core, the effect of a coflow on this structure bears directly on the analysis of nozzle flows under in-flight conditions, and has been examined in two bodies of literature reviewed below.


One body of relevant work concerns dual coaxial jets, in which the effect of the annular stream on jet mixing, spreading, screech, and, of particular interest here, the near-field shock structure has been investigated. The coflowing jets of the present study can be regarded as dual coaxial jets in the limit of large annular-stream thickness and zero nozzle-lip thickness, the limit realized by the flight-simulation experiments discussed in the following paragraph. \citet{masuda1994aerodynamic} studied underexpanded coaxial impinging jets, in which a sonic central jet is directed toward a flat plate and the annular stream is set either parallel or inclined to the inner jet; over annular pressure ratios of about 1.5 to 7, which carry the annular stream from subsonic to supersonic, they found that the annular stream reduces the diameter of the central-jet Mach disk. \citet{narayanan1993mach} studied free dual coaxial jets in which the annular stream issues obliquely, at about 26 degrees to the jet axis, and found that increasing the annular-flow NPR increases the Mach-disk diameter and shifts it downstream. \citet{rao1996near}, through their experimental study of coaxial jets, pointed out the notable compression of the inner flow by the outer one, leading to premature formation of the Mach disk compared to that of a single jet. \citet{lee2004supersonic} conducted experiments over various combinations of inner- and outer-flow NPRs using a parallel annular stream, and found that the effect of the annular flow on the Mach-disk diameter depends on the central-flow NPR: at low central NPR, the diameter grows with increasing annular NPR, while at higher central NPR it shrinks. They also reported that the annular-stream thickness has only a minor effect. \citet{srinivasarao2012effect} studied an underexpanded central jet surrounded by an annular stream directed inward at about 23 degrees, at central NPRs of 3 to 5, and found that the coflow elongates the supersonic core and, at NPRs of 4 and 5, replaces the Mach reflection of the single jet with a regular reflection. Unlike the studies above, which concern the statistically stationary jet, \citet{ahmad2022influence} investigated the transient startup of an underexpanded jet issuing into a parallel coflow, and found that increasing the coflow weakens the embedded shock and the Mach disk, with the slipstream ceasing to emanate and the counter-rotating vortex rings (CRVRs) of the starting phase suppressed at sufficiently high coflow; using a Prandtl--Meyer analysis, they attributed this weakening to a narrowing of the lip-centered expansion fan with increasing coflow.

Closer to the configuration considered in this study is the body of work on underexpanded jets in simulated flight, where the jet exits into a uniform, parallel free stream. An early study is that of \citet{avduevskii1970flow}, who examined underexpanded jets issuing into supersonic streams at very high pressure ratios and observed that the outer stream renders the pressure along the jet boundary non-uniform and reduces the dimensions of the initial segment of the jet. \citet{norum1984effects,norum1988shock} measured the shock-cell geometry and noise of underexpanded jets from convergent nozzles at flight Mach numbers up to 0.4, and reported a lengthening of the shock cells with forward speed, concentrated in the downstream cells. They also reported that the flight stream caused no perceptible change in the shock strengths. \citet{norum1993simulated} extended the measurements to a flight Mach number of 0.9, where the cell system stretches to more than twice its static extent. They also observed that the external stream modified the boundary conditions of the jet plume sufficiently for weak shocks to appear even in an initially shock-free, perfectly expanded jet. On the modeling side, \citet{morris1988note} predicted cell lengthening with an inviscid vortex-sheet analysis in which the external stream modifies the pressure condition at the jet boundary, and \citet{tam1992broadband} adopted a linear correction of the cell length in flight Mach number for broadband shock-noise prediction. \citet{morris1988note} further pointed out that the thick boundary layer developing along the nozzle forebody in the flight-simulation facilities shields the initial cells from the free stream, so that the predicted near-exit effects should emerge only when this layer is very thin or at high subsonic flight speeds. Consistent with this, \citet{norum1993simulated} measured a forebody boundary-layer thickness exceeding the nozzle-exit radius in their facility, about twice the value reported for the aircraft in flight. Indeed, when \citet{andre2016flighteffects} later employed a dual-stream arrangement with a substantially thinner outer boundary layer, they found, in contrast to the earlier measurements, that the stretching extends continuously through the shock cells and the cells weaken with increasing free-stream velocity. However, the underlying cause for weakening of shock-cell strength was not identified and left for future work. These studies mostly concentrate on the shock-cell spacing and its acoustic signature, and the shock-structure dynamics appear only incidentally.

In summary, these studies do not provide a consistent picture of how coflow affects the near-field shock structures. The dual coaxial studies differ widely in geometry, with impinging, free, or parallel jets and the annular stream introduced parallel, obliquely, or convergently to the axis, and they cover different ranges of central- and annular-flow NPR. The reported effect of the annular stream on the Mach disk varies just as widely, and even in sign: the annular stream is found to reduce or eliminate the Mach disk in some configurations \cite{masuda1994aerodynamic, srinivasarao2012effect, ahmad2022influence} and to enlarge it or hasten its formation in others \cite{narayanan1993mach, rao1996near}. The simulated-flight experiments do provide the parallel, large-thickness coflow most relevant to a propulsive nozzle in subsonic flight, but, as noted above, the response of the Mach disk to the coflow is left unexamined. What emerges from this literature is that the nozzle pressure ratio and the subsonic coflow velocity have an opposite effect on the Mach disk, with the former promoting its formation and growth and the latter weakening and suppressing it \cite{srinivasarao2012effect, ahmad2022influence, andre2016flighteffects}. Both parameters are simultaneously at play for a propulsive nozzle in flight, and their competition determines whether a Mach disk forms at all, which, as noted earlier, marks a substantial change in the state of the jet core. To our knowledge, the literature lacks a study of this combined effect of NPR and coflow velocity on the onset of Mach-disk formation, which is the main focus of this work. Moreover, we 
identify the mechanism by which the coflow weakens the embedded shock and suppresses the Mach disk formation. 
Furthermore, a simple, closed-form correction is proposed to the classical single-jet shock-cell length scaling that accounts for the subsonic coflow and distinguishes the regular and Mach-reflection regimes.

The remainder of the paper is organized as follows. Section \ref{sec:numerical-approach} presents the numerical approach, comprising the governing equations, their non-dimensionalization, and the computational setup used to simulate the coflowing jets. Section \ref{sec:jet-validation} validates the solver against established results for underexpanded jets. Section \ref{sec:parameter-space} defines the parameter space of nozzle NPR and subsonic coflow velocities considered in this study. Section \ref{sec:Numerical-results} presents the numerical results and analysis, where the effect of the coflow on the centerline properties, jet boundary shape, and near-field shock structures is examined. Finally, Section \ref{sec:Conclusions} summarizes the main conclusions.

\section{Numerical approach \label{sec:numerical-approach}}

\subsection{Governing equations \label{sec:governing-equations}}
The axisymmetric compressible Navier-Stokes (NS) equations are solved in generalized curvilinear coordinates, where the physical coordinates $(x,r)$ are mapped to the computational coordinates $(\xi,\eta)$. The governing equations are then given by
\begin{equation}
\frac{\partial \hat{\mathbf Q}}{\partial t}
+
\frac{\partial \hat{\mathbf E}}{\partial \xi}
+
\frac{\partial \hat{\mathbf F}}{\partial \eta}
+
\mathbf S_{\mathrm{inv}}
=
\frac{1}{Re}
\left(
\frac{\partial \hat{\mathbf E}_v}{\partial \xi}
+
\frac{\partial \hat{\mathbf F}_v}{\partial \eta}
\right)
+
\frac{1}{Re}\mathbf S_{\mathrm{vis}},
\label{eq:curvilinear_ns}
\end{equation}
where
\begin{equation}
\hat{\mathbf Q}
=
\frac{1}{J}
\begin{bmatrix}
\rho\\
\rho u\\
\rho v\\
\rho e
\end{bmatrix},
\qquad
\hat{\mathbf E}
=
\frac{1}{J}
\begin{bmatrix}
\rho U\\
\rho uU+\xi_x p\\
\rho vU+\xi_r p\\
(\rho e+p)U
\end{bmatrix},
\qquad
\hat{\mathbf F}
=
\frac{1}{J}
\begin{bmatrix}
\rho V\\
\rho uV+\eta_x p\\
\rho vV+\eta_r p\\
(\rho e+p)V
\end{bmatrix}.
\label{eq:inv_flux_curvilinear}
\end{equation}
In (\ref{eq:inv_flux_curvilinear}), $\rho$, $u$, $v$, $p$, and $e$ are the density, axial velocity, radial velocity, pressure, and total energy per unit mass, respectively. For a calorically perfect gas, the pressure is evaluated from
\begin{equation}
p=(\gamma-1)
\left[
\rho e-\frac{1}{2}\rho\left(u^2+v^2\right)
\right],
\label{eq:eos}
\end{equation}
where $\gamma$ is the ratio of specific heats. The transformation metrics and contravariant velocities, assuming fixed (static) grids, are
\begin{equation}
J=x_\xi r_\eta-x_\eta r_\xi,
\qquad
\xi_x=\frac{r_\eta}{J},
\qquad
\xi_r=-\frac{x_\eta}{J},
\qquad
\eta_x=-\frac{r_\xi}{J},
\qquad
\eta_r=\frac{x_\xi}{J},\\
\label{eq:metric_term}
\end{equation}

\begin{equation}
    U=\xi_x u+\xi_r v,
\qquad
V=\eta_x u+\eta_r v .
\label{eq:contravariant}
\end{equation}
respectively.
The viscous fluxes in (\ref{eq:curvilinear_ns}) are given by
\begin{equation}
\hat{\mathbf E}_v
=
\frac{1}{J}
\left(
\xi_x\mathbf E_v+\xi_r\mathbf F_v
\right),
\qquad
\hat{\mathbf F}_v
=
\frac{1}{J}
\left(
\eta_x\mathbf E_v+\eta_r\mathbf F_v
\right),
\label{eq:viscous_flux_curvilinear}
\end{equation}
where
\begin{equation}
\mathbf E_v=
\begin{bmatrix}
0\\
\tau_{xx}\\
\tau_{xr}\\
f_4
\end{bmatrix},
\qquad
\mathbf F_v=
\begin{bmatrix}
0\\
\tau_{xr}\\
\tau_{rr}\\
g_4
\end{bmatrix}.
\label{eq:physical_viscous_flux}
\end{equation}

\noindent The viscous stresses and the viscous fluxes in the energy equation are given by
\begin{align}
\tau_{xx}
&=
2\mu\left(\xi_x u_\xi+\eta_x u_\eta\right)
+\lambda\Theta,
&
\tau_{rr}
&=
2\mu\left(\xi_r v_\xi+\eta_r v_\eta\right)
+\lambda\Theta,
\nonumber\\
\tau_{xr}
&=
\mu
\left(
\xi_r u_\xi+\eta_r u_\eta
+
\xi_x v_\xi+\eta_x v_\eta
\right),
&
\tau_{\theta\theta}
&=
2\mu\frac{v}{r}
+\lambda\Theta .
\label{eq:stress_components}
\end{align}
and
\begin{align}
f_4
&=
u\tau_{xx}
+
v\tau_{xr}
+
\mu Pr^{-1}(\gamma-1)^{-1}
\left(
\xi_x \frac{\partial c^2}{\partial \xi}
+
\eta_x \frac{\partial c^2}{\partial \eta}
\right),
\nonumber\\
g_4
&=
u\tau_{xr}
+
v\tau_{rr}
+
\mu Pr^{-1}(\gamma-1)^{-1}
\left(
\xi_r \frac{\partial c^2}{\partial \xi}
+
\eta_r \frac{\partial c^2}{\partial \eta}
\right),
\label{eq:energy_viscous_flux}
\end{align}
respectively. \noindent For axisymmetric flow without swirl, the dilatation term is
\begin{equation}
\Theta
=
\left(\xi_x u_\xi+\eta_x u_\eta\right)
+
\left(\xi_r v_\xi+\eta_r v_\eta\right)
+
\frac{v}{r}.
\label{eq:dilatation}
\end{equation}
In (\ref{eq:stress_components})-(\ref{eq:energy_viscous_flux}), $\mu$ is the dynamic viscosity, $\lambda= \mu_B-2/3\mu$ is the second coefficient of viscosity, where $\mu_B$ denotes the bulk viscosity, and $c=\sqrt{\gamma p/\rho}$ is the speed of sound. The inviscid and viscous source terms in (\ref{eq:curvilinear_ns}), associated with the axisymmetric formulation, are, respectively,

\begin{equation}
    \mathbf S_{\mathrm{inv}}
    =
    \frac{1}{rJ}
    \left(
    \hat{\mathbf{F}}\, r_{\eta}
    -
    \begin{bmatrix}
    0 \\ 0 \\ p \\ 0
    \end{bmatrix}
    \right),
    \qquad
    \mathbf S_{\mathrm{vis}}
    =
    \frac{1}{rJ}
    \left(
    \hat{\mathbf{F}}_v\, r_{\eta}
    +
    \begin{bmatrix}
    0 \\ 0 \\ \tau_{\theta\theta} \\ 0
    \end{bmatrix}
    \right).\label{eq:source_terms}
\end{equation}

\subsection{Physical variables and non-dimensionalization} 
 The variables in \eqref{eq:curvilinear_ns}--\eqref{eq:source_terms} are non-dimensional, given by
\begin{align}
&x_i=\frac{x_i^*}{D_e^*},
\quad t=\frac{t^*a_\infty^*}{D_e^*},
\quad u_i=\frac{u_i^*}{a_\infty^*},
\quad \rho=\frac{\rho^*}{\rho_\infty^*},
 p=\frac{p^*}{\rho_\infty^* a_\infty^{*2}},
\quad\mu=\frac{\mu^*}{\mu_\infty^*},
\quad\lambda=\frac{\lambda^*}{\mu_\infty^*},
\quad T= \frac{T^*}{(\gamma-1)T_\infty^*},
\end{align}
where $(\cdot)^*$ denotes a dimensional quantity and the non-dimensionalization uses the nozzle-exit diameter ($D_e^*$) and ambient density ($\rho_\infty^*$), sound speed ($a_\infty^*$), temperature ($T_\infty^*$) and  viscosity ($\mu_\infty^*$). In this paper, $(x_1,x_2)\equiv(x,r)$ and $(u_1,u_2)\equiv(u,v)$ are the physical coordinates and velocity components, respectively. The Reynolds and Prandtl numbers are given by
\begin{equation}Re=\frac{\rho_\infty^* a_\infty^* D_e^*}{\mu_\infty^*},\qquad 
Pr=\frac{\mu_\infty^* C_p^*}{k_\infty^*}.
\end{equation}
All axisymmetric simulations in this study use a Reynolds number ($\textrm{Re}$) of 50{,}000; the three-dimensional runs for validation presented in Appendix~\ref{app:grid_test_3_D} use a reduced Reynolds number of $5300$, for the reasons discussed there. A power law is used to model the temperature dependence of viscosity,
\begin{equation}
    \mu = [(\gamma-1)T]^n.
\label{eq:temperature_model}
\end{equation}
For air, the bulk viscosity ($\mu_B$) is $0.6\mu$, specific heat ratio ($\gamma$) is 1.4 and Prandtl number ($Pr$) is 0.72. The non-dimensional ideal gas law is given by
\begin{equation}
    p=\dfrac{\gamma -1}{\gamma} \rho T
    \label{eq:ideal_gas}
\end{equation}



\subsection{Time-averaging and mean flow variables}
The flow variables, i.e., $\boldsymbol{\phi} = [\rho,\, \rho u,\, \rho v,\, p]$, are time-averaged over 100 non-dimensional time units using
\begin{equation}
    \bar{\phi}(x,r) = \frac{1}{N_\mathrm{t}} \sum_{i=1}^{N_\mathrm{t}} \phi(t_i; x,r),
\end{equation}
where $N_\mathrm{t}$ is the total number of samples and $t_1$ and $t_{N_t}$ are the start and end times 
for time-averaging. Time-averaged quantities, such as the density gradient magnitude, $|\nabla \bar{\rho}|$, and the Mach number, $\bar{M}$, are obtained using 
\begin{equation}
    |\nabla \bar{\rho}| =\sqrt{\left(\dfrac{\partial \bar{\rho}}{\partial x}\right)^2 + \left(\dfrac{\partial \bar{\rho}}{\partial r}\right)^2} ,\qquad \qquad \bar{M} =  \dfrac{\overline{\rho u}}{\bar{\rho}} /\sqrt{\gamma \dfrac{\bar{p}}{\bar{\rho}}}
\end{equation}


\subsection{Simulation setup \label{sec:numerical_aspects}}

\subsubsection{Computational domain and numerical methods \label{subsubsec:domain_numerics}}
\begin{figure}[hbtp]
        \includegraphics[width=1.0\textwidth]{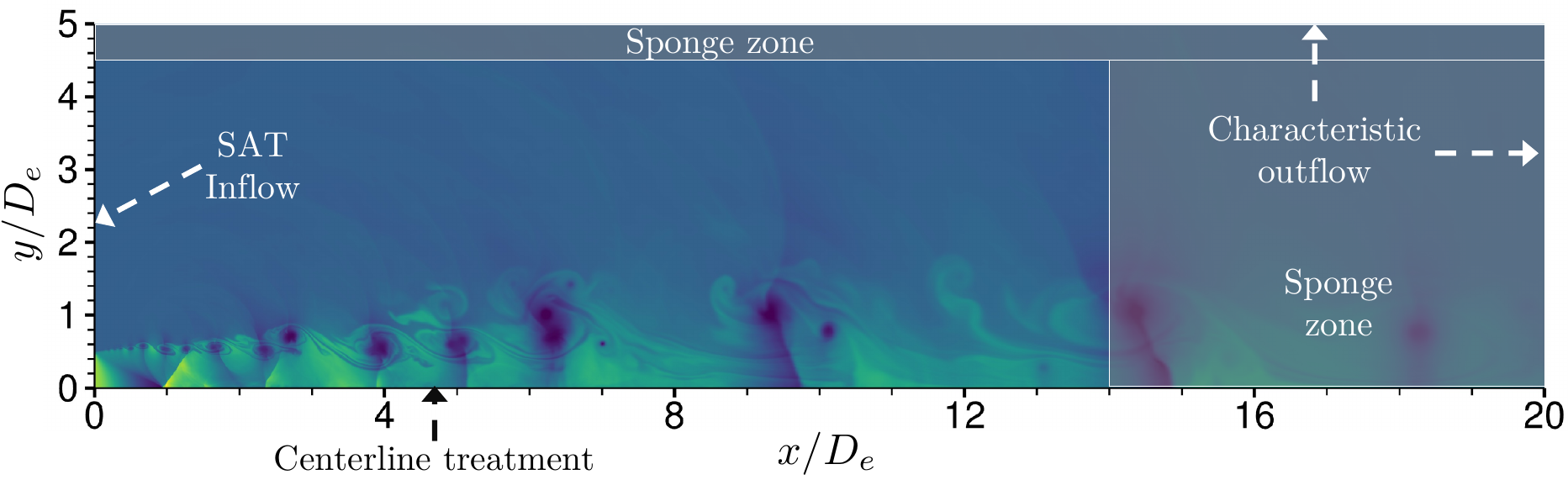}
    \caption{A schematic showing the extent of axisymmetric computational domain in axial and radial directions, and the boundary conditions applied at various boundaries.}
    \label{fig:computational_domain}
\end{figure}

Figure~\ref{fig:computational_domain} illustrates the computational domain as well as the boundary conditions (further discussed in Section \ref{subsec:initial-boundary-conditions}). The domain ranges from $0$ to $20\,D_e$ in the axial ($x$-) direction and extends up to $5\,D_e$ in the radial ($r$-)direction. The symbols $y$ and $r$ are used interchangeably in the plots.
A grid independence test using three sets of grids, as detailed in Appendix \ref{app:grid_test_3_D}, is performed to ensure that the simulations resolve all the flow scales. The mean centerline density obtained from these three grids is in good agreement; the finer grid is chosen for subsequent analysis. Moreover, a three-dimensional (3-D) simulation with a nozzle pressure ratio (NPR) of $4.26$, as elaborated in Appendix \ref{app:grid_test_3_D}, is performed to confirm that the flow statistics and shock structures obtained from the axisymmetric simulations are suitable for analyses in this study. A comparison of the centerline quantities and the shock locations shows that the axisymmetric simulation results are nearly identical to the 3-D simulation results in the first Mach disk/shock cell, hence the axisymmetric calculations are sufficient for this investigation focused on Mach-disk formation and shock-structure transitions.


Inviscid flux derivatives in \eqref{eq:curvilinear_ns} are evaluated using a sixth-order weighted essentially non-oscillatory (WENO) scheme \cite{hu2010adaptive} to provide high-order spatial accuracy in smooth regions while maintaining non-oscillatory shock-capturing capability near shocks. The viscous fluxes are approximated using the fourth-order centered finite-difference scheme and time integration uses the classical explicit fourth-order Runge–Kutta (RK4) method. To obtain the numerical solution, the conservation equations \eqref{eq:curvilinear_ns} are solved iteratively using a Courant–Friedrichs–Lewy (CFL) number of $0.5$.

\subsubsection{Initial and boundary conditions \label{subsec:initial-boundary-conditions}}
The inflow boundary condition (BC) is applied using the simultaneous approximation term (SAT) \cite{svard2007stable,sharan2018time}, which weakly imposes the nozzle-exit conditions, given by
\begin{equation}
    \dfrac{p_\mathrm{e}}{p_\mathrm{\infty}} = \left[\dfrac{2+(\gamma-1)M_j^2}{\gamma +1}\right]^{\dfrac{\gamma}{\gamma-1}},
    \qquad \dfrac{T_\mathrm{e}}{T_\mathrm{\infty}}=  \dfrac{T_\mathrm{0}}{T_\mathrm{\infty}}\left(1+\dfrac{\gamma-1}{2}\right)^{-1},
    \qquad U_\mathrm{e} =\left(\dfrac{2T_\mathrm{0}}{(\gamma+1)T_\mathrm{\infty}}\right)^{1/2},
\end{equation}
\noindent where $p_\mathrm{e}$, $T_\mathrm{e}$, $U_\mathrm{e}$, $T_\mathrm{0}$, and $M_j$ are the nozzle exit pressure, temperature, velocity, reservoir temperature,  and fully expanded Mach number, respectively. The density is obtained using Eq.~\eqref{eq:ideal_gas}. The non-dimensional ambient density, temperature and pressure are  $\rho_{\infty}=1$, $T_\mathrm{\infty}=1/(\gamma-1)$, and $p_\mathrm{\infty}=1/\gamma$, respectively. 
The velocity profile at inflow is modeled using a tanh profile as
\begin{equation}
u_e(r)=\frac{U_e}{2}\left[1-\tanh\left(\frac{|r|-r_e}{2\theta_\mathrm{0} } \right)\right],\label{eq:vel_profile}
\end{equation}
\noindent where $r_\mathrm{e}=D_\mathrm{e}/2$ is the radius of nozzle exit and $\theta_\mathrm{0}= 0.02~r_e$ is the momentum thickness. This choice of momentum thickness provides favorable 
comparisons against the experimental results, as discussed in Section \ref{sec:jet-validation}. Similar profiles as \eqref{eq:vel_profile} are used for pressure and temperature at the nozzle exit for $r\leq r_{\mathrm{e}}$. For non-zero coflow cases, the inlet velocity is specified as:
\begin{equation}
u_e(r)_\mathrm{a}=U_\mathrm{a}+\frac{U_e-U_\mathrm{a}}{2}\left[1-\tanh\left(\frac{|r|-r_e}{2\theta_\mathrm{0} } \right)\right],
\end{equation}
\noindent where $U_\mathrm{a}$ is the coflow velocity. For $r > r_{\mathrm{e}}$, the ambient flow conditions and coflow velocity are prescribed as the target state for the SAT inflow boundary treatment. Similarly, the entire computational domain is initialized with ambient conditions and the coflow velocity, except within the nozzle-exit region ($r \leq r_{\mathrm{e}}$).

Navier--Stokes characteristic boundary conditions (NSCBC) \cite{poinsot1992boundary}, combined with sponge zones \cite{bodony2006analysis}, 
are applied at the outflow boundaries. The sponge zones are implemented by adding the term $\frac{\sigma(\zeta)}{J}\left(\mathbf Q- \mathbf Q_\mathrm{ref}\right)$, where $\mathbf Q=J\hat{\mathbf Q}$ and $\mathbf Q_\mathrm{ref}$ is the target ambient state, to the right hand side of the semi-discretization of \eqref{eq:curvilinear_ns}.
The sponge strength is defined as
\begin{equation}
    \sigma(\zeta)
    =
    \sigma_\mathrm{sp}
    \left[
    \frac{\zeta-\zeta_\mathrm{min}}
    {\zeta_\mathrm{max}-\zeta_\mathrm{min}}
    \right]^{n_\mathrm{sp}},
    \qquad
    \zeta=(x,r),
    \label{eq:sponge_strength}
\end{equation}
\noindent where $(\zeta_\mathrm{min},\zeta_\mathrm{max})$ defines the extent of the sponge zone, with $\zeta_\mathrm{max}$ denoting the outflow boundary. The sponge amplitude is set to $\sigma_\mathrm{sp}=0.5$, with $n_\mathrm{sp}=5.0$. The sponge zones at the outflow boundaries have widths of $30\%$ and $10\%$ of the domain length normal to the boundary in the $x$- and $r$-directions, respectively. A symmetry boundary condition is imposed at the centerline, $r=0$, to avoid the singularity issue associated with the axisymmetric formulation, given by
\begin{equation}
    \dfrac{\partial \rho}{\partial r} =0, \qquad
    \dfrac{\partial \rho u}{\partial r} =0, \qquad
    \rho v= 0, \qquad
    \dfrac{\partial \rho e}{\partial r} =0.
    \label{eq:centerline_treatment}
\end{equation}

\section{Simulation validation for underexpanded jets \label{sec:jet-validation}}
Extensive validation of the flow solver \cite{sharan2018time,sharan2022high} was carried out for underexpanded jets. Since the study focuses on the combined effects of coflow and NPR on Mach-disk formation, we first consider two cases for jets expanded into a quiescent medium (no coflow): a moderately underexpanded jet, where regular reflection occurs at the centerline \cite{panda1999measurement, chatterjee2009screech, muraoka2022onset}, and a highly underexpanded jet, where Mach reflection occurs at the centerline \cite{panda1999measurement, sugawara2020three, muraoka2022onset}. Similarly, the onset of Mach-disk formation predicted by the current numerical method for jets expanded into a quiescent medium was compared with results from the literature \cite{muraoka2022onset, antsupov1974properties, crist1966study, addy1981effects, lee2004supersonic} by comparing the Mach-disk diameter for various NPR jets taken from \citet{muraoka2022onset}’s parameter set. For all the validation jet cases considered in this section, $\textrm{NTR }= \frac{T_0}{T_\infty} = 1.0$ is used.  

\subsection{Moderately underexpanded jets in quiescent medium \label{sec:validation-quiescent-medium-moderately-underexpanded-jets}}
\begin{figure}[htbp]
    \centering
    \begin{subfigure}[t]{0.48\textwidth}
        \centering
        \includegraphics[
            width=\linewidth
        ]{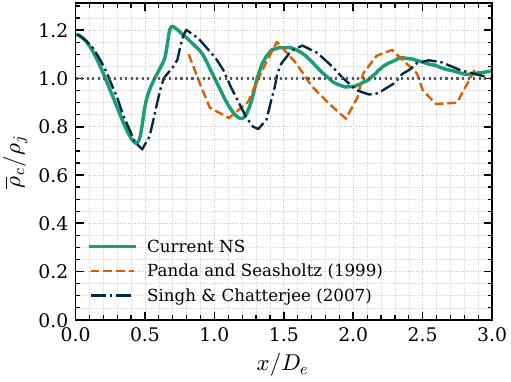}
        \caption{$M_j =1.19$}
        \label{fig:Mj-119-centerline-density}
    \end{subfigure}
    \hfill
    \begin{subfigure}[t]{0.48\textwidth}
        \centering
        \includegraphics[
            width=\linewidth
        ]{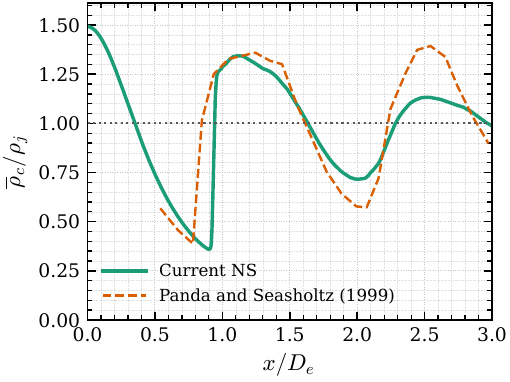}
        \caption{$M_j =1.43$}
        \label{fig:Mj-143--centerline-density}
    \end{subfigure}

    \caption{Centerline density comparison of Current Navier-Stokes (NS) solution for (a) $M_j=1.19$ with \citet{panda1999measurement} (Experiment) and \citet{singh2007numerical} (LES) profiles, and (b) $M_j=1.43$ with \citet{panda1999measurement} (Experiment) results.}
    \label{fig:validation-quiescent-medium-moderately-underexpanded-jets}
\end{figure}

Centerline density obtained from the present simulations for moderately underexpanded jets corresponding to fully expanded Mach numbers $M_j = 1.19$ and $1.43$, which correspond to NPR values of 2.39 and 3.32 for a convergent choked nozzle, are compared with the results of \citet{panda1999measurement} (experimental) and \citet{singh2007numerical} (large eddy simulation (LES)). The comparisons are shown in Fig.~\ref{fig:validation-quiescent-medium-moderately-underexpanded-jets}, where the time-averaged centerline density normalized by the fully expanded jet density ($\rho_j$) 
from the present simulations agrees well with the results reported in the literature.

\subsection{Highly underexpanded jets in quiescent medium \label{sec:validation-quiescent-medium-highly-underexpanded-jets}}
\begin{figure}[htbp]
    \centering
    \begin{subfigure}[t]{0.48\textwidth}
        \centering
        \includegraphics[
            width=\linewidth
        ]{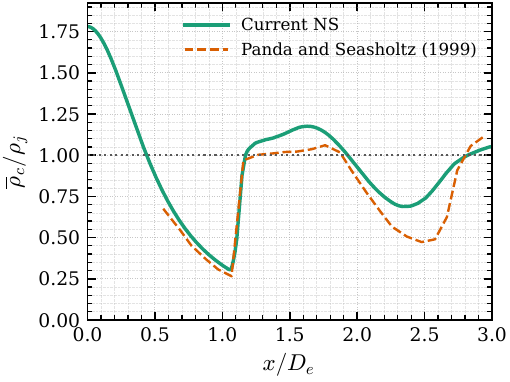}
        \caption{$M_j =1.6$}
        \label{fig:Mj-16-centerline-density}
    \end{subfigure}
    \hfill
    \begin{subfigure}[t]{0.48\textwidth}
        \centering
        \includegraphics[
            width=\linewidth
        ]{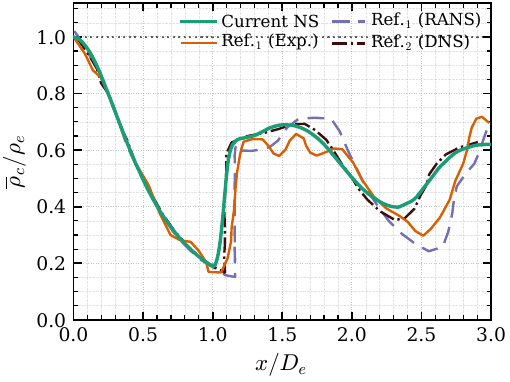}
        \caption{$M_j =1.56$}
        \label{fig:Mj-56--centerline-density}
    \end{subfigure}

    \caption{Centerline density comparison of current NS solution for (a) $M_j=1.6$ with \citet{panda1999measurement}, and (b) $M_j=1.56$ with \citet{sugawara2020three} (Ref.$_1$) and \citet{muraoka2022onset} (Ref.$_2$).}
    \label{fig:panda-sugawara-validation-quiescent-medium-highly-underexpanded-jets}
\end{figure}

For highly underexpanded jets, results from \citet{panda1999measurement} (experimental), \citet{sugawara2020three} (experimental and RANS), and \citet{muraoka2022onset} (direct numerical simulation) are considered. Centerline density results from \citet{panda1999measurement} for $M_j = 1.6$, corresponding to an NPR value of 4.25, are compared with the current numerical solution in Fig.~\ref{fig:Mj-16-centerline-density}; a good match is observed in both the location ($x/D_e = 1.1$) and the strength of the Mach disk. Similarly, a comparison of time-averaged centerline density profiles normalized by the non-dimensional nozzle-exit jet density ($\rho_e$) with experimental and RANS results from \citet{sugawara2020three} and DNS results from \citet{muraoka2022onset} is shown in Fig.~\ref{fig:Mj-56--centerline-density}, where an excellent agreement is again observed in the Mach-disk location and the density jump across it.
It should be noted that as discussed in Section \ref{subsec:initial-boundary-conditions}, agreement in normal shock strength with the experimental results was obtained for an inlet condition featuring a thin shear layer, corresponding to a momentum thickness of $2\%$ of the nozzle exit radius. Accordingly, this shear-layer thickness is adopted for all subsequent numerical simulations. For centerline results downstream of the Mach disk, there appears to be a significant discrepancy compared to experimental results, which has previously been attributed to the loss of axisymmetry downstream of the Mach disk \cite{muraoka2022onset}. However, since the present study focuses on the analysis of the shock structures in the first shock cell, the axisymmetry assumption is deemed sufficient to produce reliable results.
\begin{figure}[htbp]
    \centering
    \begin{subfigure}[t]{0.24\textwidth}
        \centering
        \includegraphics[
            width=\linewidth
        ]{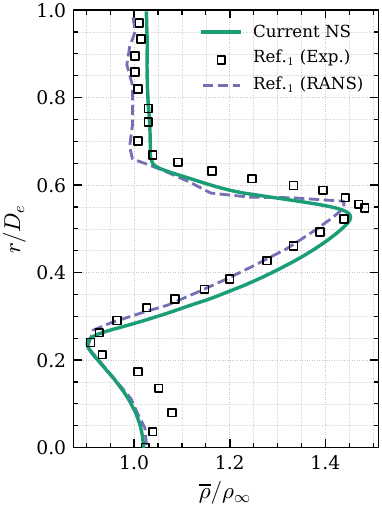}
        \caption{$x/D_e =0.6$}
        \label{fig:x_D_0_6}
    \end{subfigure}
    \hfill
    \begin{subfigure}[t]{0.24\textwidth}
        \centering
        \includegraphics[
            width=\linewidth
        ]{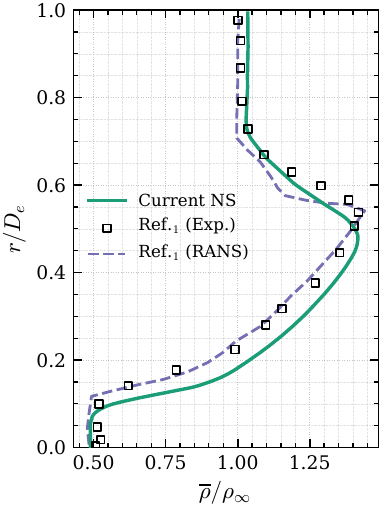}
        \caption{$x/D_e =1.0$}
        \label{fig:x_D_1_0}
    \end{subfigure}
    \hfill
    \begin{subfigure}[t]{0.24\textwidth}
        \centering
        \includegraphics[
            width=\linewidth
        ]{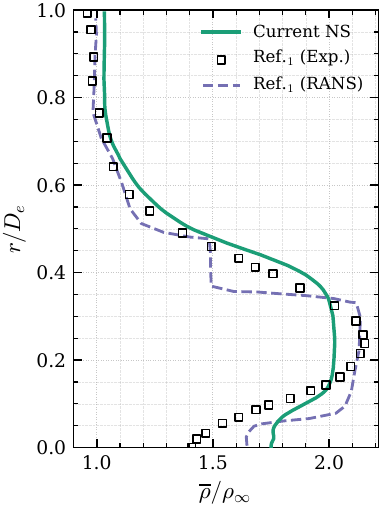}
        \caption{$x/D_e =1.5$}
        \label{fig:x_D_1_5}
    \end{subfigure}
    \hfill
    \begin{subfigure}[t]{0.24\textwidth}
        \centering
        \includegraphics[
            width=\linewidth
        ]{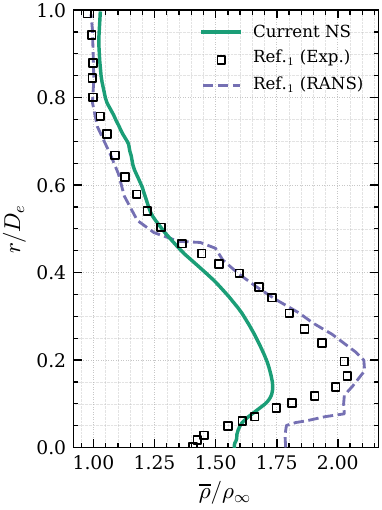}
        \caption{$x/D_e =1.8$}
        \label{fig:x_D_1_8}
    \end{subfigure}
    \caption{Radial density profile comparison at (a) $x/D_e=0.6$, (b) $x/D_e=1.0$, (c) $x/D_e=1.5$ and (d) $x/D_e=1.8$ with \citet{sugawara2020three} (Experimental and RANS solutions).}
    \label{fig:sugawara-radial-validation-quiescent-medium-highly-underexpanded-jets}
\end{figure}
In addition to the centerline density, radial density profiles normalized by ambient density ($\rho_\infty$) from the current numerical simulation are compared with those available in \cite{sugawara2020three}, as shown in Fig.~\ref{fig:sugawara-radial-validation-quiescent-medium-highly-underexpanded-jets}. A good agreement with the literature is observed for radial locations upstream of the Mach disk ($x/D_e \simeq 1.1$), while some deviations are observed for locations downstream of the Mach disk.

\subsection{Transition NPR in quiescent medium \label{sec:validation-quiescent-medium-transition-NPR}}
\begin{figure}[htbp]
    \centering
    \begin{subfigure}[t]{0.48\textwidth}
        \centering
        \includegraphics[width=\linewidth]{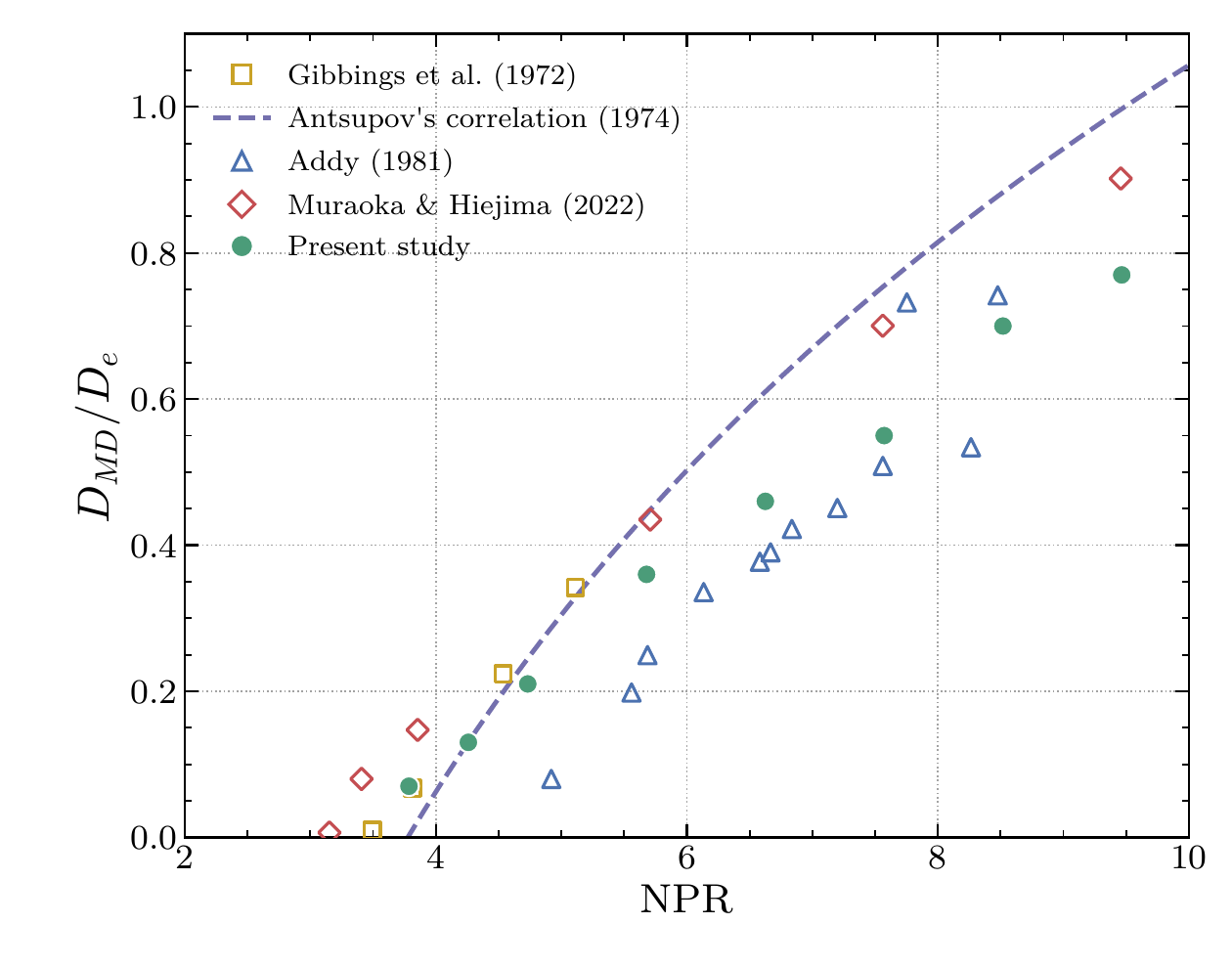}
        \caption{}
        \label{fig:mach-disk-diameter-comparison-quiescent}
    \end{subfigure}
    \hfill
    \begin{subfigure}[t]{0.48\textwidth}
        \centering
        \includegraphics[width=\linewidth]{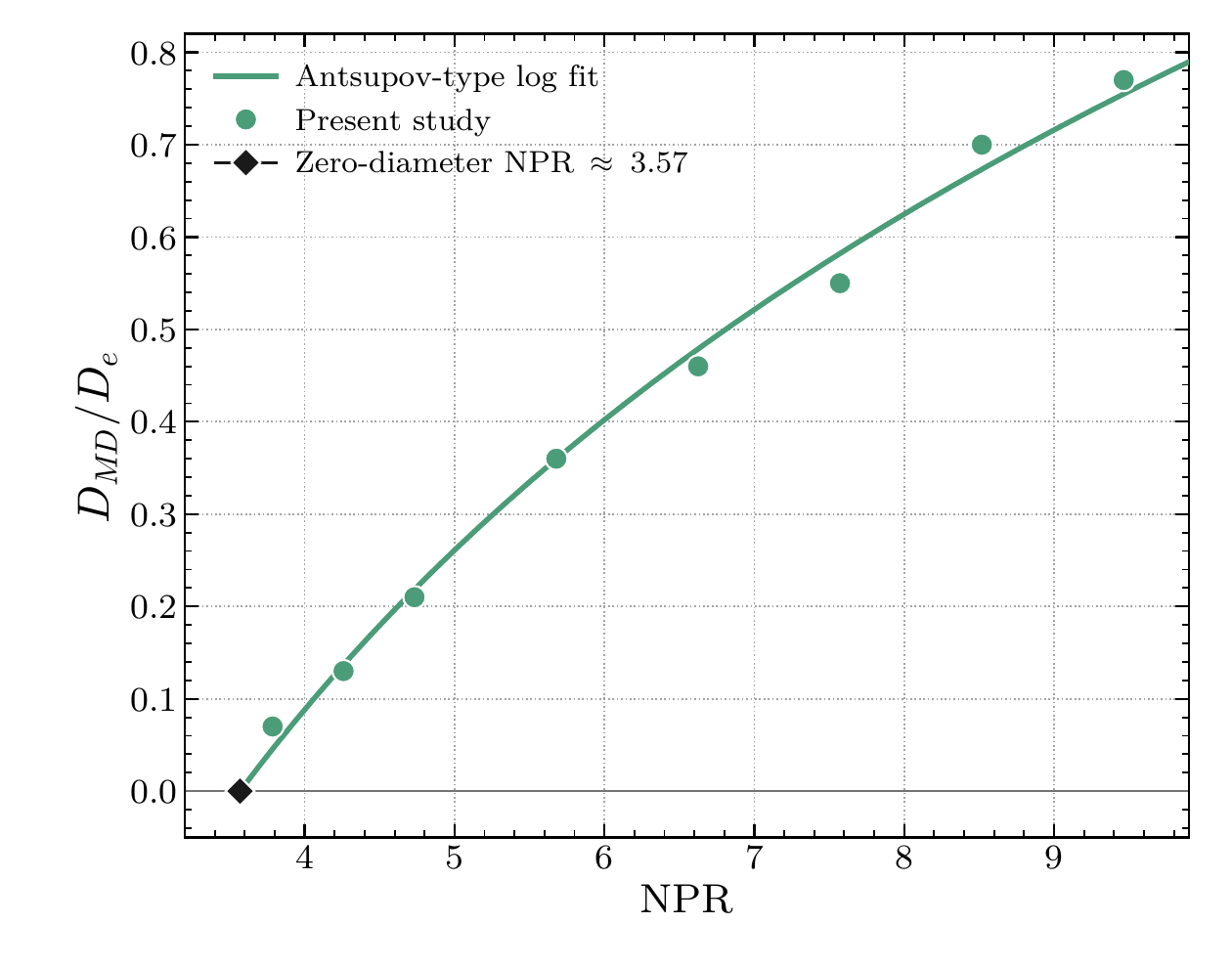}
        \caption{}
        \label{fig:transition-npr-extrapolation-quiescent}
    \end{subfigure}
    \caption{Mach-disk diameter for underexpanded jets in a quiescent medium: (a) the present $D_{MD}/D_e$ compared with the correlation of \citet{antsupov1974properties} and the data of \citet{addy1981effects}, Gibbings \textit{et al.}~\cite{gibbings1972flow}, and \citet{muraoka2022onset}; (b) extrapolation of the present data to zero Mach-disk diameter using the Antsupov-type logarithmic fit of Eq.~\eqref{eq:antsupov-type-fit}, giving a transition NPR of 3.57.}
    \label{fig:mach-disk-diameter-quiescent}
\end{figure}

The Mach-disk diameter obtained from the current simulations of the underexpanded jets in a quiescent medium is compared with the values reported in the literature \cite{antsupov1974properties, crist1966study, addy1981effects, lee2004supersonic, muraoka2022onset} in Fig.~\ref{fig:mach-disk-diameter-quiescent}(a). Transition NPR, as mentioned in Section~\ref{sec:Introduction}, is the NPR at which the regular reflection at the centerline transitions to Mach reflection, and can be quantified by the NPR where the Mach disk diameter reduces to zero (or the Mach disk forms). Based on this definition, Mach-disk formation has been observed at NPR values of 3.78 in \cite{antsupov1974properties}, 3.9 in \cite{crist1966study}, 3.67 in \cite{addy1981effects}, and 3.79 in \cite{lee2004supersonic}, among others. In contrast, \citet{muraoka2022onset}, from their direct numerical simulation of axisymmetric jets, found the transition to occur at a lower NPR range of approximately 3.08 to 3.12.

The current simulations exhibited transition between NPR $=3.31$ and $3.79$. Specifically, a regular reflection with no Mach-disk was obtained at NPR $=3.31$, whereas a Mach reflection with a finite Mach-disk appeared at NPR $=3.79$ (see the density gradient plots for the $U_c=0.0$ column in Appendix~\ref{app:coflow-results}), for the parameter space presented in Section~\ref{sec:parameter-space}. To locate the transition within this interval, the present Mach-disk diameter data is fitted with a logarithmic expression of the same form as that proposed by \citet{antsupov1974properties},
\begin{equation}
    \frac{D_{MD}}{D_e} = 1.78\,\log_{10}(\mathrm{NPR}) - 0.98, \qquad R^2 = 0.99,
    \label{eq:antsupov-type-fit}
\end{equation}
and extrapolated to zero Mach-disk diameter, as shown in Fig.~\ref{fig:mach-disk-diameter-quiescent}(b), yielding a transition NPR of $3.57$. In \eqref{eq:antsupov-type-fit}, $R^2$ denotes the coefficient of determination, whose values range from 0 to 1; a value of 0 indicates a poor model that captures none of the variability in the dependent variable, and a value of 1 indicates that the model captures 100\% of the variability, providing a perfect fit. The extrapolated NPR value (for zero Mach-disk diameter) falls in the range $[3.31,\,3.79]$, which is well within the range reported in the literature.

\section{Parameter space \label{sec:parameter-space}}
The parameter space chosen for this study is given in Table \ref{tab:parameter-space}. It spans NPR values from moderately underexpanded jet to highly underexpanded jet. A choked convergent nozzle is assumed; therefore, the Mach number at the nozzle exit is sonic (\textit{i.e.}, $M_e = 1.0$). Nozzle temperature ratio $\textrm{NTR} = 1.2$ is considered, which for a convergent choked nozzle with isentropic flow inside the nozzle and $\gamma = 1.4$ leads to 
the same nozzle-exit and ambient temperature, \textit{i.e.,} $\frac{T_e}{T_\infty}=1.0$. 


Subsonic coflows are imposed as ambient flow moving at a fraction of the nozzle exit velocity, \textit{i.e.,} $U_c = \frac{U_a}{U_e}$, where $U_a$ is the velocity of ambient flow and $U_e$ is the velocity of the jet at nozzle exit. $U_c$ varies from 0.0 to 0.8, representing nozzles in ground test to in-flight conditions.

\begin{table}[!htbp]
\centering
\caption{Parameter space considered in the present study.}
\label{tab:parameter-space}
\renewcommand{\arraystretch}{1.2}
\begin{tabular}{|c|c|c|c|c|c|c|c|c|c|c|}
\hline
\textbf{Parameter} & \multicolumn{10}{c|}{\textbf{Values}} \\ \hline
$\mathrm{PR}$ 
& 1.5 & 1.75 & 2.0 & 2.25 & 2.5 & 3.0 & 3.5 & 4.0 & 4.5 & 5.0 \\ \hline
$\mathrm{NPR}$ 
& 2.84 & 3.31 & 3.79 & 4.26 & 4.73 & 5.68 & 6.63 & 7.57 & 8.52 & 9.47 \\ \hline
$U_c=\dfrac{U_a}{U_e}$ 
& \multicolumn{10}{l|}{$0.0,\ 0.2,\ 0.4,\ 0.6,\ 0.8$} \\ \hline
\multicolumn{11}{|l|}{Combination: All possible $\mathrm{NPR}$--$U_c$ combinations} \\ \hline
\end{tabular}
\end{table}

\section{Numerical results and analysis\label{sec:Numerical-results}}
This section examines how subsonic coflow modifies the near-field shock structures of an underexpanded jet, and in particular the formation of the Mach disk. The detailed examination and the supporting analysis are presented for the representative NPR = 4.26 (PR = 2.25) jet, which shows the clearest transition from Mach reflection at zero coflow to regular reflection at high coflows and is therefore well suited to isolating the role of the coflow. The resulting trends are then quantified over the full range of NPR and coflow velocities considered in this study (see Table \ref{tab:parameter-space}). We first describe the near-field shock structures of this representative jet and define the quantities used to characterize the Mach disk. We then track how the shock structures transition, both for this jet and across the entire NPR--coflow parameter space, as the coflow velocity is increased. To explain this transition, we examine the effect of coflow on the inclination of the jet boundary at the nozzle lip 
and the pressure along the boundary. A method-of-characteristics (MOC) analysis then identifies the boundary pressure as the primary inviscid mechanism of Mach-disk suppression, which also reduces the jet boundary inclination at the nozzle lip. We then characterize the effect of coflow on the first shock-cell geometry and finally assess the role of viscous effects.

\subsection{Near-field shock structures for NPR = 4.26 (PR = 2.25) jet \label{sec:near-field-structures}}

\begin{figure}[htbp]
    \centering

    \begin{subfigure}[t]{0.48\textwidth}
        \centering
        \includegraphics[
            width=\linewidth,
            trim={1.0cm 1.0cm 8.0cm 1.0cm},
            clip
        ]{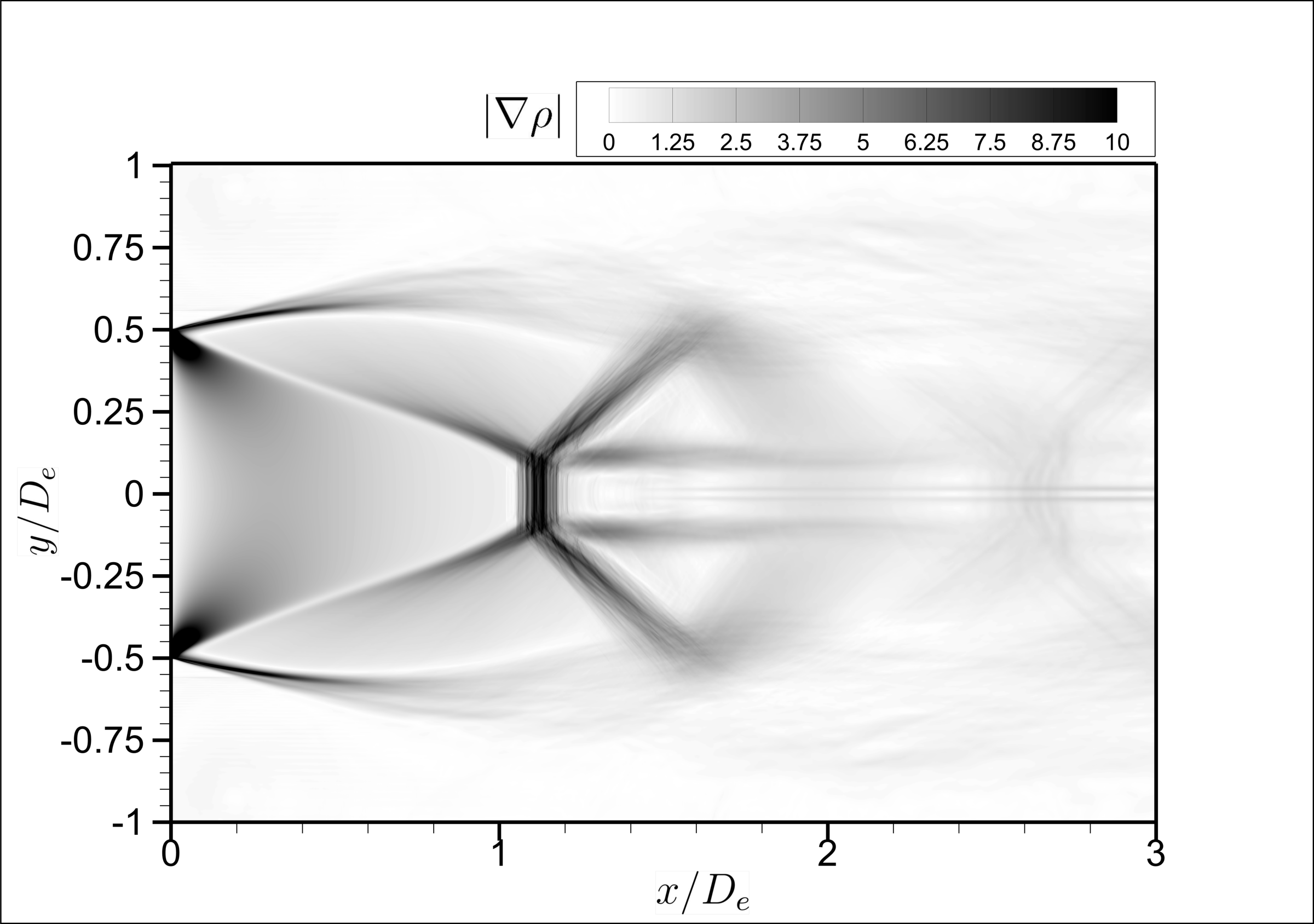}
        \caption{$U_c=0.0$}
        \label{fig:p225-densitygrad-c0}
    \end{subfigure}
    \begin{subfigure}[t]{0.48\textwidth}
        \centering
        \includegraphics[
            width=\linewidth,
            trim={1.0cm 1.0cm 8.0cm 1.0cm},
            clip
        ]{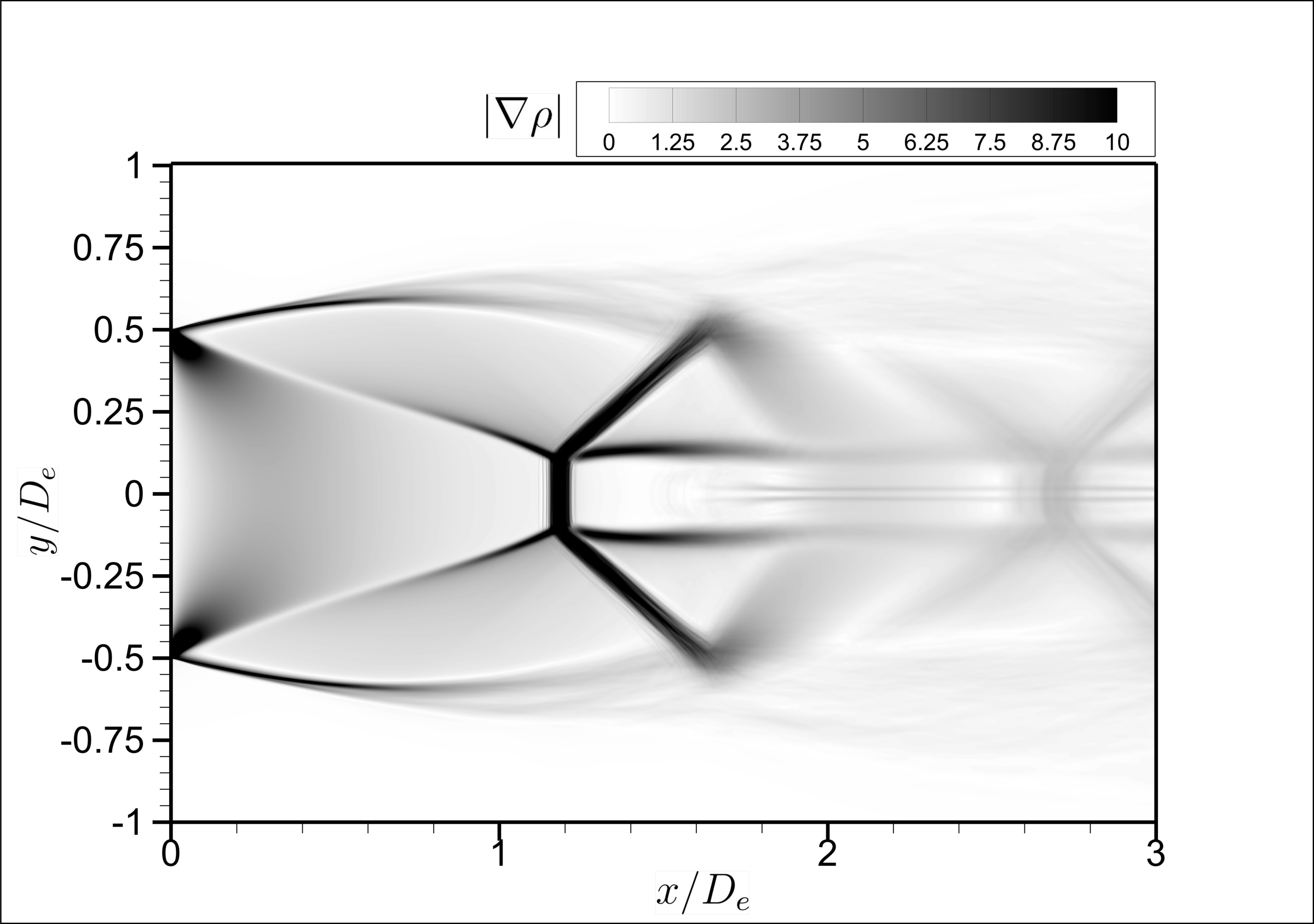}
        \caption{$U_c=0.2$}
        \label{fig:p225-densitygrad-c2}
    \end{subfigure}

    \begin{subfigure}[t]{0.48\textwidth}
        \centering
        \includegraphics[
            width=\linewidth,
            trim={1.0cm 1.0cm 8.0cm 1.0cm},
            clip
        ]{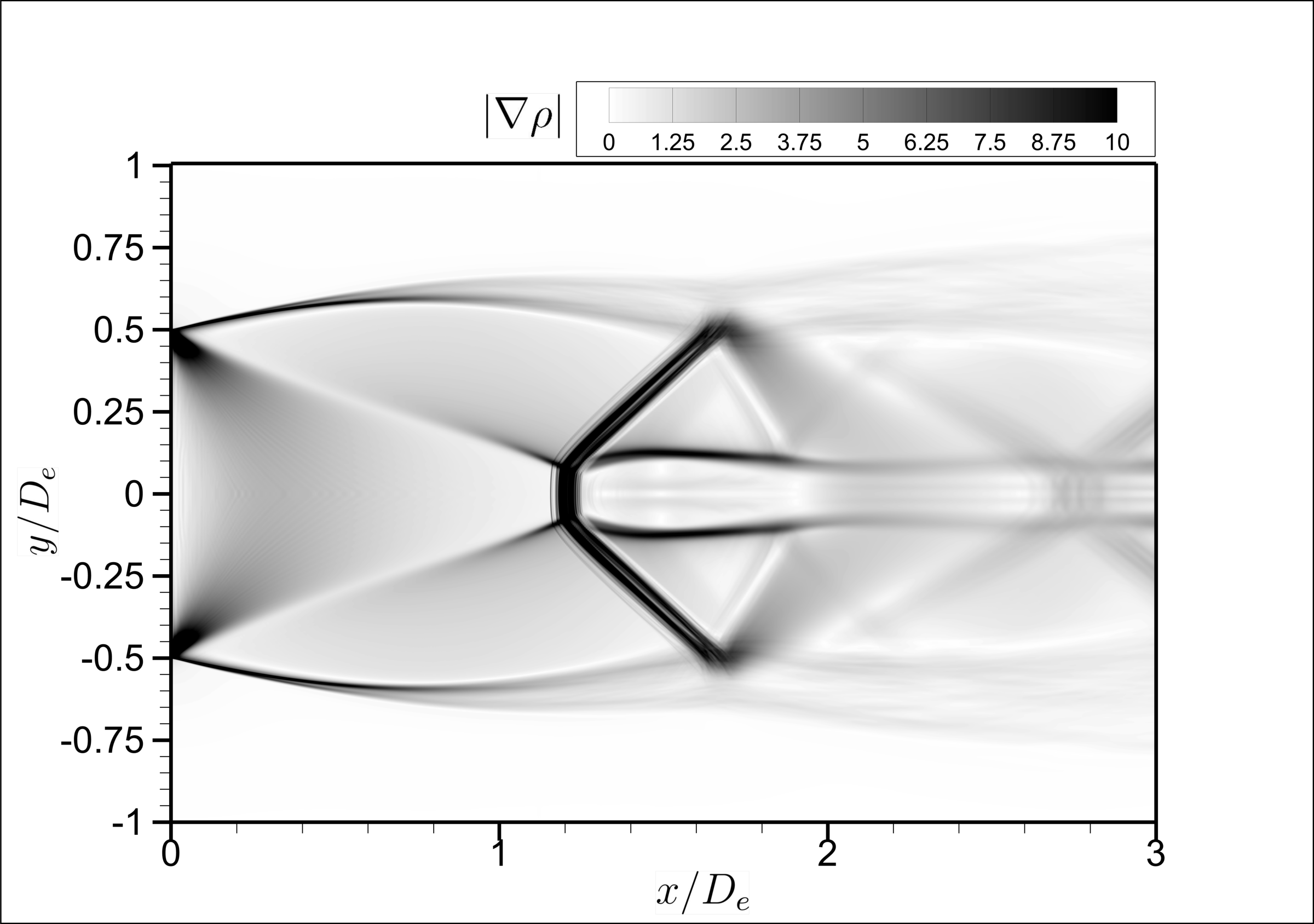}
        \caption{$U_c=0.4$}
        \label{fig:p225-densitygrad-c4}
    \end{subfigure}
    \begin{subfigure}[t]{0.48\textwidth}
        \centering
        \includegraphics[
            width=\linewidth,
            trim={1.0cm 1.0cm 8.0cm 1.0cm},
            clip
        ]{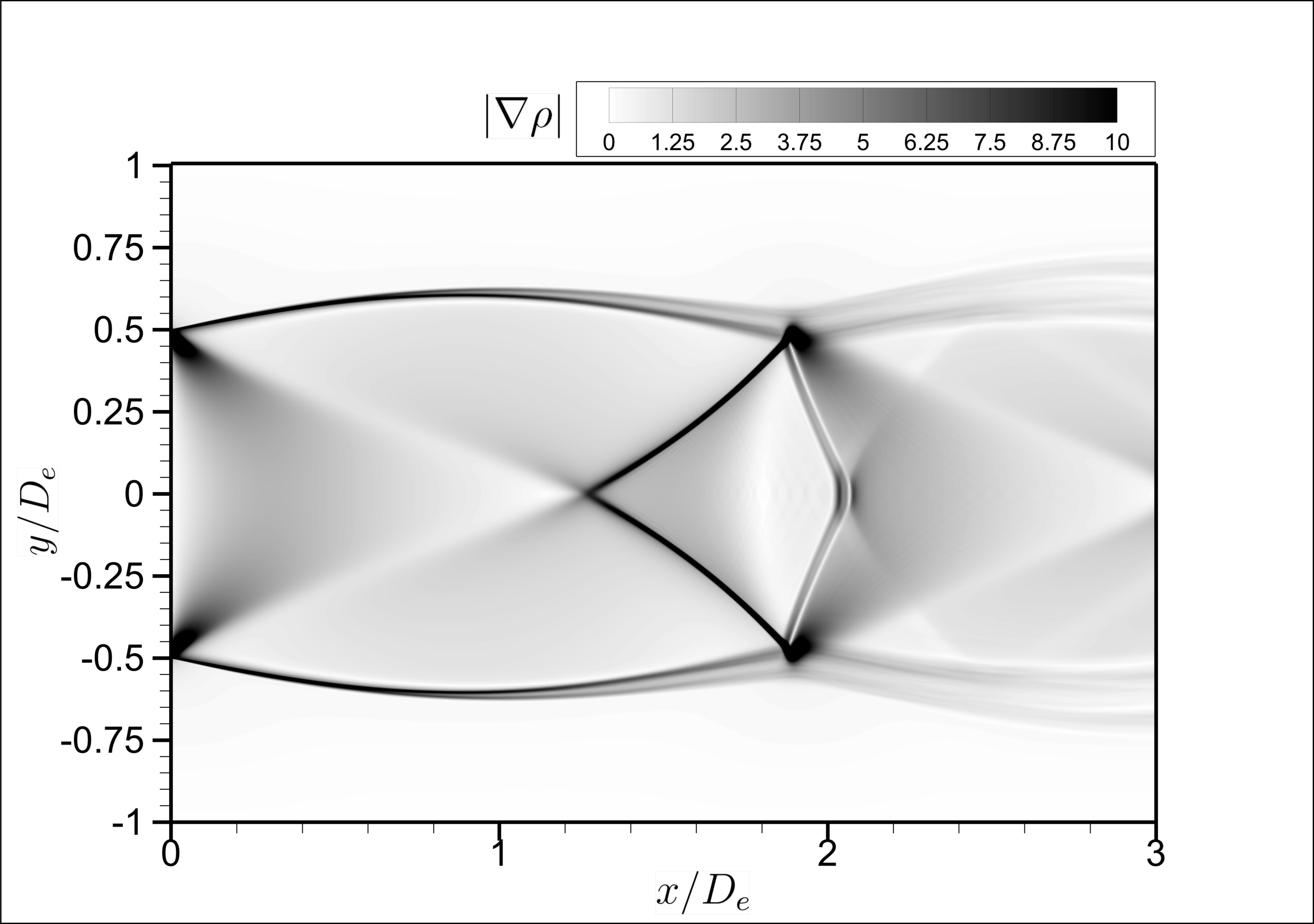}
        \caption{$U_c=0.6$}
        \label{fig:p225-densitygrad-c6}
    \end{subfigure}
    
    \begin{subfigure}[t]{0.48\textwidth}
        \centering
        \includegraphics[
            width=\linewidth,
            trim={1.0cm 1.0cm 8.0cm 1.0cm},
            clip
        ]{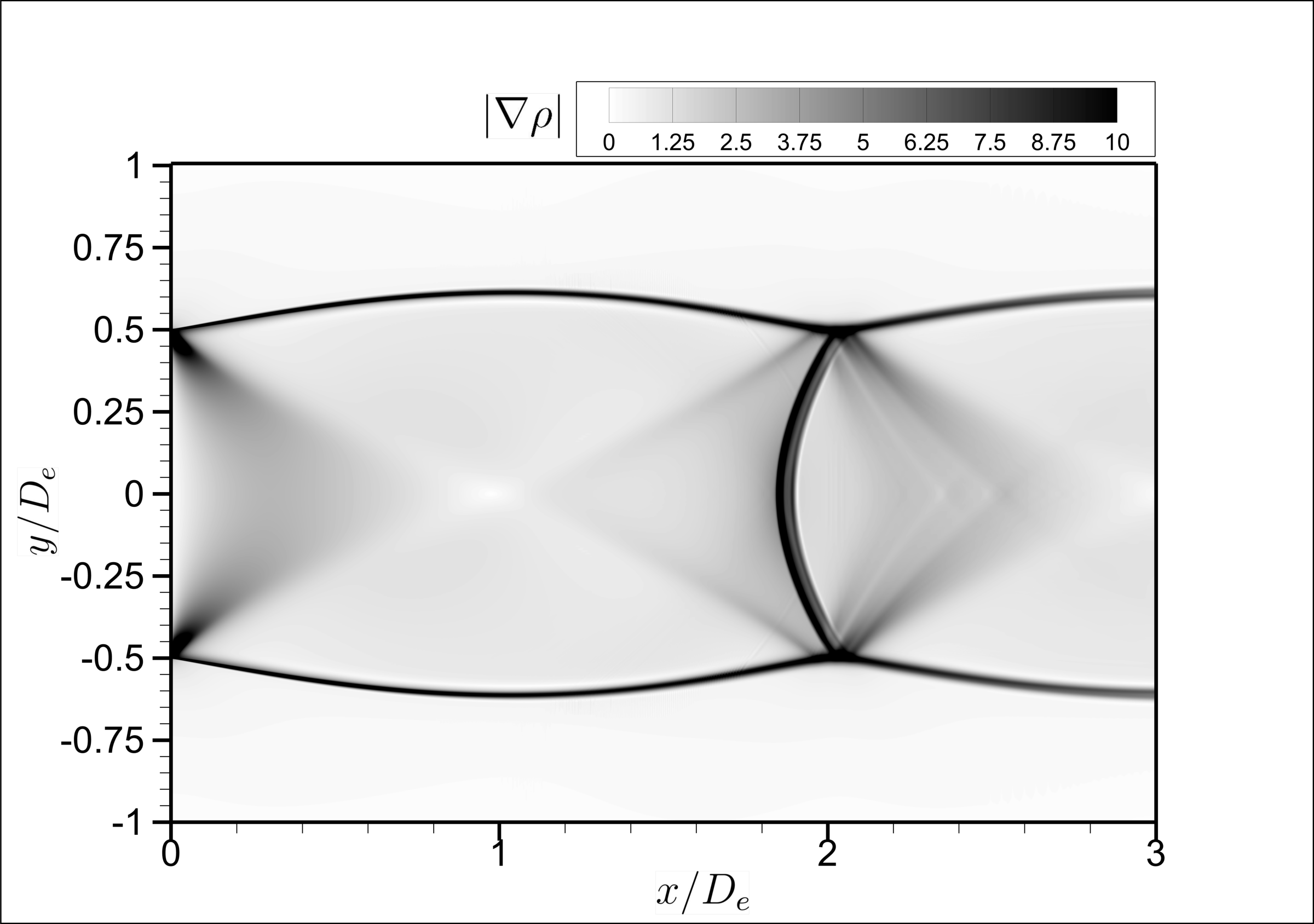}
        \caption{$U_c=0.8$}
        \label{fig:p225-densitygrad-c8}
    \end{subfigure}

    \caption{Comparison of time-averaged density-gradient contours for underexpanded jets at NPR = 4.26 (PR = 2.25) with increasing coflow ratio $U_c$.}
    \label{fig:PR2.25-coflowing-jets}
\end{figure}

\begin{figure}[htbp]
    \centering
    
    \begin{subfigure}[t]{0.48\textwidth}
        \centering
        \includegraphics[
            width=\linewidth
        ]{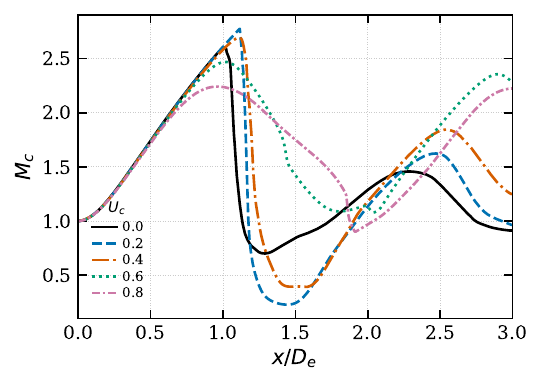}
        \caption{Centerline Mach number}
        \label{fig:p225-centerline-mach}
    \end{subfigure}
    \begin{subfigure}[t]{0.48\textwidth}
        \centering
        \includegraphics[
            width=\linewidth
        ]{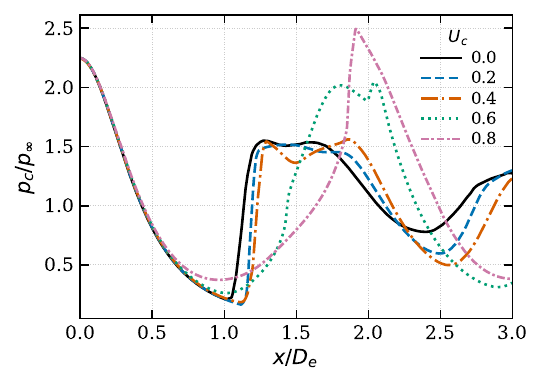}
        \caption{Centerline static pressure}
        \label{fig:p225-centerline-pressure}
    \end{subfigure}
    \begin{subfigure}[t]{0.48\textwidth}
        \centering
        \includegraphics[
            width=\linewidth
        ]{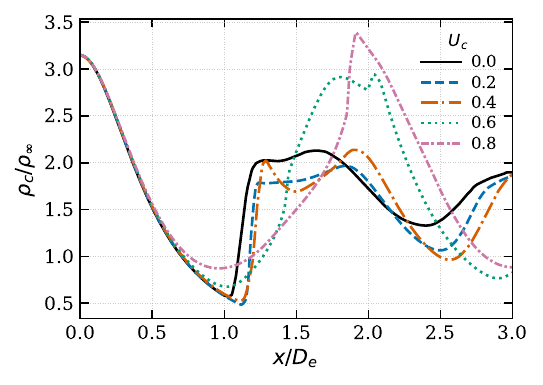}
        \caption{Centerline density}
        \label{fig:p225-centerline-density}
    \end{subfigure}

    \caption{Centerline flow properties for NPR = 4.26 (PR = 2.25) jet for coflow velocity ratios ranging from $U_c=0.0$ to $U_c=0.8$: (a) Mach number, (b) static pressure, and (c) density.}
    \label{fig:p225-centerline-properties}
\end{figure}

The effect of NPR on the transition of the near-field shock structures from regular to Mach reflection was discussed in Section~\ref{sec:validation-quiescent-medium-transition-NPR} for a jet expanded to quiescent ambient. In this section, we investigate the effect of coflow on this transition. The contours of density gradient magnitude (numerical schlieren) for jets with NPR = 4.26 (PR = 2.25) at different ambient coflows are presented in Fig.~\ref{fig:PR2.25-coflowing-jets}. A Mach disk is observed around $x/D \simeq 1.12$ for a quiescent ambient, which agrees well with the location reported by \citet{panda1999measurement} for NPR = 4.25 (PR = 2.25) jet. The centerline Mach number plot in Fig.~\ref{fig:p225-centerline-mach} further confirms the presence of the Mach disk at this location, where a large jump in Mach number from supersonic ($M_c \simeq 2.2$) to subsonic ($M_c \simeq 0.4$) is observed. The drop in Mach number across the normal shock is accompanied by a jump in both static pressure and density, as shown in Figs.~\ref{fig:p225-centerline-pressure} and \ref{fig:p225-centerline-density}, respectively. Before examining the influence of coflow conditions on this shock structure, the method used to quantify the Mach disk diameter is first described, which is then used as the main parameter to quantify the transition NPR.

\begin{figure}[htbp]
    \centering

    \begin{subfigure}[t]{0.48\textwidth}
        \centering
        \includegraphics[
            width=\linewidth
        ]{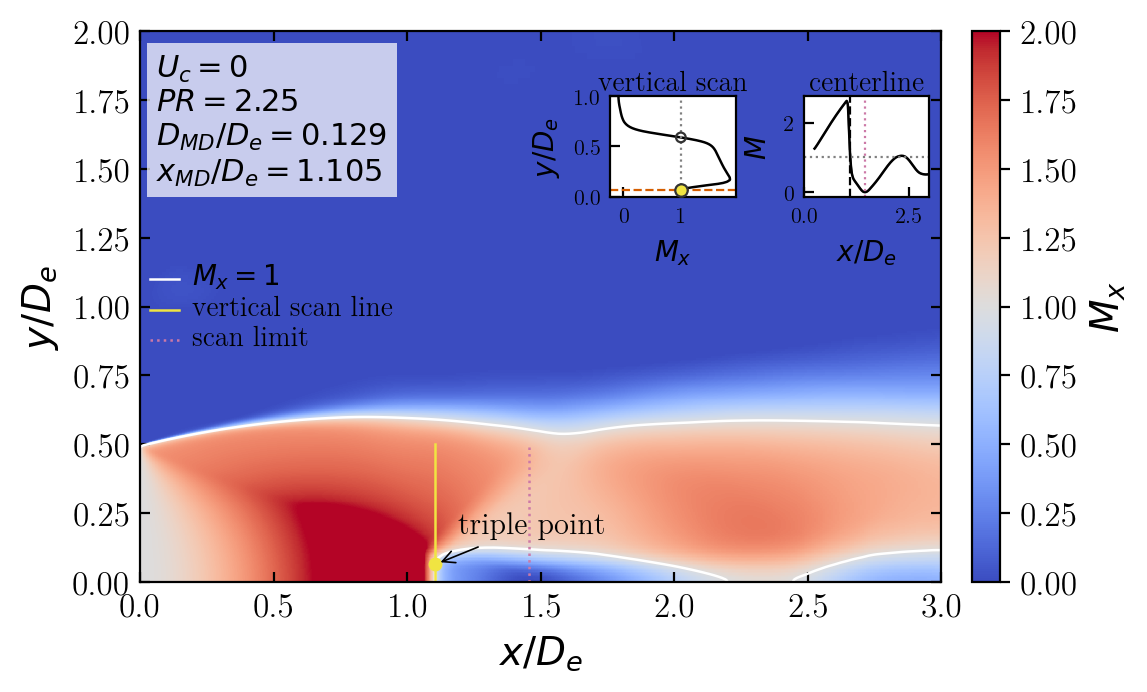}
        \caption{$U_c=0.0$}
        \label{fig:c0p225-mach-disk-detection}
    \end{subfigure}
    \begin{subfigure}[t]{0.48\textwidth}
        \centering
        \includegraphics[
            width=\linewidth
        ]{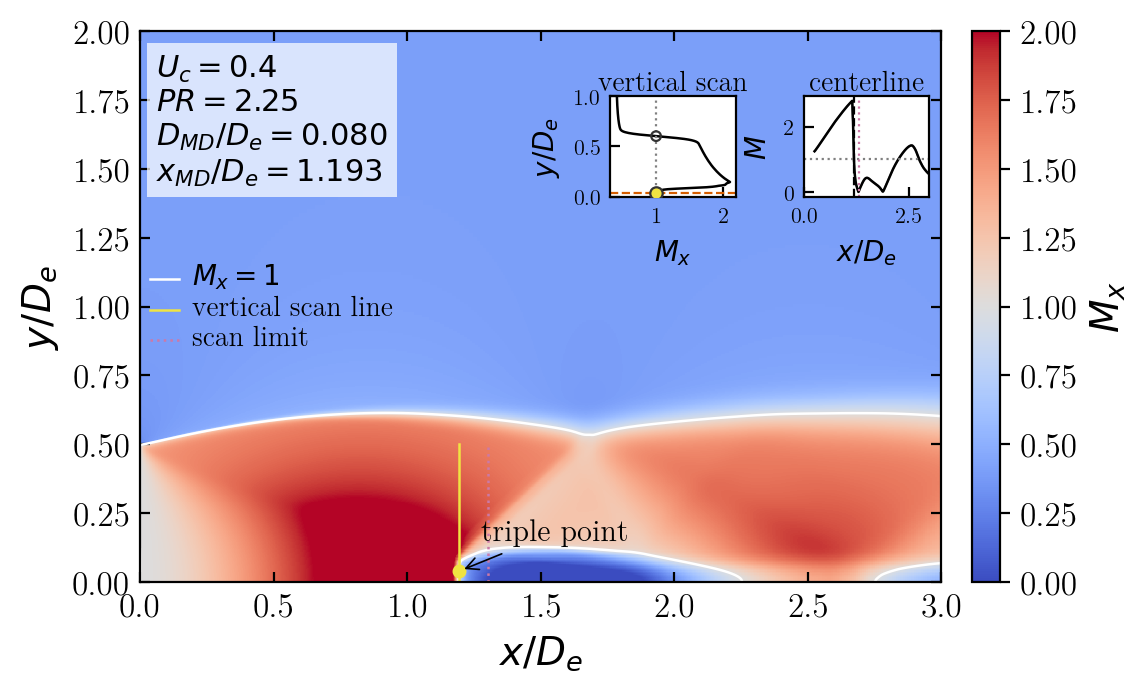}
        \caption{$U_c=0.4$}
        \label{fig:c4p225-mach-disk-detection}
    \end{subfigure}
    \begin{subfigure}[t]{0.48\textwidth}
        \centering
        \includegraphics[
            width=\linewidth
        ]{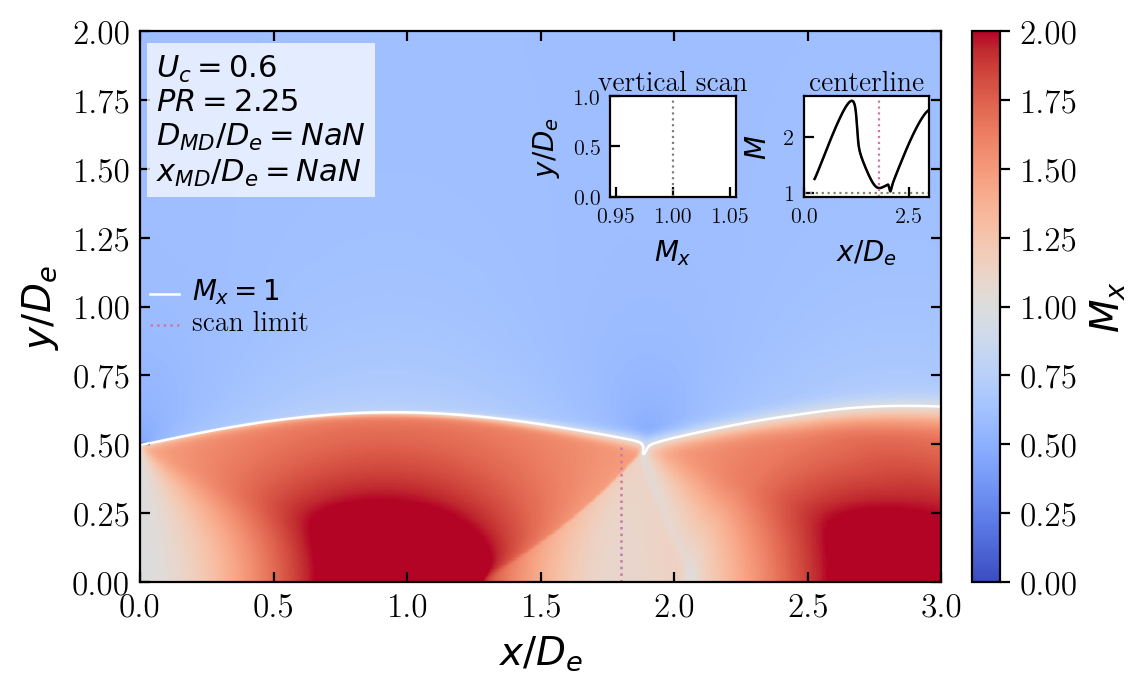}
        \caption{$U_c=0.6$}
        \label{fig:c6p225-mach-disk-detection}
    \end{subfigure}

    \caption{Mach-disk diameter detection for NPR = 4.26 (PR = 2.25) at selected coflow velocity ratios: (a) $U_c=0.0$, (b) $U_c=0.4$, and (c) $U_c=0.6$. The first two cases have Mach disks of diameters $0.129D_e$ and $0.08D_e$ at axial locations $1.105D_e$ and $1.193D_e$ respectively, while the last case has no Mach disk.}.
    \label{fig:p225-mach-disk-detection}
\end{figure}

The Mach disk location is defined as the $x-$coordinate along the jet centerline where the local axial Mach number ($M_x$) first reaches the sonic speed. The diameter of the Mach disk is then determined using the triple point location (see Fig.~\ref{fig:mach-reflection-schematic}), behind which a slipstream emerges, separating the subsonic flow behind the Mach disk from the supersonic flow behind the reflected shock. Hence, plotting $M_x$ along a vertical line drawn at an immediate downstream location of the Mach disk, the triple point is identified as the radial location where $M_x$ is sonic. This is better visualized in Fig.~\ref{fig:p225-mach-disk-detection}. In this figure, the vertical line is marked as the vertical scan line, and the axial Mach number along it is shown in the vertical scan inset, where the triple point is identified by the yellow marker as the first sonic crossing. The centerline inset shows the local Mach number along the jet axis, which is used to define the scan limit. The scan limit is taken as the first axial location at which the local Mach number attains a local minimum, and the detection of the Mach disk is confined upstream of it. The Mach disk diameter is then twice the radial distance of the triple point from the centerline. Using this method, the Mach disk for the NPR = 4.26 (PR = 2.25) jet in a quiescent ambient, as seen in Fig.~\ref{fig:c0p225-mach-disk-detection}, is found to be located at $1.105 D_e$ from the nozzle exit and has a diameter of $0.129 D_e$.

Having established the method for quantifying the Mach disk location and diameter, we now examine the effect of increasing the coflow velocity on the shock structure. As the coflow velocity increases from quiescent to higher subsonic speeds, it can be visually observed from Fig.~\ref{fig:PR2.25-coflowing-jets} that the 
Mach disk at lower coflow speeds transitions to regular reflection (with oblique shocks) at higher coflows. This is further confirmed by the centerline plots in Fig.~\ref{fig:p225-centerline-properties}. A Mach disk is observed up to $U_c = 0.4$, where a sudden drop in Mach number to a subsonic speed occurs at the Mach disk location, accompanied by a sharp jump in static pressure (Fig.~\ref{fig:p225-centerline-pressure}) and density (Fig.~\ref{fig:p225-centerline-density}). At a higher coflow speed ($U_c = 0.6$), 
the embedded shock reflects regularly from the centerline as an oblique shock instead of forming a Mach disk. The centerline Mach number in the regular reflection remains supersonic, as seen in Fig.~\ref{fig:p225-centerline-mach} for $U_c = 0.6$, further confirming the absence of a Mach disk for this case. The centerline plots in Figs.\ref{fig:p225-centerline-pressure} and \ref{fig:p225-centerline-density} for $U_c = 0.6$ show a gradual change in pressure and density around $x/D_e = 1.25$, where the reflection point lies, in contrast to the sharp jumps observed across the Mach disk in the lower coflow cases. For the highest coflow case ($U_c = 0.8$), the compression waves do not combine before reaching the centerline to form an embedded shock and therefore reflect back regularly as compression waves. Interestingly, for this case, a relatively sharp density gradient is observed at a downstream location around $x/D_e = 1.9$, as seen in Fig.~\ref{fig:p225-densitygrad-c8}. The structure attains a bow shape 
and is accompanied by jumps in both static pressure and density, as seen in Figs.\ref{fig:p225-centerline-pressure} and \ref{fig:p225-centerline-density}, respectively. The local centerline Mach number across this structure drops slightly below sonic to $\sim 0.89$, from approximately 1.25 just upstream of the structure (Fig.~\ref{fig:p225-centerline-mach}). This drop to subsonic value occurs across the entire length of the structure, suggesting that the bow-shaped structure is indeed a weak shock. Although not identical, it is worth noting that \citet{panda1998shock}, among others, reported a similar concave-shaped shock in the downstream (second or third) shock cells, attributing it to shock--turbulence interaction arising from the shock encountering the fluctuating shear layer. In the present case, however, the structure appears only at high coflow velocities, where, as shown later in Section~\ref{sec:viscous-effects}, the shear layer is confined to an extremely thin region and the corresponding instabilities are strongly suppressed. Thus, while the two structures are geometrically similar, the underlying physics is fundamentally different. To confirm that the structure is not an artifact of the axisymmetric formulation, the highest coflow case was repeated with a three-dimensional solver, which reproduces the same bow-shaped structure at the same axial location. The results from the 3D simulation for the case are shown in Fig.~\ref{fig:3D_0_8_coflow_compare} of Appendix~\ref{app:grid_test_3_D}. A possible explanation for the formation of this structure, tied to the pressure imposed on the jet boundary by coflow, is proposed in Section~\ref{sec:jet-boundary-streamline-pressure}.

\subsection{Characterization of shock-structure transitions \label{sec:transition-characterization}}
\begin{figure}[htbp]
    \centering

    \begin{subfigure}[t]{0.50\textwidth}
        \centering
        \includegraphics[
            height=0.28\textheight,
            trim={0.2cm 0.2cm 0.2cm 0.2cm},
            clip
        ]{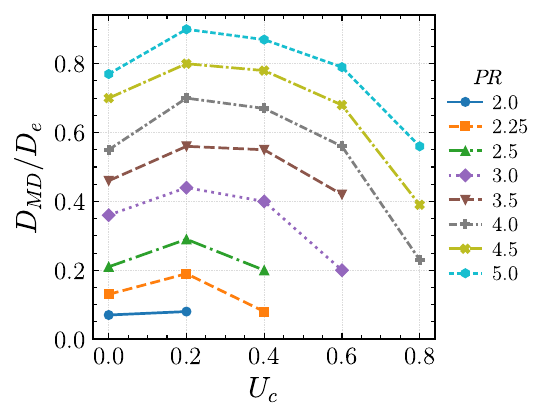}
        \caption{Mach-disk diameter variation with $U_c$ for different PR values}
        \label{fig:mach-disk-diameter-vs-coflow}
    \end{subfigure}
    \hfill
    \begin{subfigure}[t]{0.48\textwidth}
        \centering
        \includegraphics[
            height=0.28\textheight,
            trim={0.2cm 0.2cm 0.2cm 0.2cm},
            clip
        ]{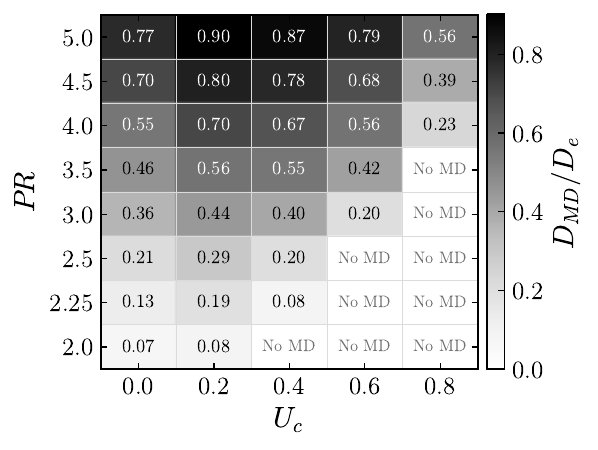}
        \caption{Mach-disk diameter map over the parameter space}
        \label{fig:mach-disk-diameter-map}
    \end{subfigure}

    \caption{Mach-disk diameter variation with coflow velocity ratio $U_c$ and PR.}
    \label{fig:mach-disk-diameter-coflow}
\end{figure}

The key observation from the above discussion is that increasing the coflow velocity leads to a transition from Mach reflection to regular reflection at the centerline for a jet that initially exhibits a Mach disk. To characterize this transition with increasing coflow, we plot the diameter of Mach disks against the coflow speed for jets at various PR values, considering only those jets that initially form a Mach disk at zero coflow \textrm{i.e.} $\mathrm{PR} \geq 2.0$ ($\mathrm{NPR} \geq 3.79$). From Fig.~\ref{fig:mach-disk-diameter-vs-coflow}, the Mach disk diameter is seen to initially increase up to around $U_c = 0.2$, independent of PR. Beyond this value of $U_c$, the diameter decreases for all pressure ratios. For jets with relatively low pressure ratios ($\mathrm{PR} \le 3.5$), the shock structure transitions from Mach to regular reflection (\textit{i.e.}, the Mach disk diameter reduces to zero) at higher coflow speeds, indicated by ``No MD'' in Fig.~\ref{fig:mach-disk-diameter-map}. 

For the quiescent ambient ($U_c = 0.0$) cases, the Mach-disk diameter in Section~\ref{sec:validation-quiescent-medium-transition-NPR} followed $D_{MD}/D_e = a\log_{10}(\mathrm{NPR})+b$ with $a = 1.78$ and $b = -0.98$. Allowing the constants $a$ and $b$ to be functions of coflow to account for its effect and refitting them at each $U_c$ shows that the slope $a$ varies linearly with $U_c$ while the intercept $b$ requires a quadratic dependence, yielding
\begin{equation}
  \frac{D_{MD}}{D_e} = \max\!\left[0,\;\left(1.78 + 1.45\,U_c\right)\log_{10}(\mathrm{NPR}) - \left(0.98 + 0.40\,U_c + 1.60\,U_c^{2}\right)\right], \quad R^2 = 0.99,
  \label{eq:coflow}
\end{equation}
which reduces identically to the no-coflow fit at $U_c = 0$ and predicts the measured points with a root mean squared error (RMSE) of $0.018$ (or $1.8\%$ of $D_e$). Coflow thus steepens the logarithmic slope linearly while lowers the intercept quadratically. A negative value is interpreted as the absence of a Mach-disk and floored to zero. Equation~\eqref{eq:coflow} correctly recovers eight of the nine ``No MD'' cases of Fig.~\ref{fig:mach-disk-diameter-map}, the exception being $(\mathrm{PR}, U_c) = (3.5, 0.8)$ where a weak disk is predicted.

It is intuitive to assume that the ambient coflow influences the centerline shock structures through the jet boundary; possibly by altering the jet boundary (reflection surface) geometry and the boundary conditions for the characteristic waves inside the supersonic core, which ultimately influence the nature of compression wave reflections from the jet boundary. These modified compression waves then determine whether regular or Mach reflection takes place at the centerline. To understand what causes the transition with coflow, we now turn to the possible mechanisms by which coflow modifies the shock structures.

\subsection{Effect of coflow on jet boundary inclination at the nozzle lip \label{sec:jet-boundary-inclination}}

\begin{figure}[htbp]
    \centering

    \begin{subfigure}[t]{0.48\textwidth}
        \centering
        \includegraphics[
            width=\linewidth
        ]{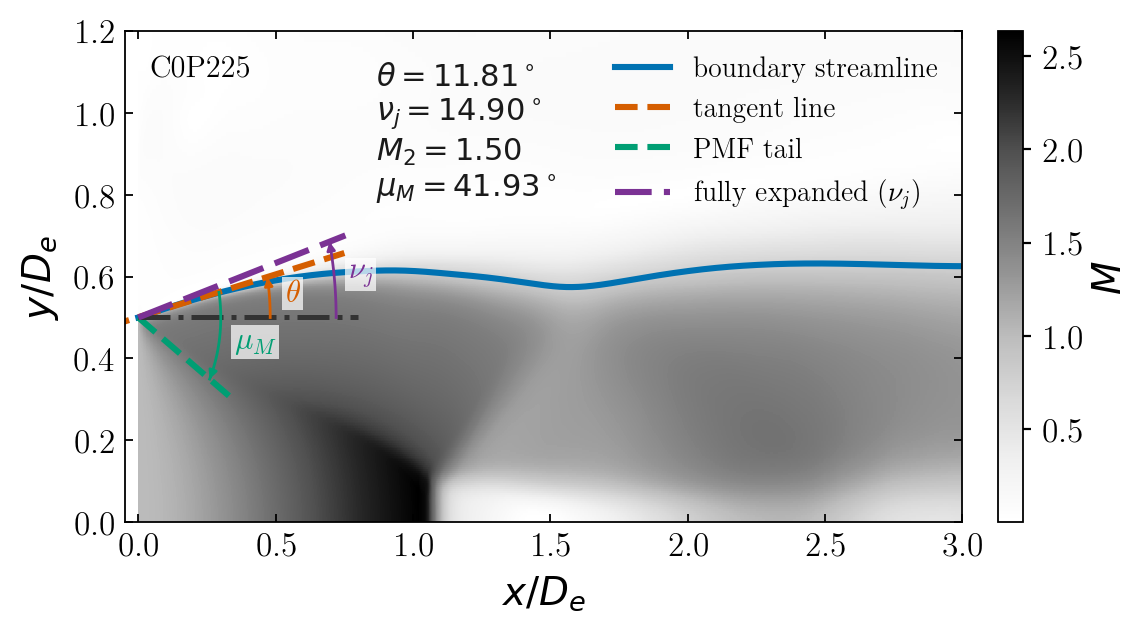}
        \caption{$U_c=0.0$}
        \label{fig:c0p225-streamline-fit-angle}
    \end{subfigure}
    \begin{subfigure}[t]{0.48\textwidth}
        \centering
        \includegraphics[
            width=\linewidth
        ]{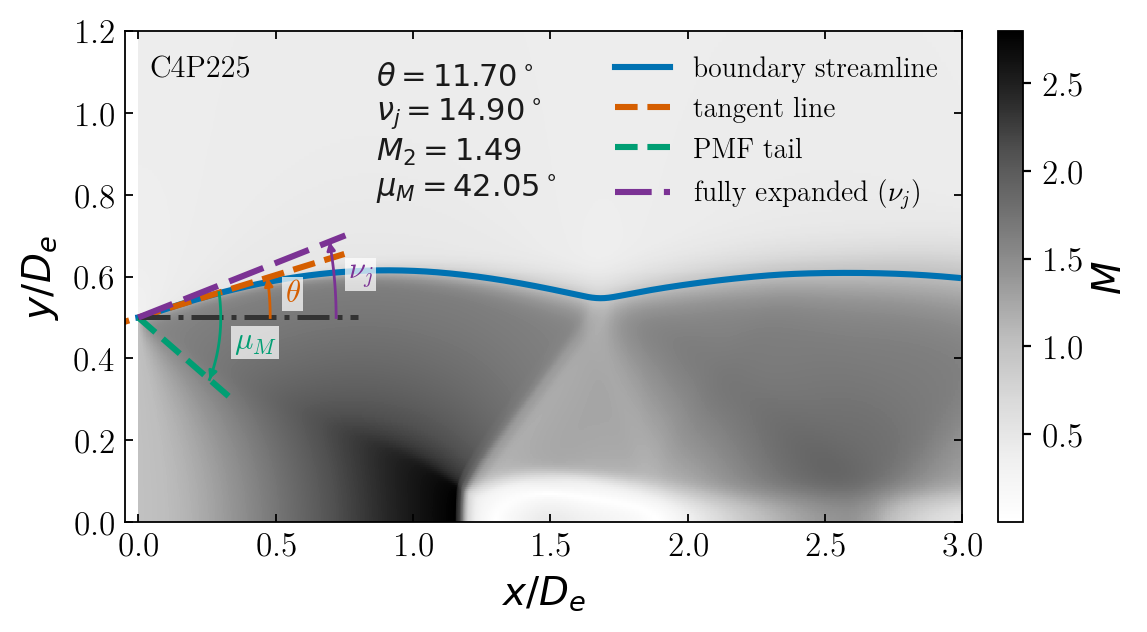}
        \caption{$U_c=0.4$}
        \label{fig:c4p225-streamline-fit-angle}
    \end{subfigure}
    \begin{subfigure}[t]{0.48\textwidth}
        \centering
        \includegraphics[
            width=\linewidth
        ]{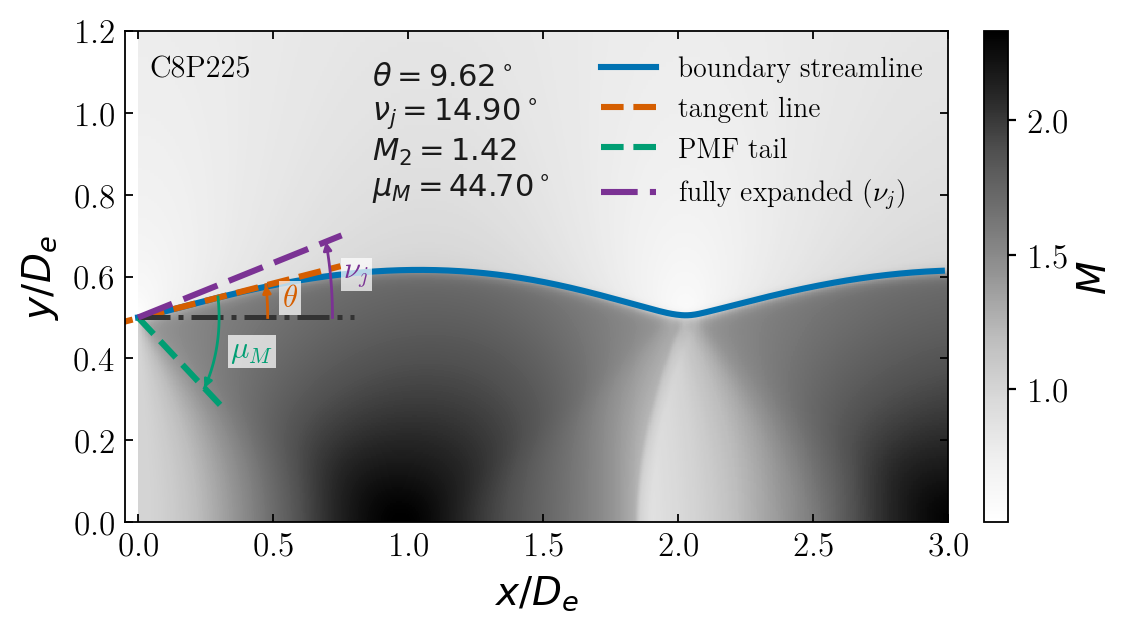}
        \caption{$U_c=0.8$}
        \label{fig:c8p225-streamline-fit-angle}
    \end{subfigure}

    \caption{Initial inclination of jet boundary and Prandtl--Meyer fan (PMF) tail for NPR = 4.26 (PR = 2.25) at selected coflow velocity ratios.}
    \label{fig:p225-boundary-initial-and-pmf-angle}
\end{figure}

\begin{figure}[htbp]
    \centering

    \begin{subfigure}[t]{0.48\textwidth}
        \centering
        \includegraphics[
            height=0.28\textheight,
            trim={0.2cm 0.2cm 0.2cm 0.2cm},
            clip
        ]{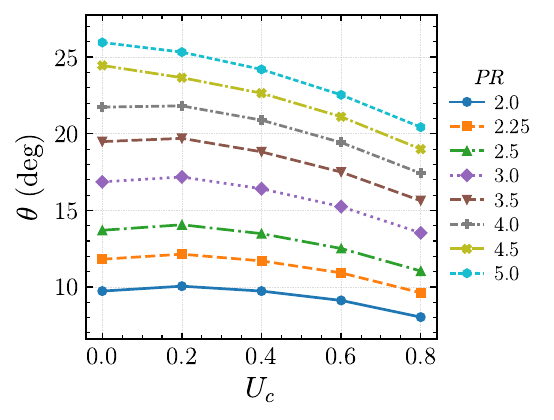}
        \caption{Initial jet-boundary inclination angle variation with $U_c$}
        \label{fig:boundary-streamline-initial-angle-vs-coflow}
    \end{subfigure}
    \hfill
    \begin{subfigure}[t]{0.48\textwidth}
        \centering
        \includegraphics[
            height=0.28\textheight,
            trim={0.2cm 0.2cm 0.2cm 0.2cm},
            clip
        ]{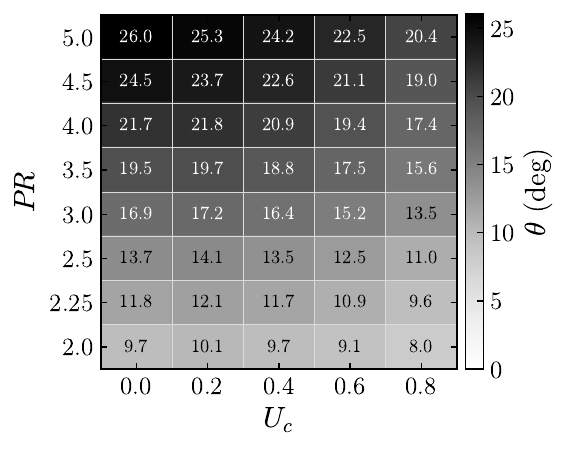}
        \caption{Initial jet-boundary inclination angle map over the parameter space}
        \label{fig:boundary-streamline-initial-angle-map}
    \end{subfigure}

    \caption{Variation of the initial jet-boundary inclination ($\theta$) with coflow velocity ratio $U_c$ at different PR.}
    \label{fig:boundary-streamline-initial-angle-coflow}
\end{figure}

Past underexpanded jet flow studies (\textit{e.g.}, ~\citep{ahmad2022influence, love1955some}) have attributed the changes in shock structures to the jet boundary inclination angle at the nozzle lip. 
To examine that, a tangent line is fitted at the nozzle lip $(x/D_e, y/D_e) = (0, 0.5)$ to the streamline passing through it (hereafter referred to as boundary streamline), as shown in Fig.~\ref{fig:p225-boundary-initial-and-pmf-angle}. The inclination (denoted by $\theta$) of the tangent line for jets at various coflow speeds and NPR values are shown in Fig.~\ref{fig:boundary-streamline-initial-angle-coflow}, where $\theta$ is seen to decrease almost monotonically with increasing coflow, independent of NPR. If viscous effects are neglected, the inclination  ($\theta$), here quantified using the boundary streamline, can be related to the expansion through the Prandtl-Meyer fan centered at the nozzle lip, given by

\begin{equation}
    \theta = \nu(M_2) - \nu(M_1)
    \label{eq:prandtl-meyer-turning-angle}
\end{equation}
where,
\begin{equation}
    \nu(M) =
    \sqrt{\frac{\gamma+1}{\gamma-1}}
    \tan^{-1}
    \left[
        \sqrt{\frac{\gamma-1}{\gamma+1}(M^2-1)}
    \right]
    -
    \tan^{-1}
    \left[
        \sqrt{M^2-1}
    \right].
    \label{eq:prandtl-meyer-function}
\end{equation}

In Eq.~\eqref{eq:prandtl-meyer-turning-angle}, $M_1$ is the local Mach number at the head of the expansion fan, corresponding to the nozzle exit Mach number. For sonic jets, $M_1 = M_e = 1.0$, so $\nu(M_1) = 0.0$, and $M_2$ is the local Mach number at the tail of the expansion fan. Equation~\eqref{eq:prandtl-meyer-turning-angle} therefore simplifies to
\begin{equation}
    \theta = \nu(M_2).
    \label{eq:prandtl-meyer-turning-angle-simplified}
\end{equation}
The inclination of the Prandtl-Meyer fan tail is then related to $M_2$ by
\begin{equation}
    \mu_{M} = \sin^{-1}\left(\frac{1}{M_2}\right).
    \label{eq:prandtl-meyer-tail-angle}
\end{equation}
Thus, the jet boundary inclination ($\theta$) for an equivalent inviscid jet is related to the post-Prandtl-Meyer-fan Mach number ($M_2$), which in turn is related to the Mach angle of the fan tail ($\mu_{M}$). 
It is observed that increasing coflow ($U_c$) leads to a decrease in $\theta$. The decreasing  $\theta$ implies that the expansion of the jet flow at the nozzle exit is reduced, and from Eq.~\eqref{eq:prandtl-meyer-turning-angle-simplified}, the Mach number downstream of the fan is also reduced. This is reflected to some extent in Fig.~\ref{fig:p225-centerline-mach}, where the maximum centerline Mach number decreases with increasing coflow. The embedded shock thus formed, when these expansion waves reflect from the jet boundary, would be weaker, and since weaker embedded shocks are more favorable to regular reflection at the centerline~\citep{chang1973mach}, this is consistent with the transition (from Mach to regular reflection) observed with increasing $U_c$. Similar behavior has previously been reported in the references ~\citep{love1955some, ahmad2022influence}, among others. While the above discussion suggests that the coflow acts on the internal shock structures, and influences the reflection type (Mach or regular), by modifying the inclination of the jet boundary at the nozzle lip ($\theta$), it does not quantify how this occurs. In the following section we therefore test this proposition directly, by examining whether an inviscid method of characteristics (MOC) analysis, with the coflow accounted for solely through a change in the initial inclination of the jet boundary, reproduces the transition from Mach to regular reflection observed in the Navier-Stokes solutions of Fig.~\ref{fig:PR2.25-coflowing-jets}.


\subsection{Method of characteristics (MOC) analysis \label{sec:method-of-characteristics}}
The method of characteristics (MOC) is applied here to underexpanded jets for the values of $\theta$ obtained in Section~\ref{sec:jet-boundary-inclination}, in order to assess whether the decrease in jet boundary inclination at the nozzle lip with increasing coflow is by itself sufficient to alter the type of reflection (Mach or regular) at the axis. The embedded shock, its reflection, and hence the type of reflection it undergoes at the centerline are not obtained explicitly from the present MOC solutions. They are inferred, instead, from the coalescence of the compression waves reflected from the jet boundary and from the location of the first coalescence, as described below.

For MOC calculation, we adapt the formulation in Chapter 13 of \cite{ferri1949elements} to predict near-field inviscid irrotational jet flow, while incorporating the effect of coflow through the corresponding $\theta$ value. In the solution, the intersection of like-characteristics indicates the formation of an embedded shock (see Figs. \ref{fig:regular-reflection-schematic} and \ref{fig:mach-reflection-schematic}).
The MOC results become invalid downstream of the location where the like-characteristics intersect \cite{chang1973mach} unless shock-capturing or shock-fitting corrections are applied \cite{anderson1990modern}. Here, we do not account for the intersection of like-characteristics, and instead allow the characteristics to fold back on top of each other; this method is thus closer to a fold-back type approach \cite{love1955some}. As a result, the formation of the embedded shock is not explicitly captured in the results, and consequently the Mach disk location cannot be directly predicted. Nevertheless, as noted in classical references \cite{ferri1949elements, shapiro1953dynamics, anderson1990modern}, embedded shocks begin to form where like characteristics coalesce and start to fold back over each other in the MOC results. The first coalescence point therefore serves as an indicator of the type of reflection at the centerline in the present MOC results. If the characteristics coalesce (\textit{i.e.}, coalescence points are observed in the MOC results), an embedded shock forms, and its reflection at the centerline may be either regular or Mach; on the other hand, if they do not coalesce (\textit{i.e.}, no coalescence points are observed in the MOC results), every incident compression wave reflects regularly, and the reflection is unambiguously regular. Accordingly, changes in the location of the coalescence point are an indicator of the expected transition in the centerline reflection. A coalescence point that lies farther downstream and closer to the axis implies a weak embedded shock, which would likely undergo a regular reflection. In contrast, a coalescence point located upstream and farther away from the axis is expected to exhibit Mach reflection \cite{chang1973mach}. The analysis here focuses on obtaining the first coalescence point; therefore, shock-fitting or shock-capturing techniques are not required. Furthermore, not accounting for the intersection of like-characteristics simplifies the problem, as the flow remains isentropic and irrotational up to the intersection point. We begin with the steady, irrotational, isentropic axisymmetric equation for the velocity potential $\phi$, given by

\begin{equation}
\left(1-\frac{u^{2}}{a^{2}}\right)\phi_{xx}
+\left(1-\frac{v^{2}}{a^{2}}\right)\phi_{yy}
-\frac{2uv}{a^{2}}\phi_{xy}+\frac{v}{y}=0,
\label{eq:axi-pde}
\end{equation}
where $(u,v)=\nabla\phi$, $a$ is the local speed of sound, and the $v/y$ term is the axisymmetric correction. The two characteristic families carry the inclination

\begin{equation}
\left.\frac{dy}{dx}\right|_{\pm}=\tan(\vartheta\pm\mu_{M}),
\label{eq:char-slope}
\end{equation}
where $\vartheta$ is the flow angle and $\mu_{M}=\arcsin(1/M)$ is the Mach angle. Along each characteristic, the $(\nu,\vartheta)$ compatibility relations read

\begin{align}
\mathrm{C}^{+}:\;
d(\vartheta-\nu) &= -\tilde{\ell}\,\frac{dx}{y},
&\tilde{\ell} &= \frac{\sin\mu_{M}\sin\vartheta}{\cos(\mu_{M}+\vartheta)},
\label{eq:Cplus}\\
\mathrm{C}^{-}:\;
d(\vartheta+\nu) &= +\tilde{m}\,\frac{dx}{y},
&\tilde{m} &= \frac{\sin\mu_{M}\sin\vartheta}{\cos(\vartheta-\mu_{M})},
\label{eq:Cminus}
\end{align}
where $\nu(M)$ is the Prandtl--Meyer function, given by Eq.~(\ref{eq:prandtl-meyer-function}). Further details of the solution procedure, using Eqs.~\eqref{eq:axi-pde}--~\eqref{eq:Cminus}, can be found in \cite{ferri1949elements}.

\begin{figure}[htbp]
    \centering
    \begin{subfigure}[t]{0.48\textwidth}
        \centering
        \includegraphics[
            height=0.20\textheight
        ]{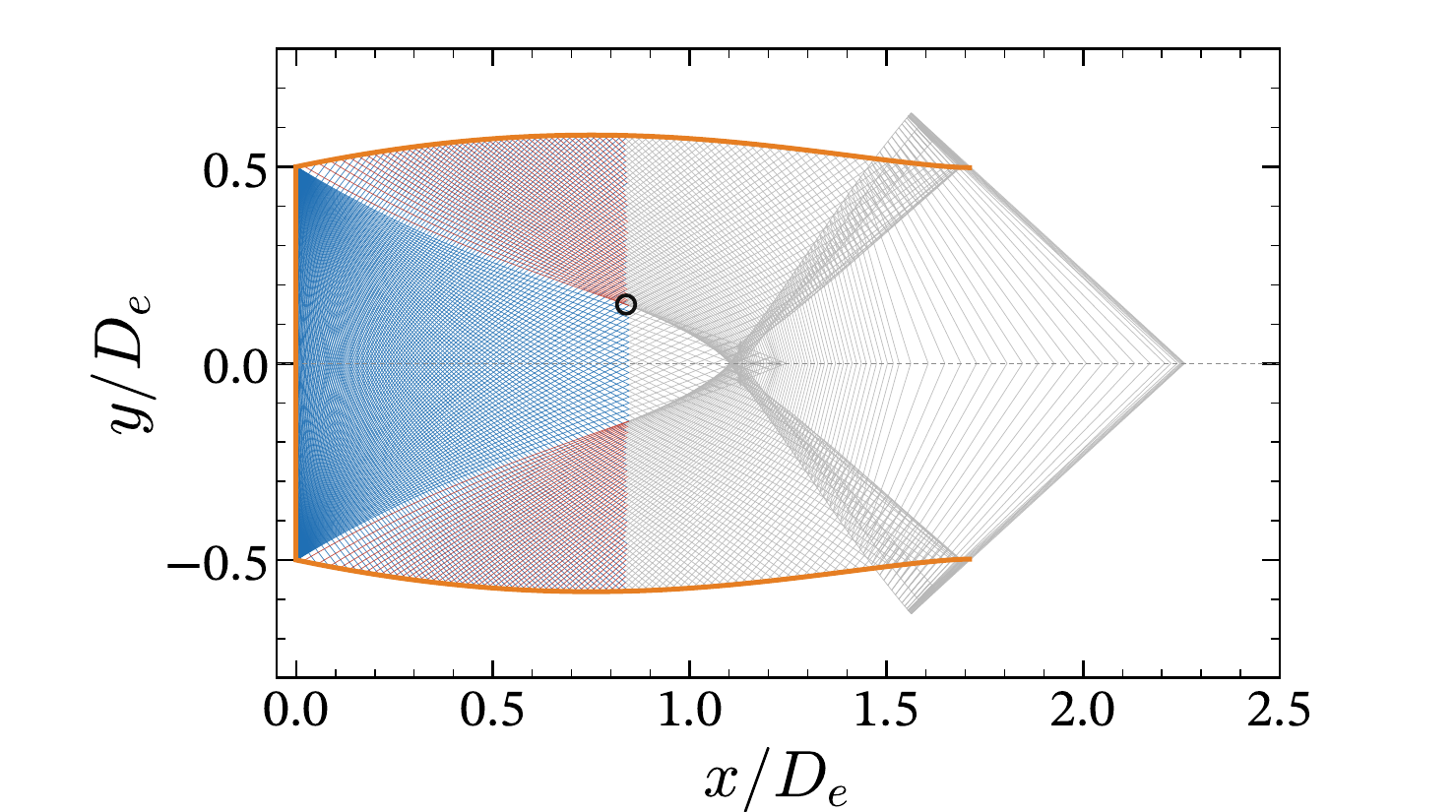}
        \caption{$U_c=0.0$}
        \label{fig:moc-thetab-c0}
    \end{subfigure}
    \begin{subfigure}[t]{0.48\textwidth}
        \centering
        \includegraphics[
            height=0.20\textheight
        ]{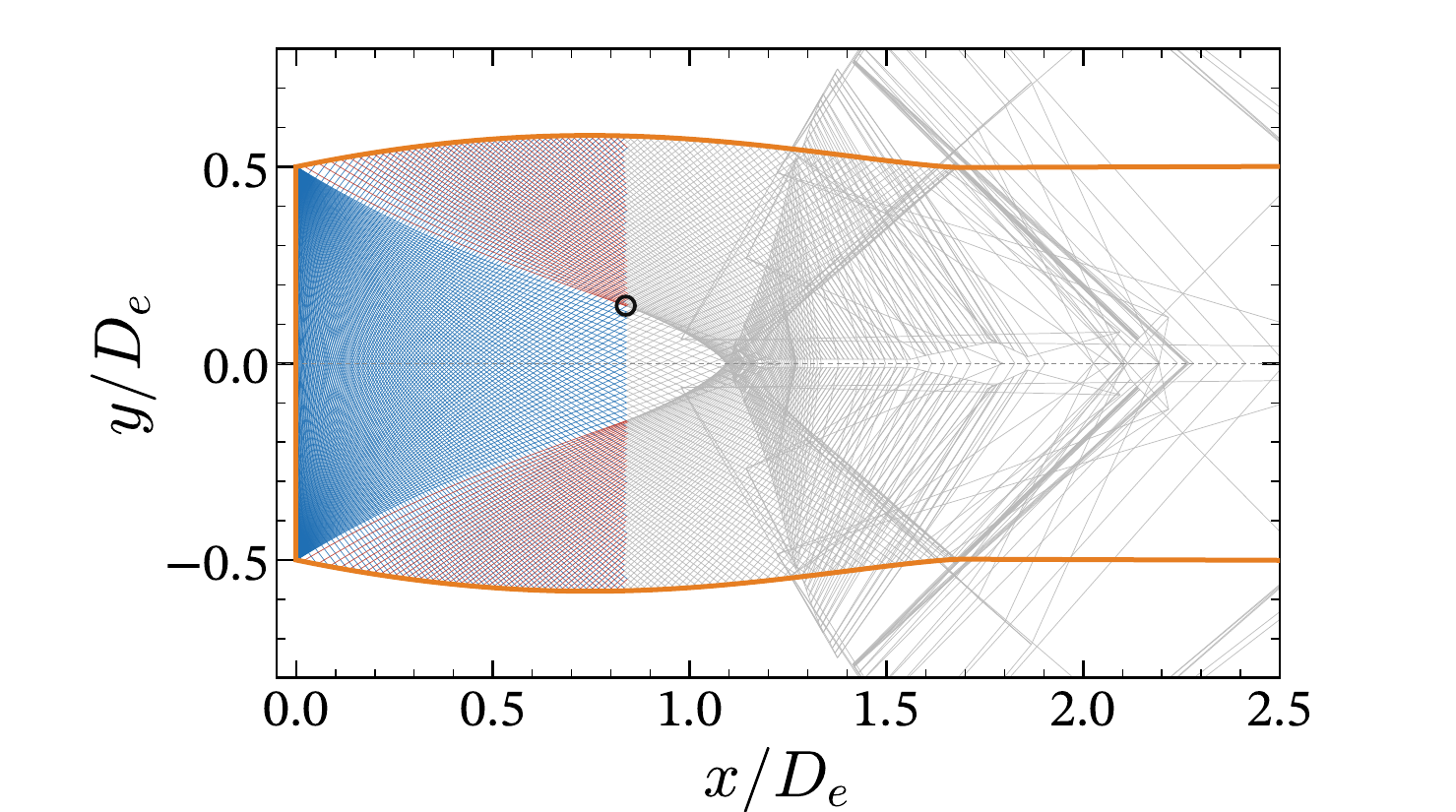}
        \caption{$U_c=0.4$}
        \label{fig:moc-thetab-c4}
    \end{subfigure}
 
    \begin{subfigure}[t]{0.48\textwidth}
        \centering
        \includegraphics[
            height=0.20\textheight
        ]{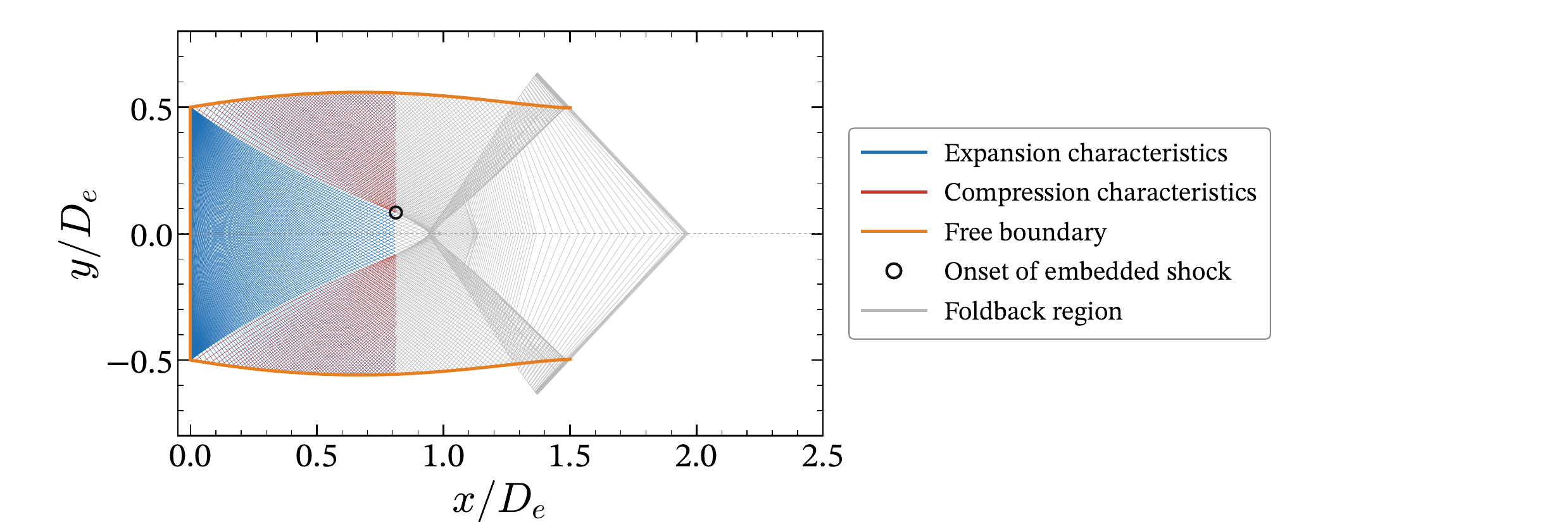}
        \caption{$U_c=0.8$}
        \label{fig:moc-thetab-c8}
    \end{subfigure}
 
    \caption{MOC solutions for $M_e=1.0$ and NPR = 4.26 (PR = 2.25) at increasing coflow ratio $U_c$, with the coflow accounted for through the initial jet boundary inclination from Fig.~\ref{fig:boundary-streamline-initial-angle-coflow} alone. The circle marks the first intersection of like characteristics, which locates the onset of the embedded shock; characteristics downstream of it are drawn in grey and constitute the foldback region, which indicates the presence of the shock and invalidates the MOC solution downstream~\citep{love1955some}}
    \label{fig:moc-thetab-cases}
\end{figure}

Five different coflow cases for the NPR = 4.26 (PR = 2.25) jet are solved using MOC. The effect of coflow is accounted for by the inclination $\theta$, which determines the angles of the Prandtl-Meyer fan launched from the nozzle lip. The values of $\theta$ are taken from Fig.~\ref{fig:boundary-streamline-initial-angle-map}. The results of the solution of~\eqref{eq:axi-pde}--\eqref{eq:Cminus} for NPR = 4.26 (PR = 2.25) jet 
are presented in Fig.~\ref{fig:moc-thetab-cases} for three representative $U_c$, where only the characteristics up to the first reflection of compression waves from the jet boundary are plotted. For the quiescent ambient case in Fig.~\ref{fig:moc-thetab-c0}, the compression waves reflected from the jet boundary begin to coalesce around $x/D_e \simeq 0.8$ (see the black circular marker in Fig.~\ref{fig:moc-thetab-c0}), indicating the onset of embedded shock formation in that region. The inability of this embedded shock to undergo regular reflection at the centerline leads to the formation of a Mach disk; however, since the intersection of like-characteristics is not accounted for in the present MOC solution, this region is simply represented as a region where compression waves fold back on one another. Downstream of this fold-back region (shown in gray in Fig.~\ref{fig:moc-thetab-c0}), the results are no longer meaningful, as the formation of shocks within the flow invalidates the assumptions of isentropy and irrotationality, which would require special treatment beyond the scope of the present analysis. Here, we focus only on the inclination of the compression waves leaving the jet boundary and their resulting coalescence, up to which the results in Fig.~\ref{fig:moc-thetab-cases} remain valid.
 
Fixing the jet boundary inclination at the nozzle lip ($\theta$), 
fixes $M_2$ through Eq.~\eqref{eq:prandtl-meyer-turning-angle-simplified}. Since the stagnation pressure is constant along the boundary streamline, $M_2$ in turn fixes the static pressure $p_s$ along the jet boundary through the isentropic relation 
\begin{align}
\frac{p_s}{p_0} = \left(1 + \frac{\gamma - 1}{2} M_2^2\right)^{-\frac{\gamma}{\gamma-1}}.
\label{eq:pressure-isentropic-relation}
\end{align}
Prescribing the inclination therefore prescribes the boundary pressure as well, and in the cases shown in Fig.~\ref{fig:moc-thetab-cases}, $p_s/p_\infty$ lies between $1.17$ (for $U_c=0$) and $1.30$ (for $U_c=0.8$) rather than the ambient value of unity required for a free surface. Each solution in Fig.~\ref{fig:moc-thetab-cases} therefore represents a jet of effective pressure ratio $\mathrm{PR}_\mathrm{eff} = \mathrm{PR}/(p_s/p_\infty)$, which takes the values $1.93$ at $U_c = 0.0$ and $1.73$ at $U_c = 0.8$, lower than the PR of $2.25$ of the coflowing jets. Hence, the effect of coflow, accounted for by $\theta$, appears to reduce the effective pressure ratio. 

To evaluate whether the effect of coflow (on the underexpanded jet shock structures) can be reproduced by reducing the pressure ratio in an underexpanded jet expanding into a quiescent ambient, as suggested above, the coalescence point of the compression waves (which indicates the formation of an embedded shock) in Fig.~\ref{fig:moc-thetab-cases} is compared with that of the NS solutions of Fig.~\ref{fig:PR2.25-coflowing-jets}. In Fig.~\ref{fig:PR2.25-coflowing-jets}, as the $U_c$ approaches 0.8, the compression waves cease to coalesce, indicated by the absence of an embedded shock in Fig.~\ref{fig:p225-densitygrad-c8}. However, this does not occur in the MOC solutions of Fig.~\ref{fig:moc-thetab-cases}, where wave coalescence is observed in each case. Therefore, imposing coflow in the MOC calculation by inclination angle ($\theta$), leads to an embedded shock formation at all coflow velocities, contrary to the NS solutions of Fig.~\ref{fig:PR2.25-coflowing-jets}. This suggests that a different mechanism through which the coflow acts on the jet boundary is required, which reproduces the shock-structure transition seen in the NS solution with increasing coflow velocity. The MOC solutions in this section impose a uniform pressure ($p_s$ = constant) along the jet boundary, as in a classical free-jet calculation. In case the coflow alters the distribution of pressure along the jet boundary, the entire downstream wave system would be affected rather than those at the nozzle lip alone. We therefore examine the boundary pressure from the NS solutions in the following section before returning to the next MOC analysis.

\subsection{Effect of coflow on jet boundary pressure \label{sec:jet-boundary-streamline-pressure}}
\begin{figure}[htbp]
    \centering

    \begin{subfigure}[t]{0.48\textwidth}
        \centering
        \includegraphics[
            height=0.28\textheight,
            clip
        ]{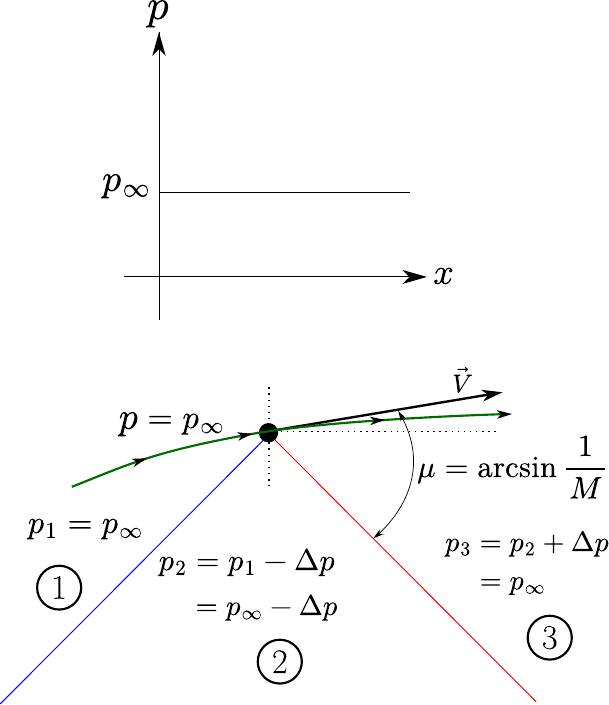}
        \caption{Jet boundary with constant pressure}
        \label{fig:compression-wave-reflection-constant-pressure}
    \end{subfigure}
    \hfill
    \begin{subfigure}[t]{0.48\textwidth}
        \centering
        \includegraphics[
            height=0.28\textheight,
            clip
        ]{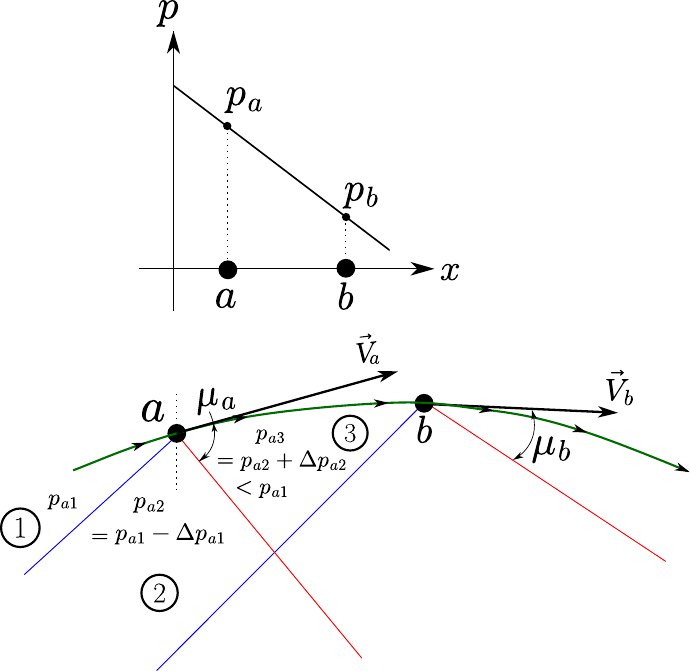}
        \caption{Jet boundary with non-uniform pressure}
        \label{fig:compression-wave-reflection-non-uniform-pressure}
    \end{subfigure}

    \caption{Schematic of characteristic waves reflection at the jet boundary. The blue lines denote the incident expansion waves, red lines denote reflected compression waves and green lines denote the jet boundary.}
    \label{fig:compression-wave-reflection}
\end{figure}

For an inviscid supersonic jet exiting into a quiescent medium, the jet boundary coincides with the boundary streamline (defined in Section \ref{sec:jet-boundary-inclination}) and has a constant static pressure equal to the ambient pressure~\citep{ferri1949elements}. Expansion waves reflect from such constant-pressure boundaries as compression waves, as illustrated schematically in Fig.~\ref{fig:compression-wave-reflection-constant-pressure}, where the green line denotes the jet boundary. In region 1, the pressure matches the ambient value, \textit{i.e.}, $p_1 = p_\infty$. As the flow passes through the expansion wave, the pressure reduces by, let's say, $\Delta p$, so that the pressure in region 2 is $p_2 = p_1 - \Delta p$. To satisfy the boundary condition and restore the pressure to the ambient value $p_\infty$, a compression wave should emerge from the boundary that increases the pressure by $\Delta p$ to provide $p_3 = p_\infty$. 

\begin{figure}[htbp]
    \centering

    \begin{subfigure}[t]{0.45\textwidth}
        \centering
        \includegraphics[
            width=\linewidth,
            trim={1.0cm 1.0cm 8.0cm 1.0cm},
            clip
        ]{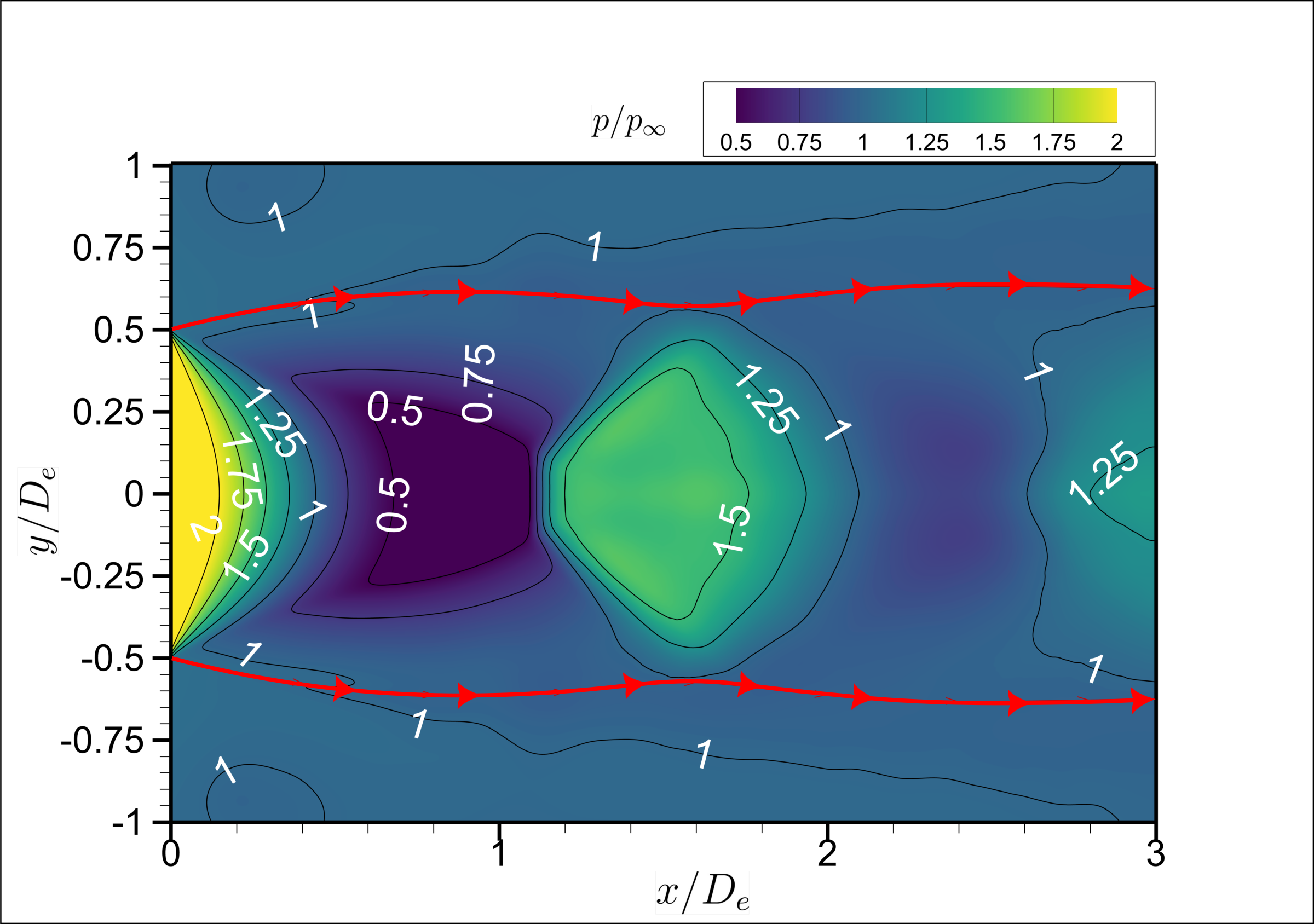}
        \caption{$U_c=0.0$}
        \label{fig:p225-pressure-c0}
    \end{subfigure}
    \begin{subfigure}[t]{0.45\textwidth}
        \centering
        \includegraphics[
            width=\linewidth,
            trim={1.0cm 1.0cm 8.0cm 1.0cm},
            clip
        ]{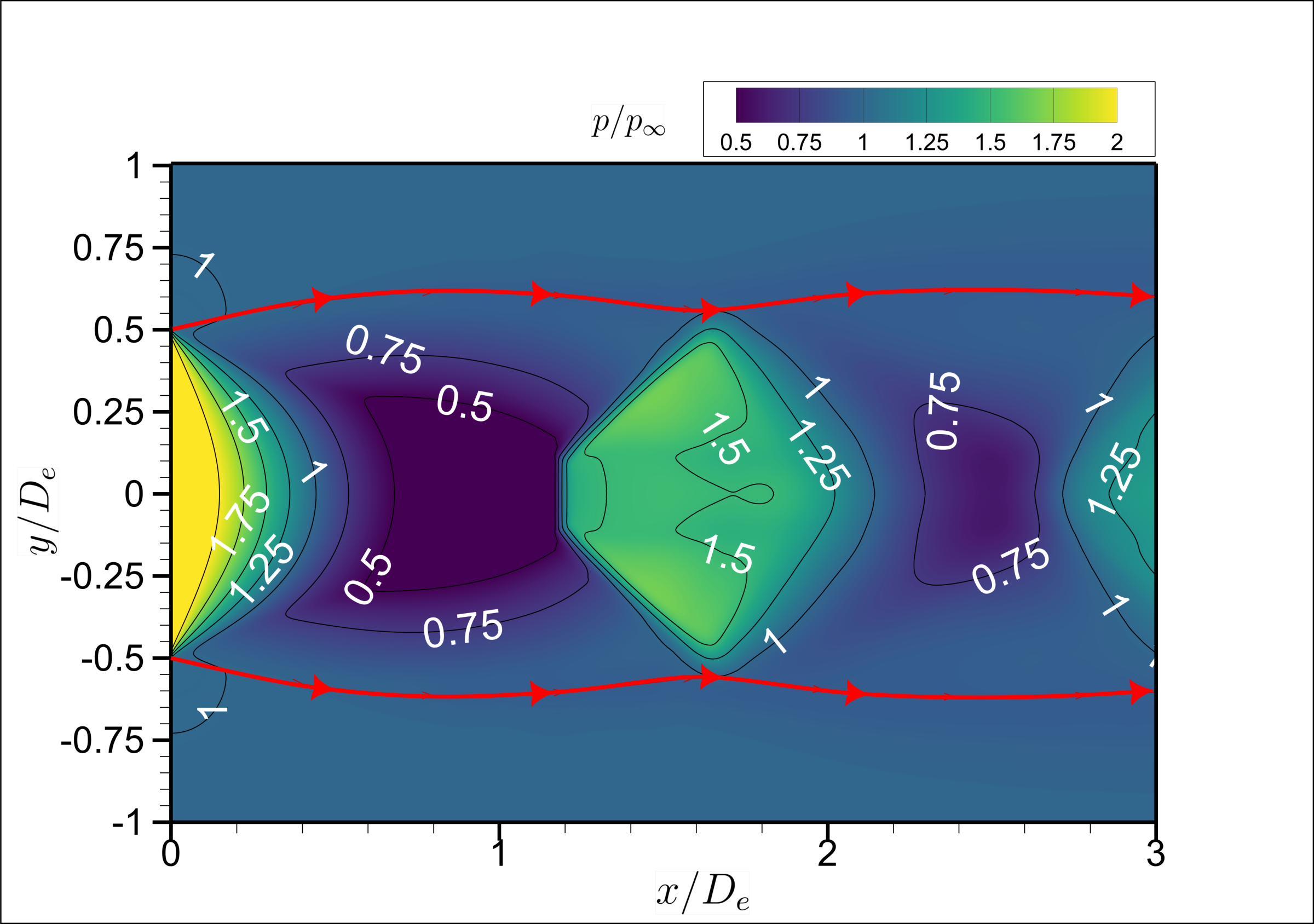}
        \caption{$U_c=0.2$}
        \label{fig:p225-pressure-c2}
    \end{subfigure}


    \begin{subfigure}[t]{0.45\textwidth}
        \centering
        \includegraphics[
            width=\linewidth,
            trim={1.0cm 1.0cm 8.0cm 1.0cm},
            clip
        ]{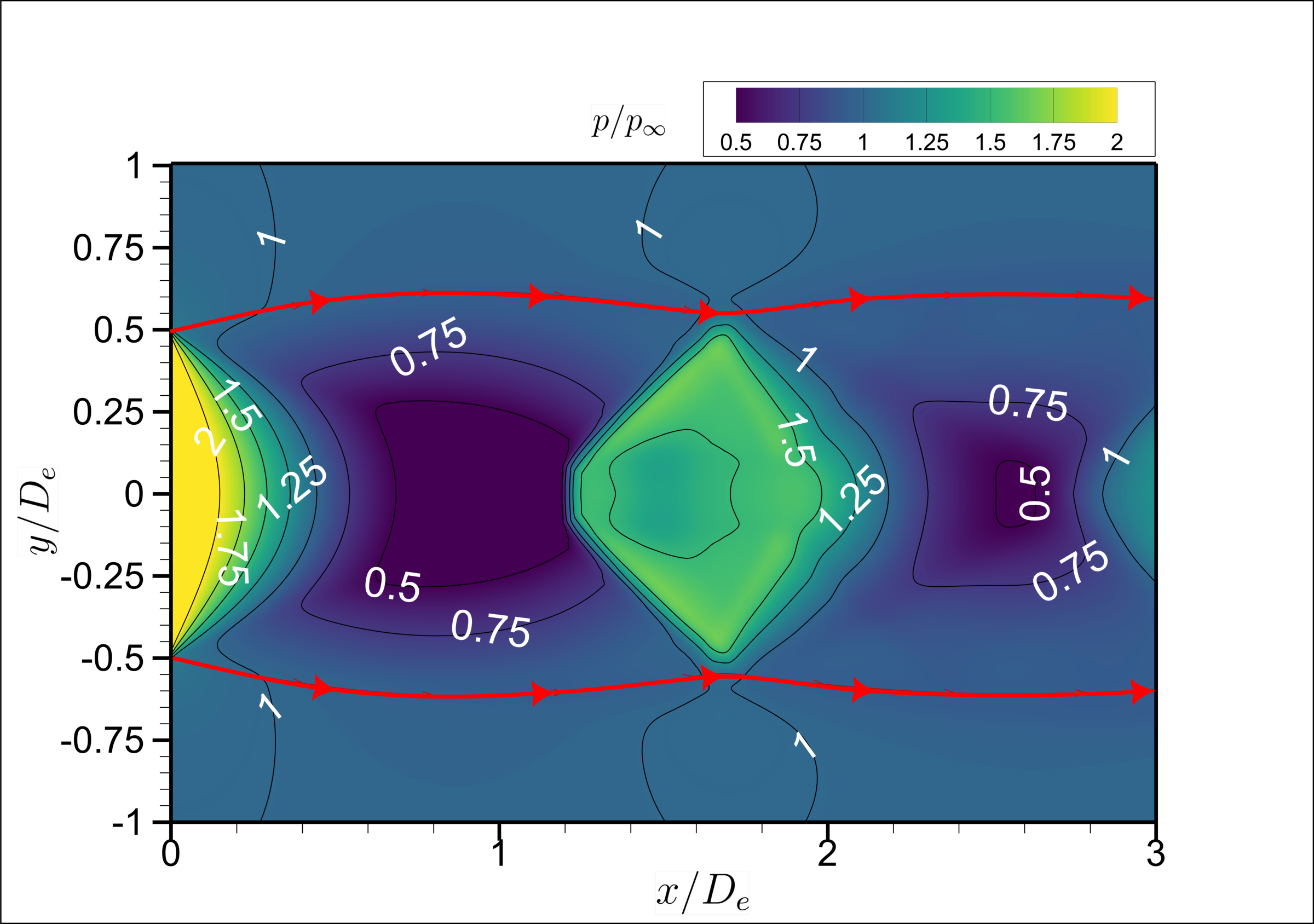}
        \caption{$U_c=0.4$}
        \label{fig:p225-pressure-c4}
    \end{subfigure}
    \begin{subfigure}[t]{0.45\textwidth}
        \centering
        \includegraphics[
            width=\linewidth,
            trim={1.0cm 1.0cm 8.0cm 1.0cm},
            clip
        ]{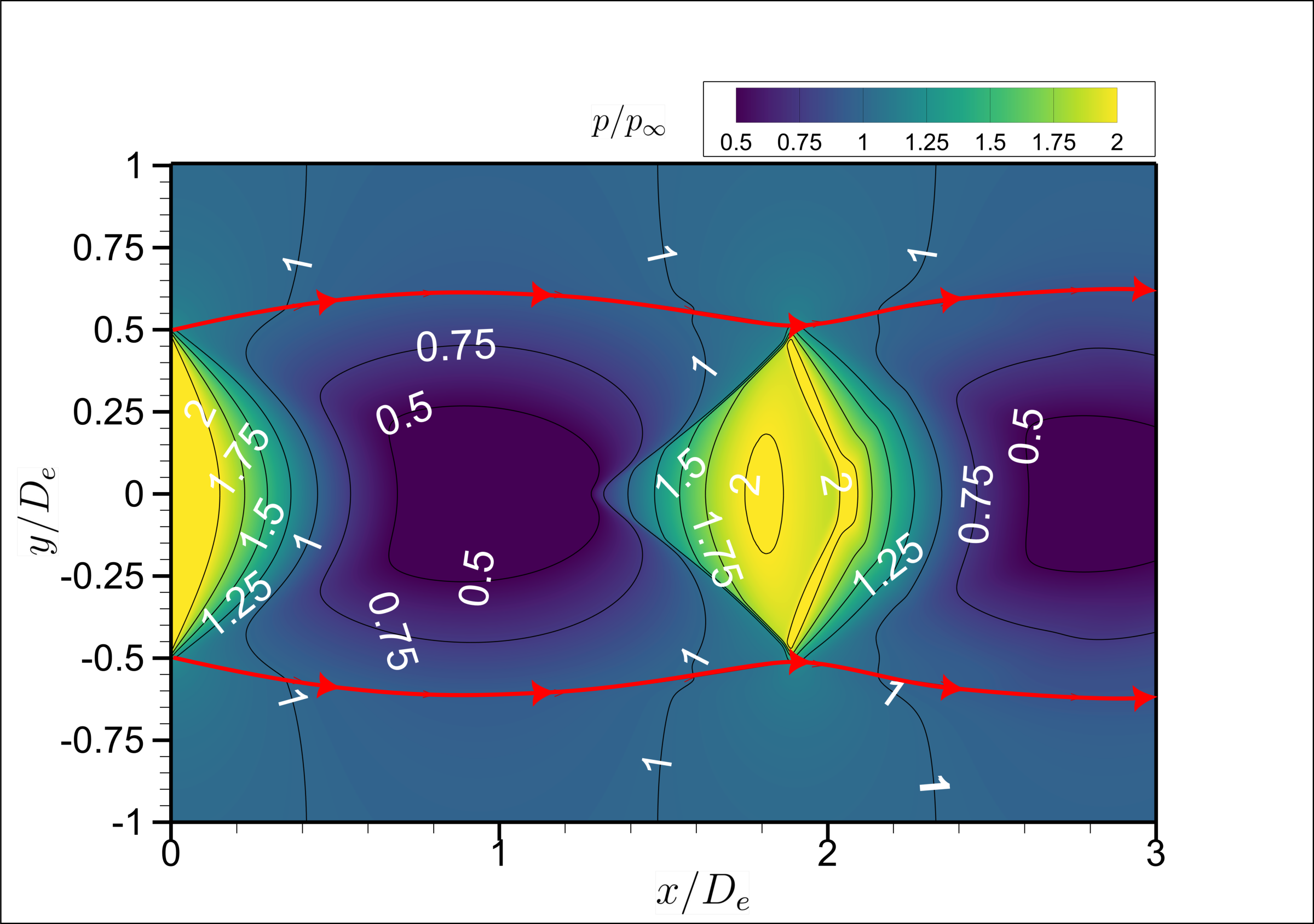}
        \caption{$U_c=0.6$}
        \label{fig:p225-pressure-c6}
    \end{subfigure}
    \begin{subfigure}[t]{0.45\textwidth}
        \centering
        \includegraphics[
            width=\linewidth,
            trim={1.0cm 1.0cm 8.0cm 1.0cm},
            clip
        ]{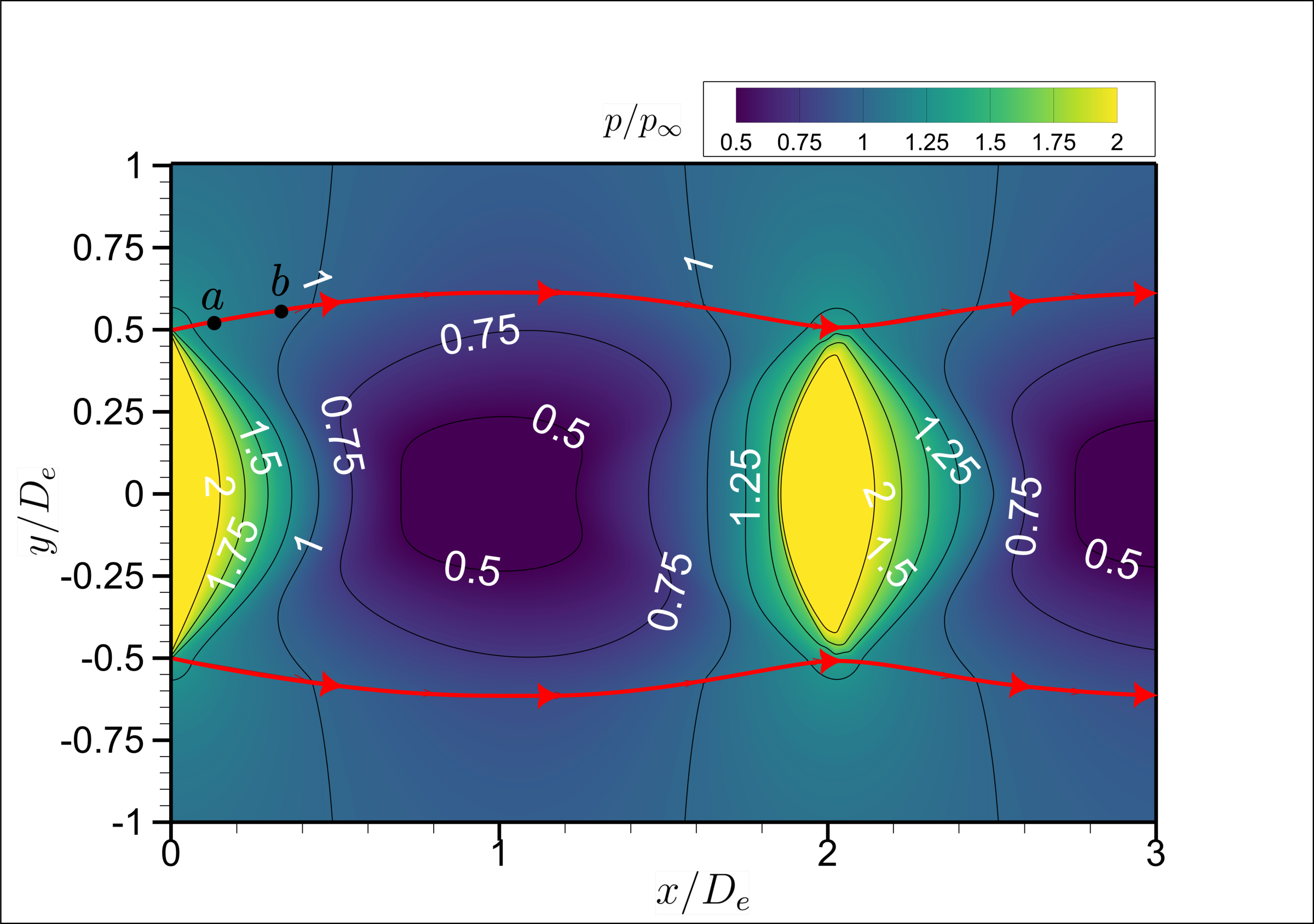}
        \caption{$U_c=0.8$}
        \label{fig:p225-pressure-c8}
    \end{subfigure}

    \caption{Comparison of time-averaged pressure contours for underexpanded jets at NPR = 4.26 (PR = 2.25) with increasing coflow ratio $U_c$. The red line represents the jet-boundary streamline launched from the nozzle lip at $(x/D_e, y/D_e) = (0, \pm 0.5)$.}
    \label{fig:PR2.25-pressure-contours-coflowing-jets}
\end{figure}

\begin{figure}[htbp]
    \centering
    \includegraphics[
        width=0.75\textwidth
    ]{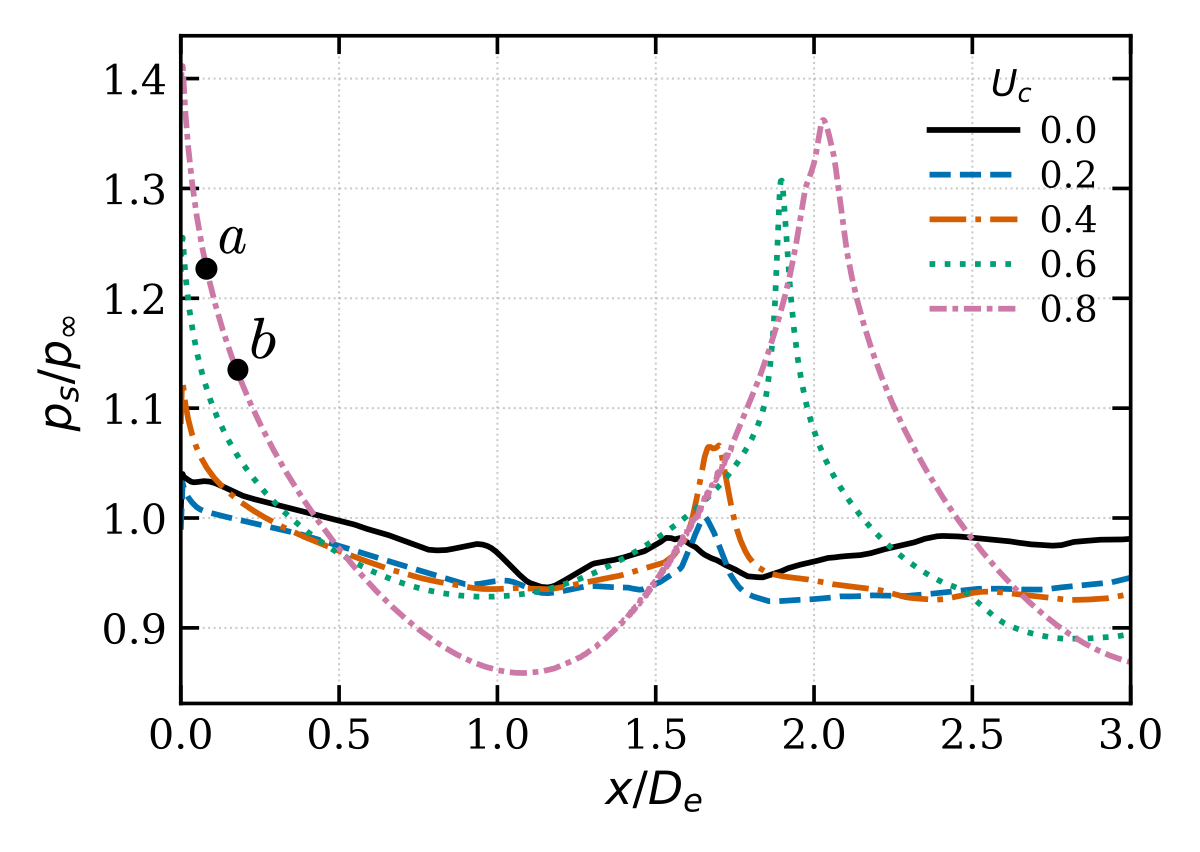}

    \caption{Boundary streamline pressure variation for NPR = 4.26 (PR = 2.25).}
    \label{fig:p225-boundary-streamline-pressure}
\end{figure}

When the ambient fluid has a subsonic velocity, the pressure along the jet boundary is no longer constant, as evident from Fig.~\ref{fig:PR2.25-pressure-contours-coflowing-jets}, where the contours of static pressure from NS solutions are plotted for the NPR = 4.26 (PR = 2.25) jet at various coflow speeds. From the figure, the iso-line of static pressure equal to the ambient pressure ($p/p_\infty = 1.0$) is seen to move progressively away from the boundary streamline with increasing coflow; the isoline $p/p_\infty = 1.0$ lies close to the boundary streamline in the quiescent case, but runs at a steep angle across it at higher coflow speeds. This departure of pressure from the ambient value arises from both viscous and inviscid effects, but can be largely explained by an inviscid argument. The subsonic ambient responds to the area variation imposed by the local jet diameter in a manner similar to a subsonic streamtube contracting over the initially increasing jet diameter, so that the ambient fluid accelerates and the pressure reduces. This creates a favorable pressure gradient that persists up to the maximum-diameter point of the jet, beyond which the jet diameter reduces, resulting in ambient fluid deceleration and pressure increase. This is confirmed in Fig.~\ref{fig:p225-boundary-streamline-pressure}, where the pressure along the boundary streamline is plotted for the  cases shown in Fig.~\ref{fig:PR2.25-pressure-contours-coflowing-jets}, showing a decrease until the maximum-diameter point followed by an increase, both becoming more pronounced with increasing coflow. To confirm that this non-uniform pressure is not an artifact of the domain extent (\textit{i.e.}, numerical reflections from the domain boundary), the $U_c = 0.8$ case, which exhibits the largest pressure variation, is repeated with radially extended domains, up to $y=7.5D_e$ and $10D_e$, and the results are discussed in Appendix~\ref{app:domain-extent}. The effects of the pressure variation at the jet boundary on the wave system within the jet core are evaluated next. 

We restrict attention here to the portion of the boundary upstream of the maximum jet diameter point in the first shock cell, along which the pressure falls (see Fig.~\ref{fig:p225-boundary-streamline-pressure}). The compression waves reflected from this portion are the ones whose coalescence forms the embedded shock, and its strength, curvature, and alignment in turn determine whether it goes through regular or Mach reflection at the centerline \cite{chang1973mach}.
Consider two successive points $a$ and $b$ on this portion of the boundary (see Figs.~\ref{fig:p225-pressure-c8} and~\ref{fig:p225-boundary-streamline-pressure}) at which the reflected compression waves leave the boundary, with $b$ downstream of $a$ so that $p_a > p_b$, where $p_a$ and $p_b$ denote the local static pressures at the two points. These points and the corresponding characteristics are shown schematically in Fig.~\ref{fig:compression-wave-reflection-non-uniform-pressure}, while Fig.~\ref{fig:compression-wave-reflection-constant-pressure} contains the schematic of waves reflecting from a constant-pressure boundary. Along the boundary streamline (shown by the green line in Fig.~\ref{fig:compression-wave-reflection-non-uniform-pressure}) the jet total pressure is conserved (with $p_0$ the jet reservoir pressure), 
so the local Mach number $M$, and hence the Mach angle $\mu_{M}$ at which the reflected waves are emitted follow from the local static pressure $p$ as functions of the axial coordinate $x$ along the boundary,
\begin{equation}
  M(x)=\sqrt{\frac{2}{\gamma-1}\left[\left(\frac{p_0}{p(x)}\right)^{\frac{\gamma-1}{\gamma}}-1\right]},
  \qquad
  \mu_{M}(x)=\sin^{-1}\left(\frac{1}{M(x)}\right).
  \label{eq:boundary-mach-angle}
\end{equation}
The decreasing boundary pressure affects the reflected compression waves in two ways. First, it weakens them. In Fig.~\ref{fig:compression-wave-reflection-constant-pressure}, as mentioned earlier, the boundary pressure is constant and equal to the ambient pressure, so the reflected compression wave must exactly cancel the pressure drop across the incident expansion wave, providing $p_{3} = p_{1}$. In Fig.~\ref{fig:compression-wave-reflection-non-uniform-pressure}, however, this is not the case; at any point (say, the point $a$ in Fig.~\ref{fig:compression-wave-reflection-non-uniform-pressure}), the decreasing boundary pressure now requires $p_{a3} < p_{a1}$. Consequently, the pressure drop across the incident expansion wave, $\Delta p_{a1}$, must exceed the pressure recovery across the reflected compression waves, $\Delta p_{a2}$, \textit{i.e.} $\Delta p_{a1} > \Delta p_{a2}$, in order to satisfy the new boundary condition. The reflected compression waves are therefore weaker in this case. Second, the decreasing boundary pressure spreads the emission angles of the reflected compression waves. At the two successive points $a$ and $b$ of Fig.~\ref{fig:compression-wave-reflection-non-uniform-pressure}, with corresponding values $M_a$, $\mu_{M,a}$ and $M_b$, $\mu_{M,b}$, Eq.~\eqref{eq:boundary-mach-angle} yields $M_a < M_b$ and, hence, $\mu_{M,a} > \mu_{M,b}$ for $ p_a > p_b$. Each compression wave is therefore emitted at a shallower angle to the local flow than the one upstream of it, as seen in Fig.~\ref{fig:compression-wave-reflection-non-uniform-pressure}. The reflected waves of Fig.~\ref{fig:compression-wave-reflection-non-uniform-pressure} consequently leave the boundary with a spread in their emission angles in contrast to the quiescent case of Fig.~\ref{fig:compression-wave-reflection-constant-pressure}, where any two points $a$ and $b$ on the boundary satisfy $p_a = p_b$, providing $\mu_{M,a} = \mu_{M,b}$, and the waves are emitted at a uniform angle to the local flow. This spread of emission angle, together with the reduced wave strengths, inhibits the coalescence of the reflected compressions into an embedded shock. Establishing whether it does, however, requires the local flow angle along the boundary in addition to the Mach angle, and hence the resolution of the upstream wave field, for which we return to the MOC analysis in the following Section~\ref{sec:method-of-characteristics-pressure-boundary}.

Independent evidence that a subsonic coflow modifies the jet interior through the boundary condition, rather than through an initial pressure mismatch, is provided by the work of \citet{norum1993simulated}. In their perfectly expanded nozzle, which exhibits no nozzle-exit pressure mismatch and would therefore be shock-free in a quiescent ambient, weak waves nonetheless appeared within the supersonic core as the external stream velocity was increased, indicating that the external stream alone can perturb the supersonic core through the boundary condition it imposes. The mechanism described above also offers a possible explanation for the weakening of the first shock cells measured by \citet{andre2016flighteffects}, the cause of which was left open in that work. It also subsumes the mechanism identified by \citet{ahmad2022influence}, whose narrowing of the lip-centered expansion fan corresponds to the reduced jet-boundary inclination at the nozzle lip of Section~\ref{sec:jet-boundary-inclination}. That narrowing is recovered here as the local effect of the imposed pressure distribution, which acts along the entire boundary rather than at the nozzle lip alone.

\subsection{MOC results for NPR = 4.26 (PR = 2.25) jet with coflow effect accounted for by pressure boundary condition\label{sec:method-of-characteristics-pressure-boundary}}

\begin{figure}[htbp]
    \centering

    \begin{subfigure}[t]{0.48\textwidth}
        \centering
        \includegraphics[
            height=0.21\textheight,
            trim={0.2cm 0.2cm 0.2cm 0.2cm},
            clip
        ]{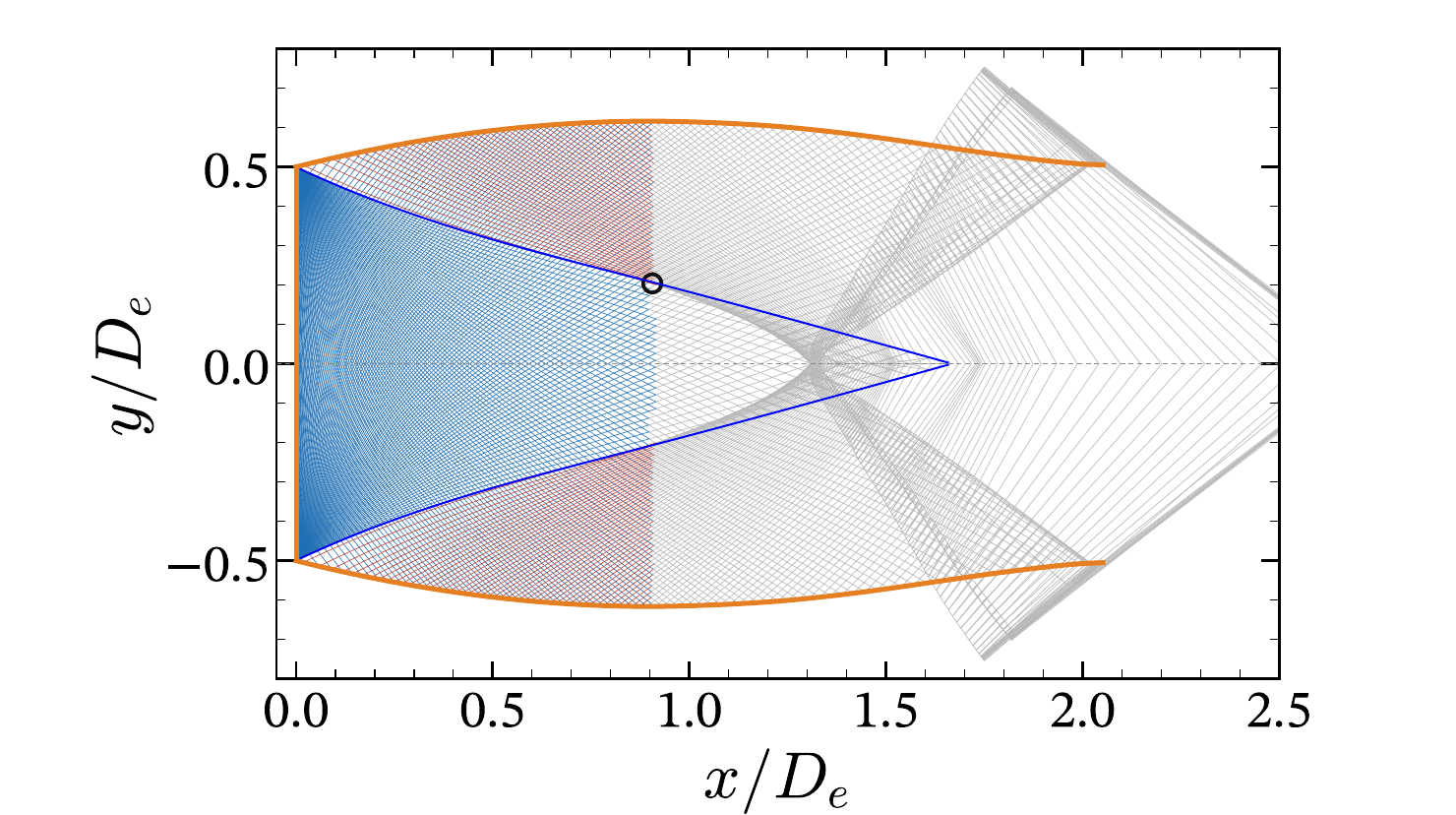}
        \caption{$U_c=0.0$}
        \label{fig:moc-pr225-c0}
    \end{subfigure}
    \begin{subfigure}[t]{0.48\textwidth}
        \centering
        \includegraphics[
            height=0.21\textheight,
            trim={0.2cm 0.2cm 0.2cm 0.2cm},
            clip
        ]{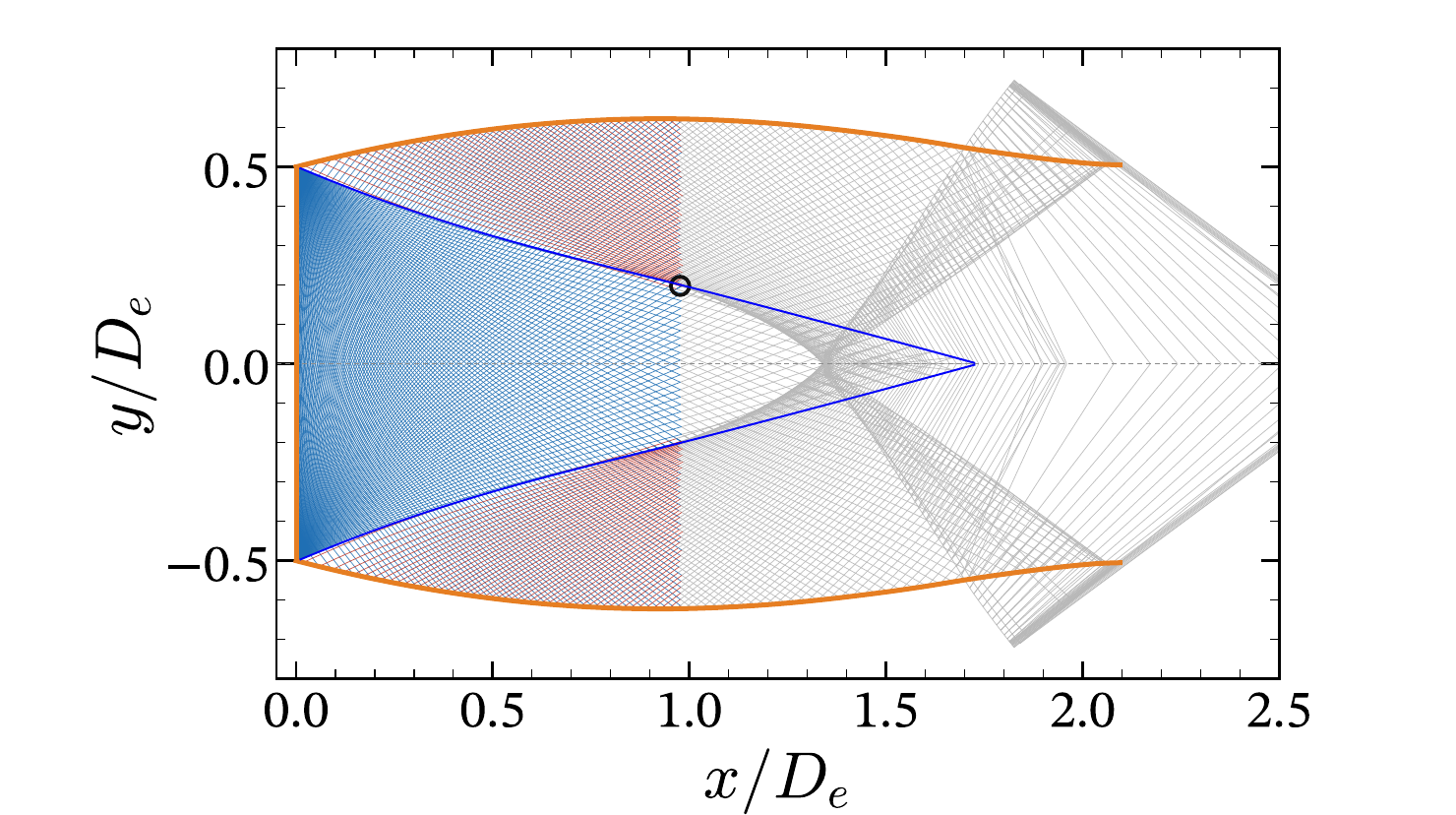}
        \caption{$U_c=0.2$}
        \label{fig:moc-pr225-c2}
    \end{subfigure}


    \begin{subfigure}[t]{0.48\textwidth}
        \centering
        \includegraphics[
            height=0.21\textheight,
            trim={0.2cm 0.2cm 0.2cm 0.2cm},
            clip
        ]{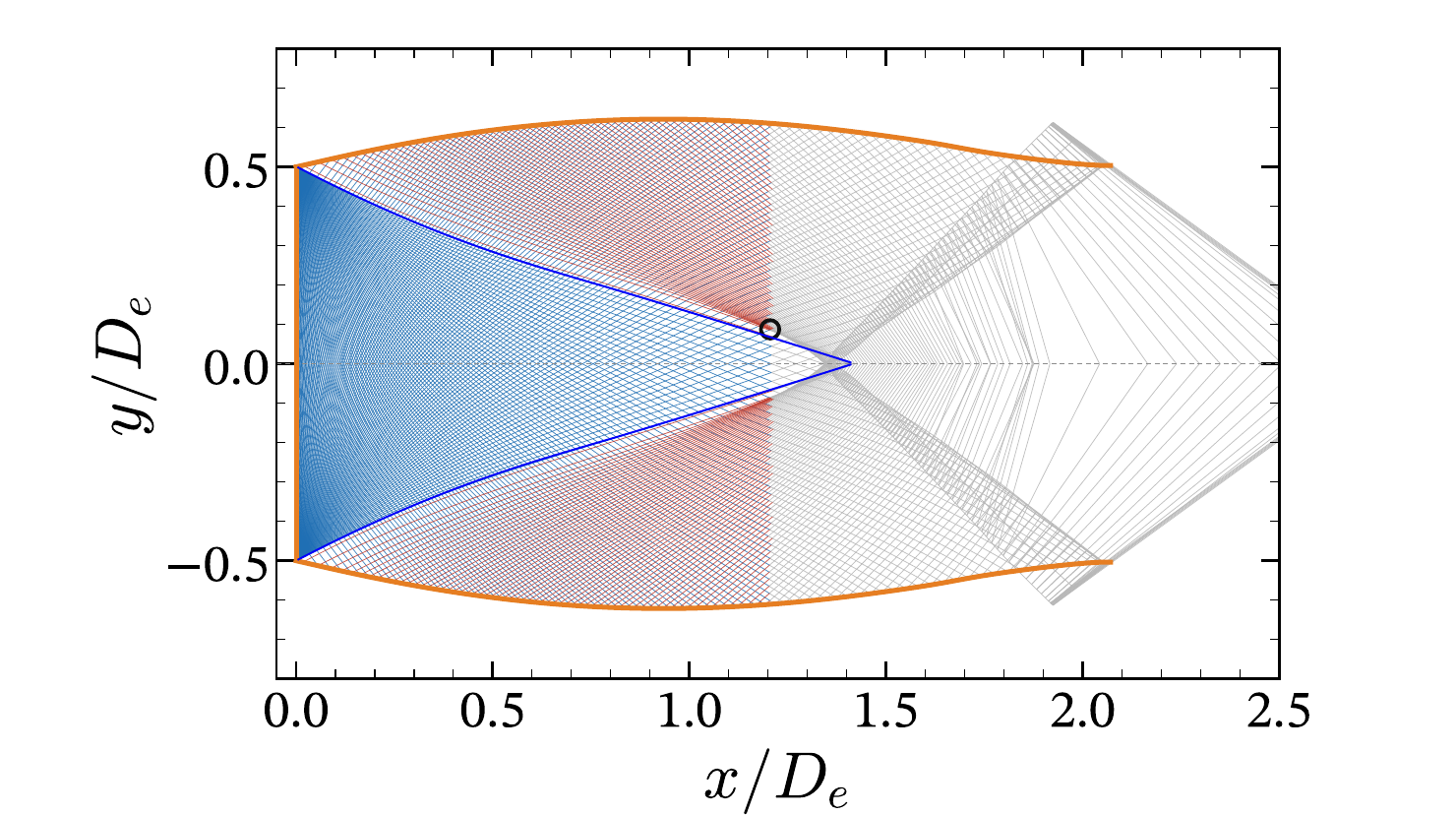}
        \caption{$U_c=0.4$}
        \label{fig:moc-pr225-c4}
    \end{subfigure}
    \begin{subfigure}[t]{0.48\textwidth}
        \centering
        \includegraphics[
            height=0.21\textheight,
            trim={0.2cm 0.2cm 0.2cm 0.2cm},
            clip
        ]{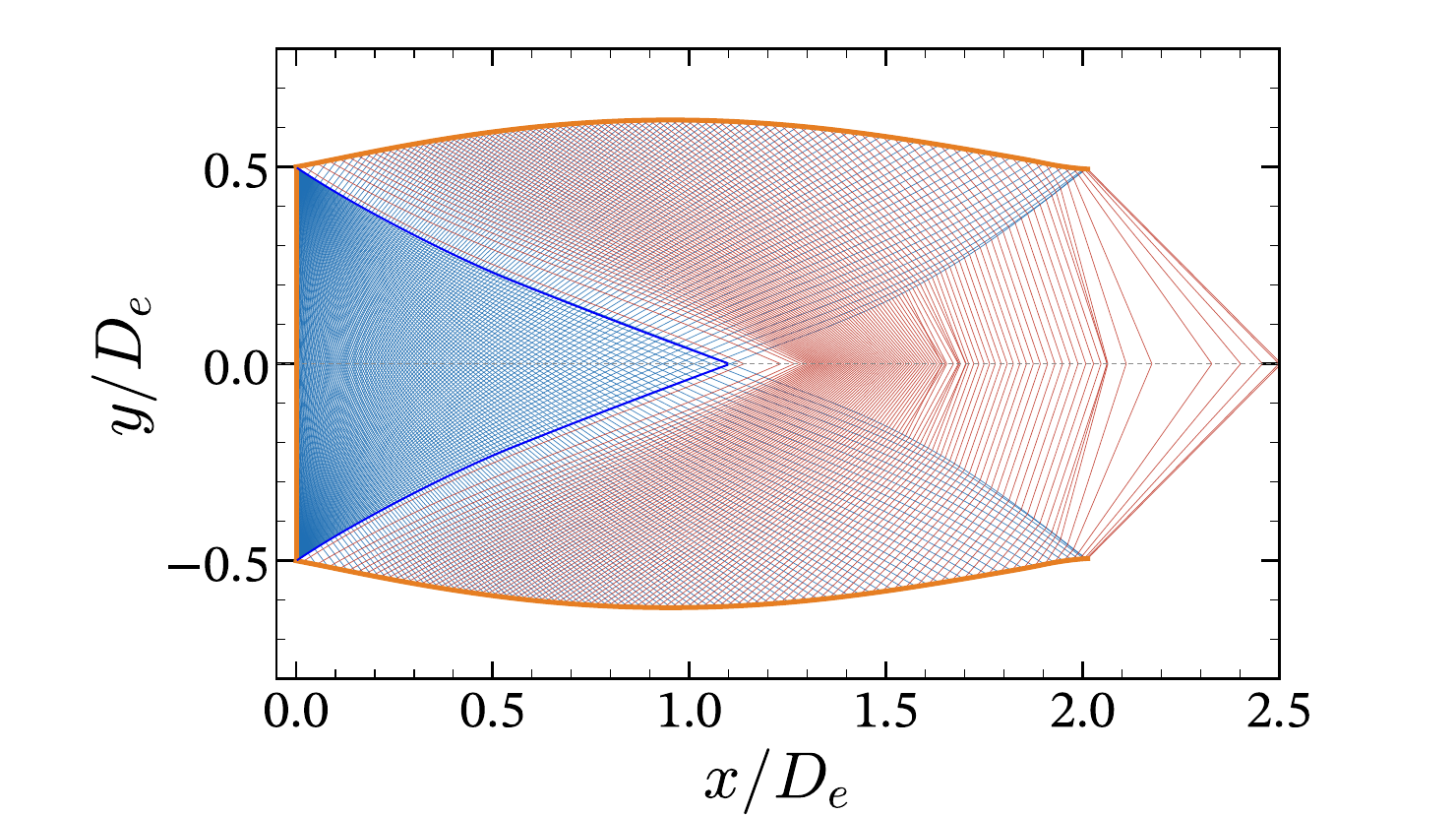}
        \caption{$U_c=0.6$}
        \label{fig:moc-pr225-c6}
    \end{subfigure}
    \begin{subfigure}[t]{0.48\textwidth}
        \centering
        \includegraphics[
            height=0.21\textheight,
            trim={0.2cm 0.2cm 0.2cm 0.2cm},
            clip
        ]{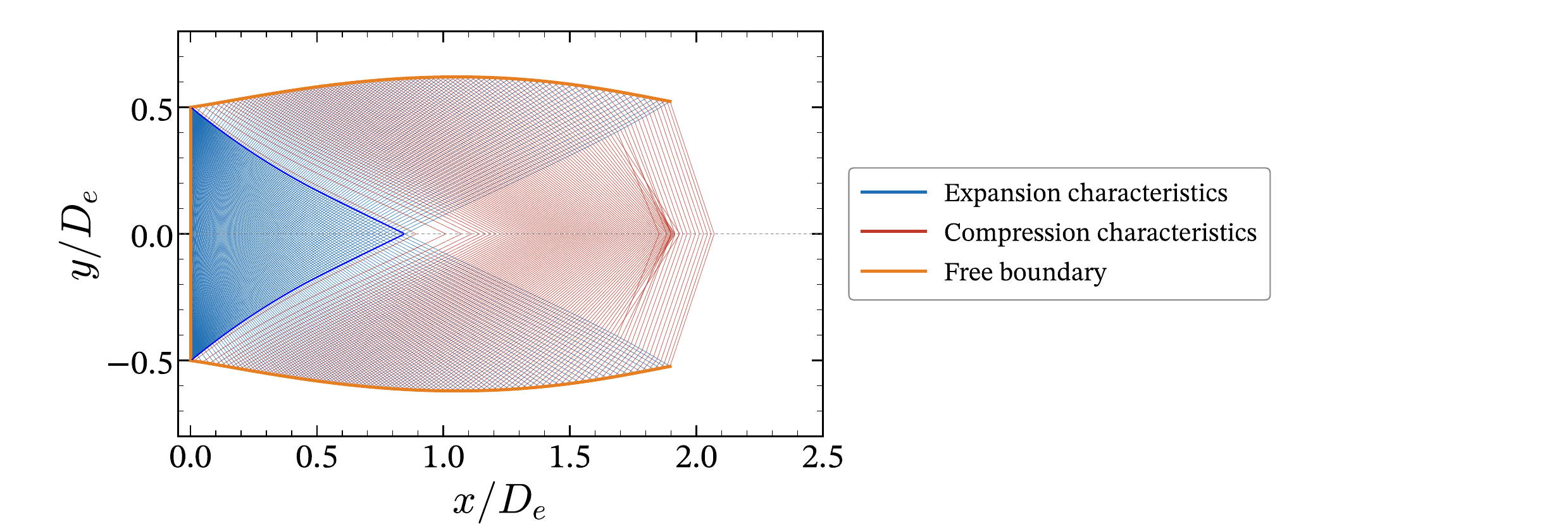}
        \caption{$U_c=0.8$}
        \label{fig:moc-pr225-c8}
    \end{subfigure}

    \caption{MOC solutions for $M_e=1.0$ and NPR = 4.26 (PR = 2.25) at increasing coflow ratio $U_c$, with the coflow accounted for through pressure from Fig.~\ref{fig:p225-boundary-streamline-pressure} imposed at boundary. The circle marks the first intersection of like characteristics, which locates the onset of the embedded shock; characteristics downstream of it are drawn in grey and constitute the foldback region, which indicates the presence of the shock and invalidates the solution downstream~\citep{love1955some}}
    \label{fig:moc-pr225-coflow-cases}
\end{figure}

In contrast to the MOC solutions of Section~\ref{sec:method-of-characteristics}, the coflow is now accounted for by the static pressure imposed along the jet boundary, taken from the NS solutions of Fig.~\ref{fig:p225-boundary-streamline-pressure}. This pressure is no longer uniform but varies with axial distance. The initial inclination is not prescribed separately. As noted in Section~\ref{sec:method-of-characteristics}, $\theta$ and $p_s$ are linked at the lip through Eqs.~\eqref{eq:prandtl-meyer-turning-angle-simplified} and~\eqref{eq:pressure-isentropic-relation}, so the fan angle follows from the value of the imposed pressure profile at the lip. The higher local static pressure at the nozzle lip seen in Fig.\ref{fig:p225-boundary-streamline-pressure} again reduces the effective pressure ratio, from $\mathrm{PR}_\mathrm{eff} = 2.24$ at $U_c = 0.0$ to $1.62$ at $U_c = 0.8$, but the reduction now originates from the imposed pressure profile instead of a separately prescribed inclination. The pressure profile continues to vary downstream, so that the coflow acts along the entire jet boundary and not at the nozzle lip alone. The treatment of Section~\ref{sec:method-of-characteristics}, where only the local effect at the lip is retained and the boundary pressure is thereafter maintained uniform, is a limiting case of the treatment discussed in this section. The resulting characteristic nets from the MOC solution are shown in Fig.~\ref{fig:moc-pr225-coflow-cases}. The axial extent over which these solutions follow the Navier--Stokes results is examined in Appendix~\ref{app:moc-validity}, where the flow Mach-number and pressure from the two are compared at two radial locations.

In Fig.~\ref{fig:moc-pr225-coflow-cases}, we first note the intersection point of the tail of the Prandtl-Meyer fan with the centerline. With increasing coflow, the intersection point moves upstream. For example, for $U_c = 0.0$, the point lies at around $x/D_e = 1.7$, as seen in Fig.~\ref{fig:moc-pr225-c0}, while for $U_c = 0.8$, it moves upstream to around $x/D_e = 0.8$, as seen in Fig.~\ref{fig:moc-pr225-c8}. This follows from the reduced $\mathrm{PR}_\mathrm{eff}$ due to the higher local pressure at the nozzle lip with increasing coflow in the imposed pressure profile (Fig.~\ref{fig:p225-boundary-streamline-pressure}), and the consequent reduction in expansion at the nozzle lip, as discussed earlier. The Mach disk location and diameter depend on the formation of the embedded shock, which in turn depends on the point where the compression waves begin to coalesce. For $U_c = 0.0$ and $U_c = 0.2$, the coalescing begins at nearly the same location, at around $x/D_e \approx 0.9$ and $1.0$ respectively as seen in Figs.~\ref{fig:moc-pr225-c0} and~\ref{fig:moc-pr225-c2}, indicating that the Mach disk formed in the two cases would be of similar nature. This is however at odds with the observations in Figs.~\ref{fig:PR2.25-coflowing-jets} and~\ref{fig:mach-disk-diameter-vs-coflow}, where the Mach-disk diameter for the two cases differ substantially ($D_{MD}/D_e = 0.13$ for $U_c = 0.0$ and 0.19 for $U_c = 0.2$). This discrepancy, tentatively tied to pronounced viscous effects at low coflow velocities, is discussed in Section~\ref{sec:viscous-effects}. For the higher coflow case of $U_c = 0.4$, the first coalescing point shifts downstream to around $x/D_e \approx 1.2$, indicating that the embedded shock formed for this case would be of shorter length, and consequently, the Mach-disk if formed would also be of smaller diameter. This is consistent with the observations in Navier-Stokes solutions of Fig.~\ref{fig:PR2.25-coflowing-jets}. Furthermore, for $U_c = 0.6$ and 0.8, the compression waves do not coalesce at all, and hence would result in regular reflection at the centerline, both of which are consistent with the regular reflections observed in Navier-Stokes solutions of Fig.~\ref{fig:PR2.25-coflowing-jets} for the corresponding coflows. Interestingly, in the highest coflow case ($U_c = 0.8$), at a downstream distance of around $x/D_e \simeq 1.9$, the compression waves are seen to recombine, leading to the formation of a shock-like structure, again consistent with what was observed in the Navier-Stokes solutions in Fig.~\ref{fig:p225-densitygrad-c8}. However, since the reflection of compression waves from the centerline is not accounted for in the present MOC solution, the full bow shape is not reproduced. 

The coalescence of compression waves seen in the MOC solution of Fig.~\ref{fig:moc-pr225-c8}, at around the same location where the weak curved shock exists in the Navier-Stokes solution of Fig.~\ref{fig:p225-densitygrad-c8} for the ($U_c = 0.8$) case, suggests that the same boundary-pressure mechanism offers a plausible explanation for the structure. As discussed in Section~\ref{sec:jet-boundary-streamline-pressure}, along the section of the jet boundary with increasing jet diameter (or the diverging section), the coflow imposes a favorable pressure gradient, and the compression waves reflected there are weakened and inclined at shallower angles, so that they spread apart rather than converge and do not coalesce into an embedded shock as seen in the MOC solution for this case in Fig.~\ref{fig:moc-pr225-c8}. Further downstream beyond the maximum-diameter point of the jet, the boundary enters the convergent portion of the first cell, where the imposed pressure gradient becomes adverse and the boundary-streamline pressure rises steeply near $x/D_e \approx 1.9$ (Fig.~\ref{fig:p225-boundary-streamline-pressure}). The Mach number along this portion of the boundary decreases with increasing pressure, so the compression waves launched from successive points are inclined at progressively steeper angles to one another. In contrast to the diverging section, these waves converge and are likely to coalesce to form an embedded shock, consistent with the recombination of the characteristics near $x/D_e \approx 1.9$ seen in the MOC solution for this case (Fig.~\ref{fig:moc-pr225-c8}). The embedded shock, when reflected from the centerline may produce a shock connecting the centerline to the jet boundary, similar to the bow-shaped structure observed in Navier-Stokes solution of Fig.~\ref{fig:p225-densitygrad-c8}. With this, we finally note that while viscous effects are indeed important and, as will be shown in Section~\ref{sec:viscous-effects}, are more pronounced at low coflow velocities, an inviscid analysis is sufficient to explain the major changes due to coflow in the shock structures in the near-field, provided the pressure imposed at the jet boundary by the coflow is taken into account. This holds in particular for the transition from Mach to regular reflection, which occurs at relatively higher coflow velocities.

\subsection{Effect of coflow on the shock-cell length}
\label{sec:shock-cell-length}
\begin{figure}[htbp]
    \centering
    \begin{subfigure}[t]{0.48\textwidth}
        \centering
        \includegraphics[
            height=0.26\textheight,
            trim={0.2cm 0.2cm 0.2cm 0.2cm},
            clip
        ]{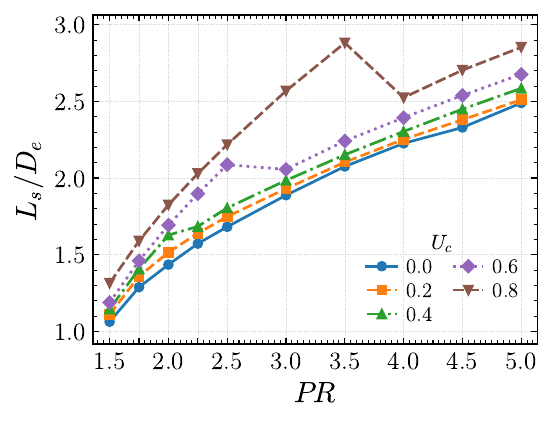}
        \caption{Shock-cell length variation with PR}
        \label{fig:cell-length-vs-PR}
    \end{subfigure}
    \hfill
    \begin{subfigure}[t]{0.48\textwidth}
        \centering
        \includegraphics[
            height=0.26\textheight,
            trim={0.2cm 0.2cm 0.2cm 0.2cm},
            clip
        ]{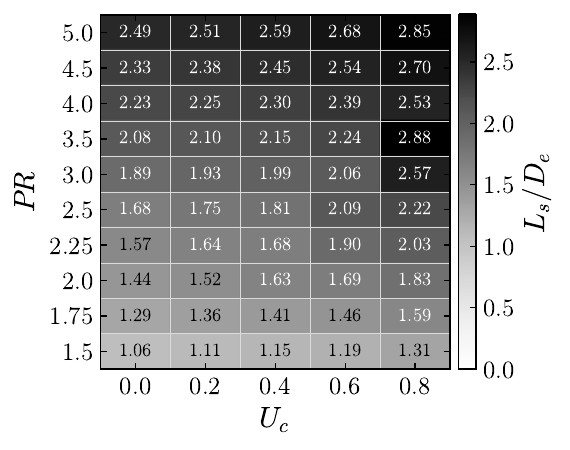}
        \caption{Shock-cell length map over the parameter space}
        \label{fig:cell-length-map}
    \end{subfigure}
    \caption{Variation of the first shock-cell length ($L_s/D_e$) with PR at different coflow velocity ratios $U_c$.}
    \label{fig:shock-cell-length}
\end{figure}

This section discusses the effect of coflow on the streamwise extent of the first shock cell, quantified by the length of the shock cell, $L_s/D_e$. Here $L_s$ is taken as the streamwise location of the first saddle of the jet boundary, \textit{i.e.}, the first local minimum in the boundary streamline radius. 
The cell lengths extracted over the PR--$U_c$ parameter space are shown in Fig.~\ref{fig:cell-length-vs-PR}, and in Fig.~\ref{fig:cell-length-map}. 
In Fig.~\ref{fig:cell-length-vs-PR}, we plot $L_s/D_e$ against PR rather than $U_c$, since the variation with PR follows the classical shock-cell-length scaling with a coflow-dependent correction, as will be discussed in Section~\ref{sec:boundary-shape}.

The shock-cell length increases with PR for every coflow ratio, but the increase is not uniform across the parameter space and depends on the reflection regime (Mach or regular). This dependency is seen in Fig.~\ref{fig:cell-length-vs-PR}, where for $U_c = 0.6$ and $0.8$ the gradually increasing cell length curve falls suddenly at PR values of 2.5 and 3.5 respectively. These combinations of $U_c$ and PR coincide with the transition values identified in Fig.~\ref{fig:mach-disk-diameter-map}, where, at these coflow ratios, increasing PR from identified values takes the centerline shock structure from regular reflection to Mach reflection. The same effect appears more weakly at $U_c = 0.4$, where the transition occurs near PR of 2.0 and the cell length at this combination of (PR, $U_c$) = (2.0, 0.4) increases only marginally across it in Fig.~\ref{fig:cell-length-vs-PR}. The portion of the curves in the figure lying left to each of these (PR, $U_c$) points (\textit{i.e.} (2.0, 0.4), (2.5, 0.6), (3.5, 0.8)) are therefore in regular reflection and those right to them are in Mach reflection.

The sudden fall in shock-cell length at these transition (PR, $U_c$) points may be attributed to the truncation of the shock-cell once the Mach disk forms. In regular reflection, the embedded shock reaches the centerline and the reflected oblique shock is launched from the axis (see Fig.~\ref{fig:regular-reflection-schematic}), whereas in Mach reflection it is launched from the triple point (see Fig.~\ref{fig:mach-reflection-schematic}), which lies off the axis. The reflected shock therefore returns to the jet boundary over a shorter distance from the nozzle exit, closing the cell earlier and shortening it. This can be seen clearly in the $U_c = 0.6$ and $0.8$ columns of Figs.~\ref{fig:app-densitygrad-sweep-pr150-pr300} and \ref{fig:app-densitygrad-sweep-pr350-pr500} in Appendix~\ref{app:coflow-results} where the transition shifts the saddle point of the jet boundary in upstream direction. This regime dependence of the cell length is made explicit through the coflow-corrected cell-length model presented in the next section.

\subsection{A coflow-corrected model for the near-field geometry}
\label{sec:boundary-shape}
The classical description of the shock-cell geometry of underexpanded-jets is attributed to \citet{prandtl1904stationaren} (interpreted through \citet{powell2010prandtl}) and \citet{pack1950note}, among others, who showed that the streamwise cell length of a supersonic jet in a quiescent medium scales as $L_s\propto\sqrt{M_j^2-1}$, where $M_j$ is the fully-expanded jet Mach number related to NPR by
\begin{equation}
  M_j=\sqrt{\frac{2}{\gamma-1}\left[\mathrm{NPR}^{\frac{\gamma-1}{\gamma}}-1\right]},
  \qquad \nu_j \equiv \nu(M_j),
  \label{eq:bs-Mb}
\end{equation}
where $\nu$ is the Prandtl--Meyer function, given in Eq.~\eqref{eq:prandtl-meyer-function}. The classical result models the cell length in the absence of an outer (ambient) stream (zero coflow). For jets in streaming ambient (non-zero coflow), the flight-effects studies reviewed in the Introduction (Section \ref{sec:Introduction}) describe the lengthening through single-coefficient linear corrections. \citet{tam1992broadband} adopted freestream Mach-number based linear correction to shock-cell length given by $L_s^f = L_s(1+0.625\,M_f)$, 
restricted to flight Mach numbers $M_f<0.5$, while \citet{andre2016flighteffects} measured fitted slopes of $0.12$ to $0.22$, increasing with $M_j$, for slightly underexpanded jets up to $M_f=0.39$. Here, with the chosen parameter space in the current study, coflow velocity ratio ($U_c$) coincides with the flight Mach number ($M_f$) \textit{i.e.} $U_c \equiv M_f$, and the notation $M_f$ is retained when quoting results from the literature. These coflow-based corrections are reported with a single coefficient, without reference to the internal reflection regime, and also their validity are reported to be limited over moderate flight speeds ($M_f \lesssim 0.5$). Building on the inclination ($\theta$) and cell-length ($L_s$) observations of Sections~\ref{sec:jet-boundary-inclination} and \ref{sec:shock-cell-length}, respectively, we propose compact expression for both quantities. In particular, the correlation for $L_s$ in the following paragraphs extends the classical scaling to include coflow effect, with separate coefficients in the expression for the two (regular and Mach) reflection regimes.

In the following fits, the coefficient of determination $R^2$ is evaluated on the normalized quantity, $\theta/\nu_j$ for the inclination and $L_s/\sqrt{M_j^2-1}$ for the cell length. Evaluated this way, $R^2$ measures the quality of the fitted coflow correction alone, isolated from the classical scaling. The fitted correlations along with their corresponding $R^2$ values are summarized in Table~\ref{tab:shape-model-coeffs}.

\subsubsection{Inclination angle} 
The inclination of the jet boundary at the nozzle lip ($\theta$), in the quiescent limit, is related to the lip-centered Prandtl--Meyer expansion angle $\nu_j$, as discussed in Section~\ref{sec:jet-boundary-inclination}. As noted there, coflow weakens the expansion, and the measured inclinations (Fig.~\ref{fig:boundary-streamline-initial-angle-coflow}) collapse onto the relation
\begin{equation}
  \theta = 0.85\,\nu_j\left(1-\tfrac13 U_c^2\right),
  \qquad R^2 = 0.77.
  \label{eq:bs-theta}
\end{equation}
This expression is independent of the regime (regular or Mach reflection), and a single curve covers the full PR--$U_c$ parameter space. The prefactor $0.85$ indicates that even in the quiescent limit, the boundary turns roughly through $85\%$ of $\nu_j$, the turning angle of a fully expanded jet, 
which is annotated alongside the measured inclination in Fig.~\ref{fig:p225-boundary-initial-and-pmf-angle}. The factor $(1-\tfrac13U_c^2)$ represents the reduction in turning with increasing coflow. The prefactor (0.85) is not strictly constant. The measured ratio $\theta/\nu_j$ in the quiescent limit increases with pressure ratio, from about $0.76$ at $\mathrm{PR}=1.5$ to about $0.88$ at $\mathrm{PR}=5$, so that Eq.~\eqref{eq:bs-theta} with a single prefactor slightly overpredicts the inclinations at low PRs and underpredicts it at high PRs. The resulting deviations remain within $3\%$ on average over the parameter space. 
The inclination decreases as the square of the coflow velocity. The quadratic form implies near-insensitivity at low coflow, the correction being below $2\%$ for $U_c \leq 0.2$.

\subsubsection{Shock-cell length}
\begin{figure}[t]
  \centering
  \begin{subfigure}[t]{0.48\textwidth}
    \centering
    \includegraphics[
                height=0.24\textheight,
                trim={0.2cm 0.2cm 0.2cm 0.2cm},
                clip
            ]{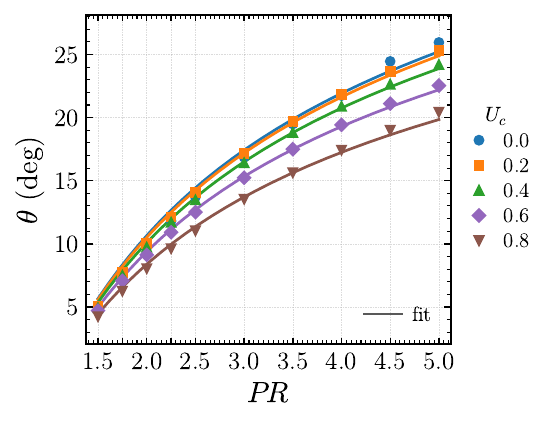}
    \caption{Jet-boundary inclination, Eq.~\eqref{eq:bs-theta}}
    \label{fig:bs-theta-fit}
  \end{subfigure}
  \hfill
  \begin{subfigure}[t]{0.48\textwidth}
    \centering
    \includegraphics[
                height=0.24\textheight,
                trim={0.2cm 0.2cm 0.2cm 0.2cm},
                clip
            ]{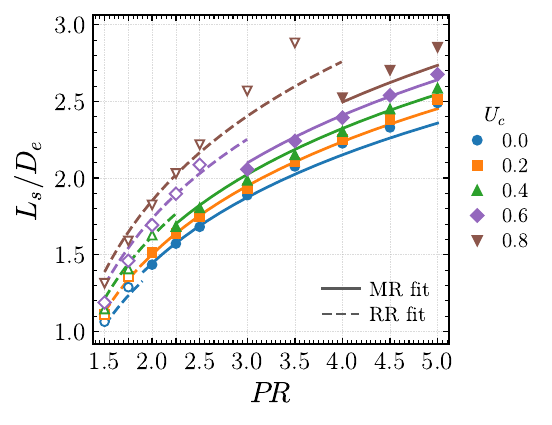}
    \caption{First shock-cell length, Eqs.~\eqref{eq:bs-Ls-MR} and \eqref{eq:bs-Ls-RR}}
    \label{fig:bs-Ls}
  \end{subfigure}
  \caption{Measured near-field shock-cell geometry against the fitted correlations}
  \label{fig:bs}
\end{figure}

The cell-length variation characterized in Section~\ref{sec:shock-cell-length} and plotted in Fig.~\ref{fig:shock-cell-length} is captured by retaining the classical $\sqrt{M_j^2-1}$ scaling modified through an added linear coflow dependence. The cell lengths scale as
\begin{align}
  L_s^{\rm MR} &= 1.26\,\sqrt{M_j^2-1}\,\left(1+0.20\,U_c\right),
  \qquad R^2=0.85,
  \label{eq:bs-Ls-MR}\\[2pt]
  L_s^{\rm RR} &= 1.21\,\sqrt{M_j^2-1}\,\left(1+0.42\,U_c\right),
  \qquad R^2=0.79,
  \label{eq:bs-Ls-RR}
\end{align}
for the Mach-reflection (MR) and regular-reflection (RR) regimes, respectively, and are plotted in Fig.~\ref{fig:bs-Ls}. Both reduce to Prandtl's scaling $L_s\propto\sqrt{M_j^2-1}$ as $U_c$ goes to 0. 
The two prefactors (1.26 and 1.21) are close to each other and to the classical value of 1.22 \cite{pack1950note}. 
The coflow factor $(1+c_{\rm reg}U_c)$ lengthens the cell, and does so more strongly in the regular-reflection regime. For comparison, the slopes ($c_{\rm reg}$) measured by \citet{andre2016flighteffects}, $0.12$ to $0.22$, are of the same order as the present MR coefficient, while the value $0.625$ adopted by \citet{tam1992broadband} is closer to the present RR coefficient. However, a one-to-one comparison is not possible because the jets of \citet{andre2016flighteffects} are only slightly underexpanded and do not form a Mach disk, so they nominally correspond to the regular-reflection branch, yet their fitted slopes lie below the present RR value of $0.42$. The difference plausibly reflects their much lower pressure ratios, their slopes representing an average over several upstream cells (the second to fifth) rather than the first cell alone, and the residual outer boundary layer of their dual-stream rig, which, though much thinner than in the early flight-simulation facilities~\citep{norum1993simulated,norum1984effects,norum1988shock}, may still partially shield the near-exit cells in a way the present clean-inlet configuration does not. The regime split also offers an explanation for the limited ranges of validity reported for single-coefficient corrections~\citep{norum1993simulated, tam1992broadband}. As the coflow increases, the jet transitions from Mach to regular reflection, and the coflow sensitivity of the cell length roughly doubles, so that a single coefficient calibrated at low coflow underestimates the lengthening at higher coflow, precisely where those fits are reported to break down.

\begin{table}[htbp]
\centering
\caption{Fitted correlations for the near-field shock-cell geometry. The coefficient of determination $R^2$ is evaluated on the normalized quantity ($\theta/\nu_j$ and $L_s/\sqrt{M_j^2-1}$) to reflect the values over coflow correction alone.}
\label{tab:shape-model-coeffs}
\begin{tabular}{llccc}
\toprule
Quantity & Fitted correlation & Regime & Eq. & $R^2$ \\
Initial inclination, $\theta$ & $0.85\,\nu_j\left(1-\tfrac13 U_c^2\right)$      & both & \eqref{eq:bs-theta} & $0.77$ \\
Cell length, $L_s^{\rm MR}$   & $1.26\,\sqrt{M_j^2-1}\,\left(1+0.20\,U_c\right)$ & MR   & \eqref{eq:bs-Ls-MR} & $0.85$ \\
Cell length, $L_s^{\rm RR}$   & $1.21\,\sqrt{M_j^2-1}\,\left(1+0.42\,U_c\right)$ & RR   & \eqref{eq:bs-Ls-RR} & $0.79$ \\
\end{tabular}
\end{table}

\subsection{Effect of coflow on viscous mixing at the jet boundary}
\label{sec:viscous-effects}

The discussion in the preceding sections has shown that the principal structural changes in the near-field shock system with increasing coflow, \textit{i.e.}, the modification of the pressure distribution along the jet boundary (Section~\ref{sec:jet-boundary-streamline-pressure}), and the consequent weakening, reorientation, and eventual disappearance of the embedded shock (Section~\ref{sec:method-of-characteristics}) are well captured by inviscid arguments. Viscous effects (diffusion) do not appear to be the primary driver of these large-scale shock structure changes. 

It should be emphasized that the present simulations enforce axisymmetry, so the shear layer here develops only through axisymmetric (ring-mode) roll-up; the fully three-dimensional azimuthal instability and subsequent turbulent breakdown from which the canonical mixing-layer scalings of Refs.~\cite{brown1974density,abramovich1963theory,dimotakis1986entrainment} were derived are structurally excluded by this formulation. The comparisons drawn below to that scaling should accordingly be read as evidence that the layer responds to the local velocity difference in the manner expected of a shear layer, not as a quantitative test of turbulent mixing-layer theory.

Viscosity governs the structure of the thin shear layer that forms at the interface between the supersonic jet core and the surrounding fluid. This layer is visible in the density-gradient contours of Fig.~\ref{fig:PR2.25-coflowing-jets} as a narrow band of elevated $|\nabla \bar{\rho}|$ bounding the jet, and this band is seen to become progressively thinner with increasing coflow ratio $U_c$. This thinning is consistent with the reduction in the velocity (or momentum) difference across the layer as the coflow velocity approaches the nozzle-exit velocity. A smaller velocity difference weakens the Kelvin--Helmholtz-type instability of the shear layer and the growth of the associated vortical structures responsible for entrainment, mixing, and spreading, thereby slowing the growth of the shear-layer thickness along the jet~\cite{brown1974density,abramovich1963theory,dimotakis1986entrainment}.

\begin{figure}[htbp]
    \centering
    \begin{subfigure}[t]{0.48\textwidth}
        \centering
        \includegraphics[height=0.24\textheight]{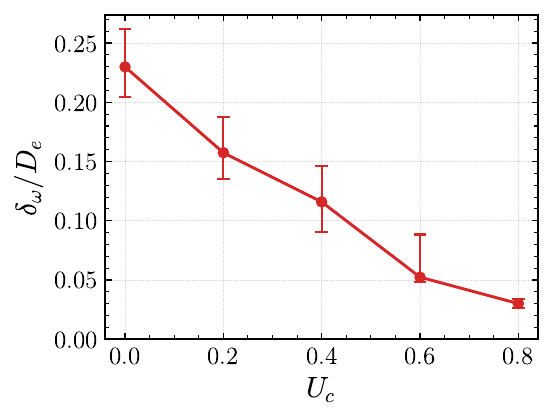}
        \caption{Variation with coflow velocity ratio $U_c$}
        \label{fig:vorticity-vs-coflow}
    \end{subfigure}
    \hfill
    \begin{subfigure}[t]{0.48\textwidth}
        \centering
        \includegraphics[height=0.24\textheight]{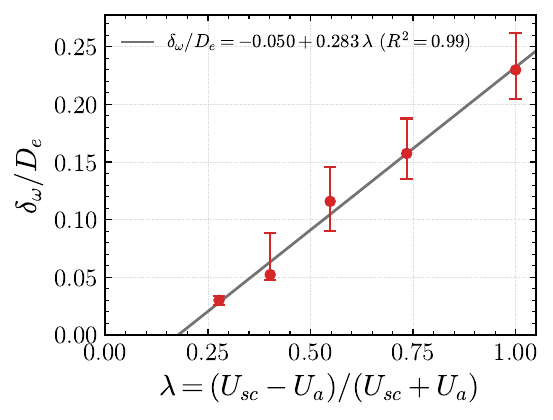}
        \caption{Collapse onto $\lambda=(U_{sc}-U_a)/(U_{sc}+U_a)$}
        \label{fig:vorticity-vs-lambda}
    \end{subfigure}
    \caption{Vorticity thickness $\delta_{\omega}$ of the shear layer for the NPR~$=4.26$ (PR~$=2.25$) jet, evaluated as the median over the region $1.5 \le x/D_e \le 3$ (error bars: the interquartile range for median). (a)~$\delta_{\omega}$ decreases monotonically with coflow; 
    (b)~$\delta_{\omega}$ varies linearly with the velocity-ratio parameter $\lambda$, consistent with the canonical free-shear-layer scaling.}
    \label{fig:shear-layer-thickness-vs-coflow}
\end{figure}

To quantify this trend, we characterize the shear layer by its vorticity thickness at a given axial location~\citep{brown1974density},
\begin{equation}
    \delta_{\omega}(x) =
    \frac{U_{sc} - U_a}{\left| \partial \bar{u}/\partial r \right|_{\max}},
    \label{eq:vorticity-thickness}
\end{equation}
where $U_{sc}$ is the local supersonic-core velocity, 
$U_a$ is the ambient flow velocity, and $\left| \partial \bar{u}/\partial r \right|_{\max}$ is the peak radial gradient of the mean axial velocity $\bar{u}(r)$ across the layer. Because the shock-cell structure modulates the layer along its length, $\delta_{\omega}$ is evaluated as the median over the developed region $1.5 \le x/D_e \le 3$. Figure~\ref{fig:vorticity-vs-coflow} shows $\delta_{\omega}$ as a function of $U_c$ for the NPR~$=4.26$ (PR~$=2.25$) jet. The vorticity thickness is found to decrease monotonically with increasing coflow confirming the qualitative trend observed in Fig.~\ref{fig:PR2.25-coflowing-jets} and indicating that increasing coflow progressively suppresses viscous mixing at the jet boundary.
 
This thinning follows the scaling expected for a free shear layer. When plotted against the velocity-ratio parameter $\lambda = (U_{sc} - U_a)/(U_{sc} + U_a)$, formed from the local supersonic-core and coflow velocities across the layer, the vorticity thickness collapses onto an approximately straight line (Fig.~\ref{fig:vorticity-vs-lambda}), $\delta_{\omega}/D_e \approx 0.28\,\lambda$ ($R^2 = 0.99$). The parameter $\lambda$ is precisely the one that governs the growth rate of a canonical turbulent mixing layer~\cite{brown1974density,abramovich1963theory,dimotakis1986entrainment}, and indicates that the layer's thinning is a direct consequence of the reduced shear as the coflow velocity approaches the core velocity.

At low coflow velocities, where $\delta_{\omega}$ is larger, viscous effects at the jet boundary are more pronounced, and this may be relevant to a feature of Fig.~\ref{fig:mach-disk-diameter-vs-coflow} that is not immediately accounted for by the inviscid analysis of Section~\ref{sec:method-of-characteristics-pressure-boundary}. Over the range $0 \le U_c \le 0.2$, while the MOC analysis indicated that the Mach disks for the $\mathrm{PR}=2.25$ jet considered in this section would be of similar nature, from the Navier-Stokes solutions of Fig.~\ref{fig:mach-disk-diameter-vs-coflow} the Mach-disk diameters in the two cases are seen to change substantially. One possible, though still tentative, explanation is that at these low coflow ratios the shear layer is thick enough to itself modify the effective inviscid boundary felt by the supersonic core, for instance, by smearing the pressure field that the reflected compression waves respond to. Such an effect is not captured by the inviscid analysis of Section~\ref{sec:method-of-characteristics}, which uses the pressure profile along the streamline launched from the nozzle lip as boundary pressure. As $U_c$ increases beyond $U_c=0.2$ and $\delta_{\omega}$ decreases substantially, the viscous effects diminish, and the solutions from inviscid analysis start to agree with Navier-Stokes solutions. We note that this explanation has not been independently verified. Establishing it more firmly would require directly relating the displacement effect of the shear layer to the jet-boundary pressure field used in the MOC analysis, which we leave to future work.

\section{Conclusions}
\label{sec:Conclusions}
The near-field shock structures of underexpanded coflowing jets exiting from a convergent nozzle ($M_e=1$)  were studied over a two-parameter space of nozzle pressure ratio (NPR) and subsonic coflow velocity ratio ($U_c$). 
Axisymmetric numerical solutions, supported by an inviscid method-of-characteristics (MOC) analysis, were used to obtain the flow statistics and explain the observed trends. The main conclusions of the study are as follows.
\begin{enumerate}
\item Increasing the subsonic coflow suppresses Mach-disk formation: At a fixed pressure ratio, increasing the coflow favors the transition from Mach reflection (with Mach-disk) at the centerline to regular reflection (with oblique shocks). Hence, this shrinks the Mach-disk diameter of a highly underexpanded jet until the Mach-disk vanishes. Equivalently, coflow increases the pressure ratio for transition from regular to Mach reflection. To the authors' knowledge, this is the first systematic mapping of the subsonic-coflow effect on Mach-disk formation across the NPR-$U_c$ space.

\item A single mechanism was identified that is sufficient to explain the suppression of Mach-disk formation with increasing coflow. The coflow imposes a non-uniform static pressure along the jet boundary, replacing the constant-pressure condition of a jet exiting into a quiescent medium. This acts on the wave system in the near-field region of the jet in two ways:
  \begin{enumerate}
  \item At the nozzle lip, coflow raises the local static pressure to a higher value. This then relaxes the need for strong expansion at the nozzle lip, so the lip Prandtl--Meyer fan is confined to a smaller angle and consequently the inclination of the jet boundary at the nozzle lip decreases. This reduced inclination is the cause previously reported in the literature for Mach-disk suppression by coflow, and here we have shown that it is a local consequence of the imposed pressure rather than an independent cause.
  \item Downstream of the nozzle lip, along the remainder of the boundary, the non-uniform pressure weakens the boundary-reflected compression waves and spread them apart at shallower angles. For sufficiently large coflows, these waves do not coalesce to form an embedded shock.
  \end{enumerate}
  The embedded shock is thus weakened or its formation is suppressed altogether. The centerline reflection is then regular, instead of the Mach reflection that produces a Mach disk.

\item The mechanism was verified by the (inviscid) MOC analysis, which imposed pressure profiles from the Navier-Stokes simulations as the boundary condition. The analysis reproduces the downstream shift, and eventual disappearance, of the compression-wave coalescence as the coflow increases. This confirms that the transition from Mach- to regular-reflection with increasing coflow are governed primarily by inviscid dynamics, and that the coflowing ambient communicates with the supersonic core through the modified pressure at the jet boundary.

\item Coflow effects on the first shock cell were characterized. The jet-boundary inclination ($\theta$) decreases quadratically with coflow, and the first shock-cell length ($L_s$) increases linearly with it. The classical Prandtl-Pack scaling~\citep{prandtl1904stationaren,pack1950note} for $L_s$ in underexpanded jets with no coflow is extended for coflowing jets using the factor $(1+c_{\mathrm{reg}}U_c)$. The slope $c_{\mathrm{reg}}$ was found to be reflection-regime dependent, with $c_{\mathrm{reg}} = 0.20$ in the Mach-reflection regime and $0.42$ in the regular-reflection regime, reducing to the classical result in the quiescent ambient limit.

\item Viscous effects were found to be pronounced for low coflow jets, while they get confined to a thin shear layer with increasing coflow velocities. Therefore, these effects are not the primary driver of the shock-structure transitions, which occurs at relatively higher coflow velocities. Furthermore, the shear-layer thickness decreases with increasing coflow, as expected for a free shear layer.
\end{enumerate}

\setcounter{section}{0} 
\renewcommand{\thesection}{\Alph{section}}

\makeatletter
\providecommand{\theHsection}{\thesection}
\renewcommand{\theHsection}{app.\Alph{section}}
\makeatother

\section*{Appendix}
\newcommand{\JetSweepImg}[2]{%
    \includegraphics[
        width=0.155\linewidth,
        trim={0 0.5cm 0 0},
        clip
    ]{figures/contours/density_grad_sweep/C#1P#2_TIMAVG.png}%
}
\section{Grid-independence results and comparisons with three-dimensional simulation}
\label{app:grid_test_3_D}
\begin{figure}[htbp]
    \centering
    \begin{subfigure}[t]{0.48\textwidth}
        \centering
        \includegraphics[
            width=\linewidth
        ]{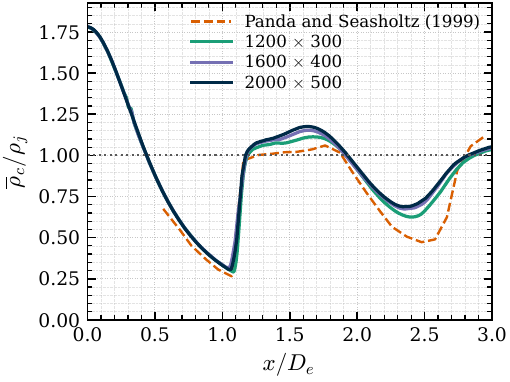}
        \caption{}
        \label{fig:Grid_independence}
    \end{subfigure}
    \hfill
    \begin{subfigure}[t]{0.48\textwidth}
        \centering
        \includegraphics[
            width=\linewidth
        ]{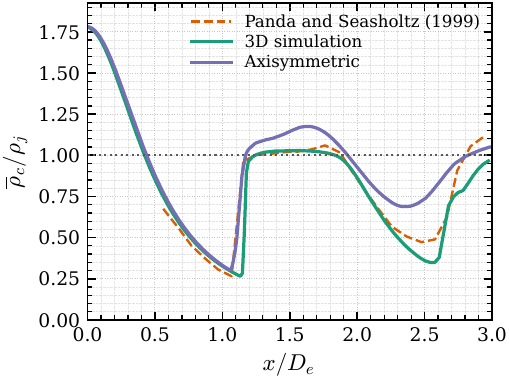}
        \caption{}
        \label{fig:3D_axi_compare}
    \end{subfigure}
    \label{fig:Grid_independence_plus_3D_compare}
    \caption{(a) Grid independence test showing centerline density for different grids. (b) Comparison between the axisymmetric and 3-D simulation results.}
\end{figure}

To ensure grid convergence, the flow statistics are compared from coarse (\(1200\times300=1.0\times10^6\)), intermediate (\(1600\times400=6.4\times10^5\)), and fine (\(2000\times500=1.0\times10^6\)) grid simulations.
The comparison of the mean centerline density, shown in Fig.~\ref{fig:Grid_independence}, indicates that the solution becomes grid-independent as the resolution is refined from the intermediate (\(1600\times400\)) to the fine (\(2000\times500\)) grid.

\begin{figure}[htbp]
    \centering
    \begin{subfigure}[t]{0.48\textwidth}
        \centering
        \includegraphics[
            width=\linewidth
        ]{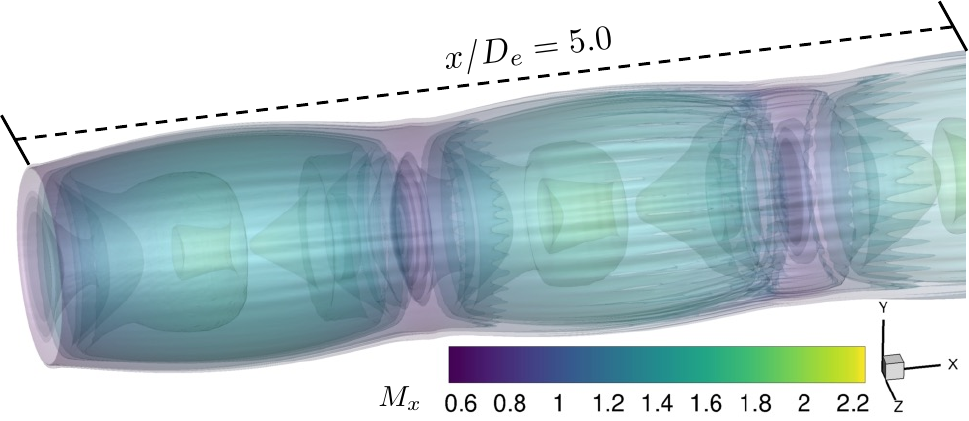}
        \caption{}
        \label{fig:3D_0_8_coflow}
    \end{subfigure}
    \hfill
    \begin{subfigure}[t]{0.48\textwidth}
        \centering
        \includegraphics[
            width=\linewidth
        ]{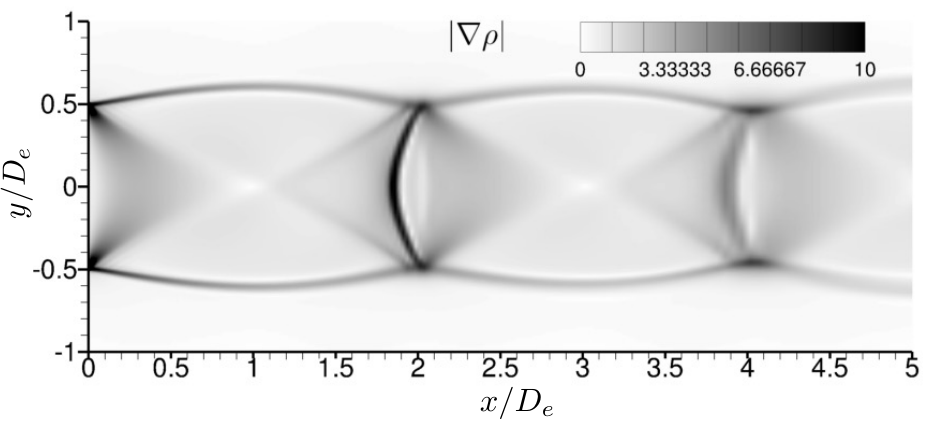}
        \caption{}
        \label{fig:2D_slice_0_8_coflow}
    \end{subfigure}
    \caption{Three-dimensional simulation of the $U_c=0.8$ jet at $\textrm{PR}=2.25$: (a) three-dimensional density gradient isosurface coloured by axial Mach number; (b) density gradient in the $x$--$y$ plane through $z=0$, showing the weak bow-shaped shock structure.}
    \label{fig:3D_0_8_coflow_compare}
\end{figure}

Furthermore, to justify the use of axisymmetric simulations in this study to evaluate the shock-structure transitions in the first shock cell, the results of a three-dimensional (3-D) simulation are compared with the axisymmetric results.
Figure~\ref{fig:3D_axi_compare} provides the mean centerline density obtained from the axisymmetric and 3-D simulations for $M_j=1.6$. The centerline density profiles show very good agreement up to the first Mach disk, supporting the use of the axisymmetric solver for studying Mach-disk formation at different coflow velocities.
Both three-dimensional simulations presented in this appendix -- this comparison (Figure~\ref{fig:3D_axi_compare}) and the confirmation of the bow-shock structure below (Figure~\ref{fig:3D_0_8_coflow_compare}) -- are performed at a reduced Reynolds number of $\mathrm{Re}=5300$, compared with $\mathrm{Re}=50{,}000$ used in the axisymmetric production runs, for computational tractability. The quantities compared here, namely the on-axis (centerline) density up to the first Mach disk and the existence and location of the embedded-shock features, are governed primarily by the inviscid core wave system, a conclusion supported independently by the close agreement between the inviscid MOC analysis and the Navier--Stokes solutions documented in Appendix~\ref{app:moc-validity}. These centerline and shock-topology comparisons are accordingly expected to be comparatively insensitive to this reduction in Reynolds number.

The axisymmetric solutions for the highest coflow case ($U_c=0.8$) revealed a distinctive weak, concave (bow-shaped) shock structure around $x/D_e \simeq 1.9$ (Section~\ref{sec:near-field-structures}). A three-dimensional direct numerical simulation (DNS) of the $U_c=0.8$, $\textrm{PR}=2.25$ ($\textrm{NPR}=4.26$) jet 
is performed to confirm that it is a genuine feature of the flow and not just an axisymmetric artifact. Figure~\ref{fig:3D_0_8_coflow} shows the three-dimensional density gradient isosurface coloured by axial Mach number, and Fig.~\ref{fig:2D_slice_0_8_coflow} shows the density gradient in the $x$--$y$ plane through $z=0$. The three-dimensional solution reproduces the same concave structure at around the same axial location as the axisymmetric solution, confirming that it is a genuine feature of the coflowing jet.

\section{Domain independence results \label{app:domain-extent}}
\begin{figure}
    \centering
    \begin{subfigure}[t]{0.48\textwidth}
        \centering
        \includegraphics[
            width=\linewidth,
            trim={1.0cm 1.0cm 8.0cm 1.0cm},
            clip
        ]{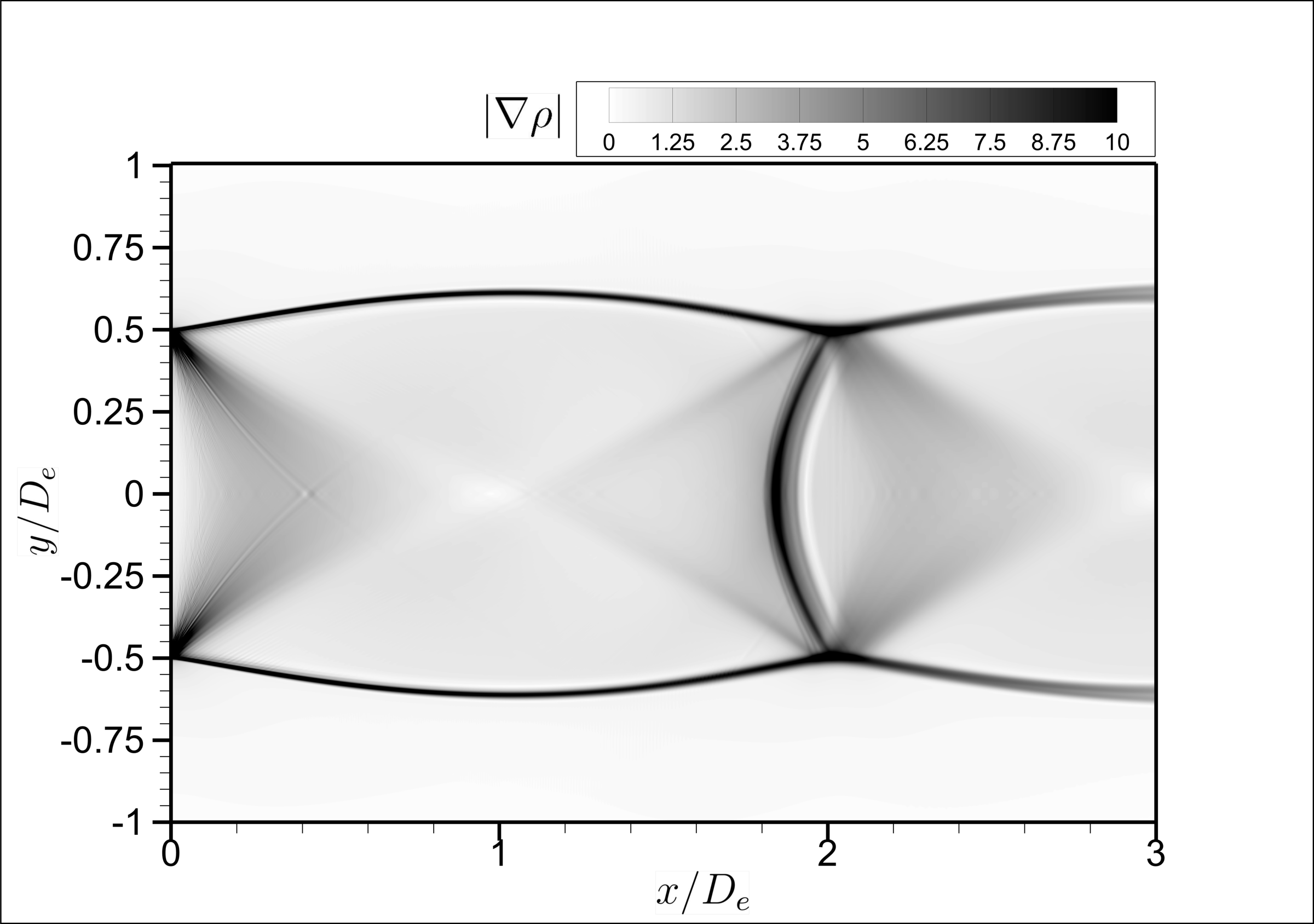}
        \caption{}
        \label{fig:c8p225-domain-extent-75}
    \end{subfigure}
    \hfill
    \begin{subfigure}[t]{0.48\textwidth}
        \centering
        \includegraphics[
            width=\linewidth,
            trim={1.0cm 1.0cm 8.0cm 1.0cm},
            clip
        ]{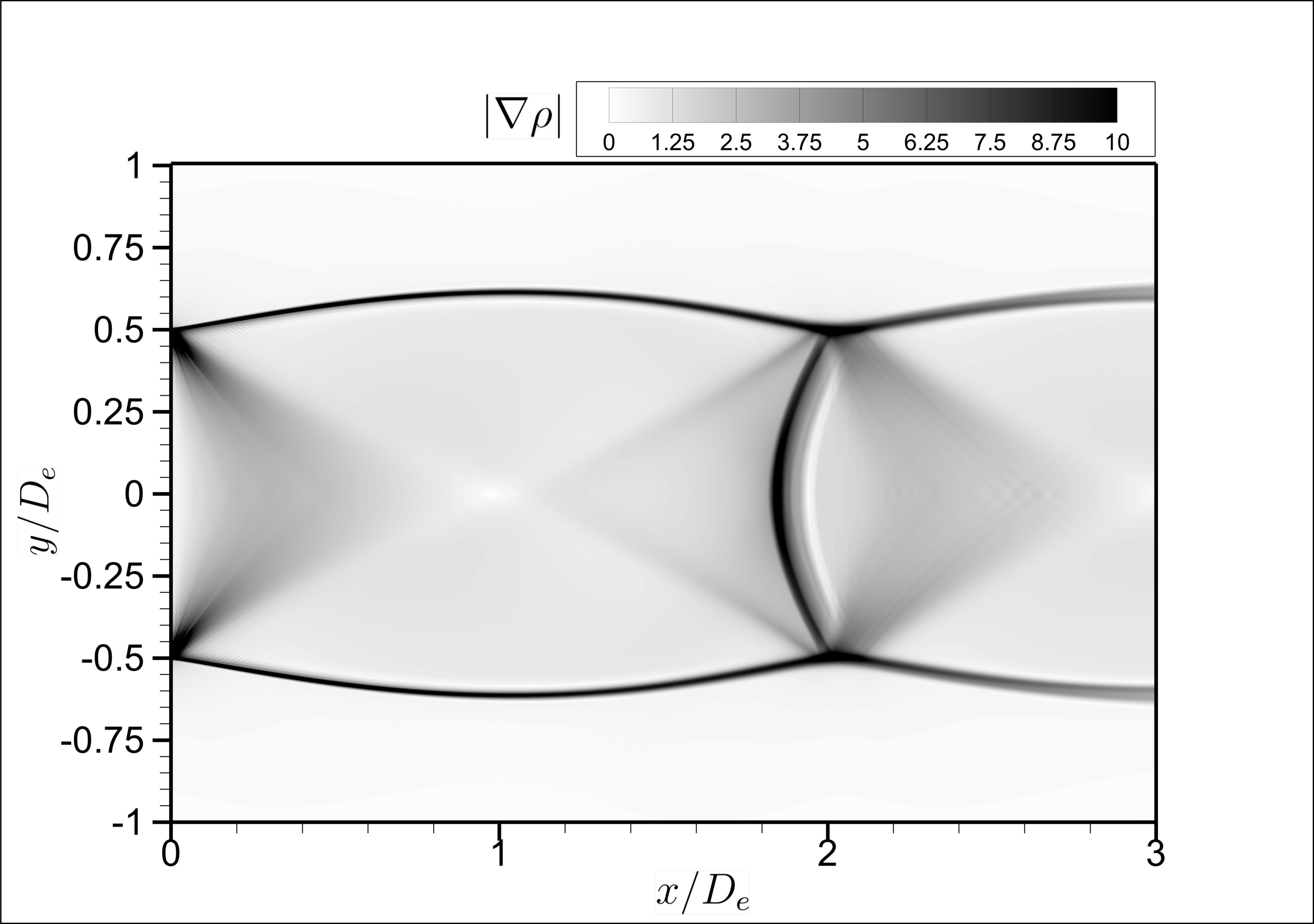}
        \caption{}
        \label{fig:c8p225-domain-extent-10}
    \end{subfigure}
    \caption{Density gradient contours for the $\mathrm{PR} = 2.25$ ($\mathrm{NPR} = 4.26$) jet at $U_c=0.8$ with extended domains: (a) $\left(x/D_e\right)_{\max} = 20$, $\left(y/D_e\right)_{\max} = 7.5$; and (b) $\left(x/D_e\right)_{\max} = 20$, $\left(y/D_e\right)_{\max} = 10$.}
    \label{fig:densitygrad-domain-extent}
\end{figure}

\begin{figure}
    \centering
    \includegraphics[width=0.75\textwidth]{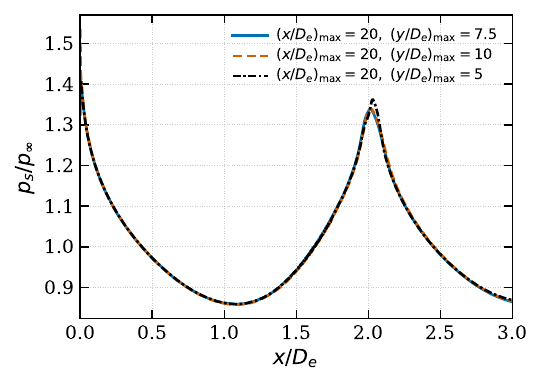}
    \caption{Comparison of the boundary streamline pressure variation for the $\mathrm{NPR} = 4.26$ ($\mathrm{PR} = 2.25$) jet at $U_c=0.8$ across different domain extents.}
    \label{fig:boundary-streamline-pressure-domain-extent}
\end{figure}

To ensure that the non-uniform pressure profiles observed in Section~\ref{sec:jet-boundary-streamline-pressure} are not dependent on the extent of the radial domain used in the study, simulations are repeated for PR = 2.25 (NPR = 4.26) jet for $U_c = 0.8$ coflow case over two larger domains with radial domains up to $y = 7.5D_e$ and $10D_e$. The corresponding density gradient contours are shown in Fig.~\ref{fig:densitygrad-domain-extent} and the pressure probed along the boundary streamline of the new domains are compared against the results of the domain used in the current study in Fig.~\ref{fig:boundary-streamline-pressure-domain-extent}. The pressure profiles along the boundary streamline across the three domains almost overlap with each other, confirming that the profiles are not dependent on the extent of the domain used in the current simulations.

\begin{landscape}
\section{Time-averaged density-gradient contours}
\label{app:coflow-results}

\noindent
\begin{minipage}{\linewidth}
\centering
\scriptsize
\setlength{\tabcolsep}{1.5pt}
\renewcommand{\arraystretch}{0.65}

\begin{tabular}{c c c c c c}
    & $U_c=0.0$ & $U_c=0.2$ & $U_c=0.4$ & $U_c=0.6$ & $U_c=0.8$ \\

    \raisebox{6.0\height}{$\textrm{PR}=1.50$} &
    \JetSweepImg{0}{15} &
    \JetSweepImg{2}{15} &
    \JetSweepImg{4}{15} &
    \JetSweepImg{6}{15} &
    \JetSweepImg{8}{15} \\

    \raisebox{6.0\height}{$\textrm{PR}=1.75$} &
    \JetSweepImg{0}{175} &
    \JetSweepImg{2}{175} &
    \JetSweepImg{4}{175} &
    \JetSweepImg{6}{175} &
    \JetSweepImg{8}{175} \\

    \raisebox{6.0\height}{$\textrm{PR}=2.00$} &
    \JetSweepImg{0}{20} &
    \JetSweepImg{2}{20} &
    \JetSweepImg{4}{20} &
    \JetSweepImg{6}{20} &
    \JetSweepImg{8}{20} \\

    \raisebox{6.0\height}{$\textrm{PR}=2.50$} &
    \JetSweepImg{0}{25} &
    \JetSweepImg{2}{25} &
    \JetSweepImg{4}{25} &
    \JetSweepImg{6}{25} &
    \JetSweepImg{8}{25} \\

    \raisebox{6.0\height}{$\textrm{PR}=3.00$} &
    \JetSweepImg{0}{30} &
    \JetSweepImg{2}{30} &
    \JetSweepImg{4}{30} &
    \JetSweepImg{6}{30} &
    \JetSweepImg{8}{30} \\
\end{tabular}

\vspace{0.15cm}


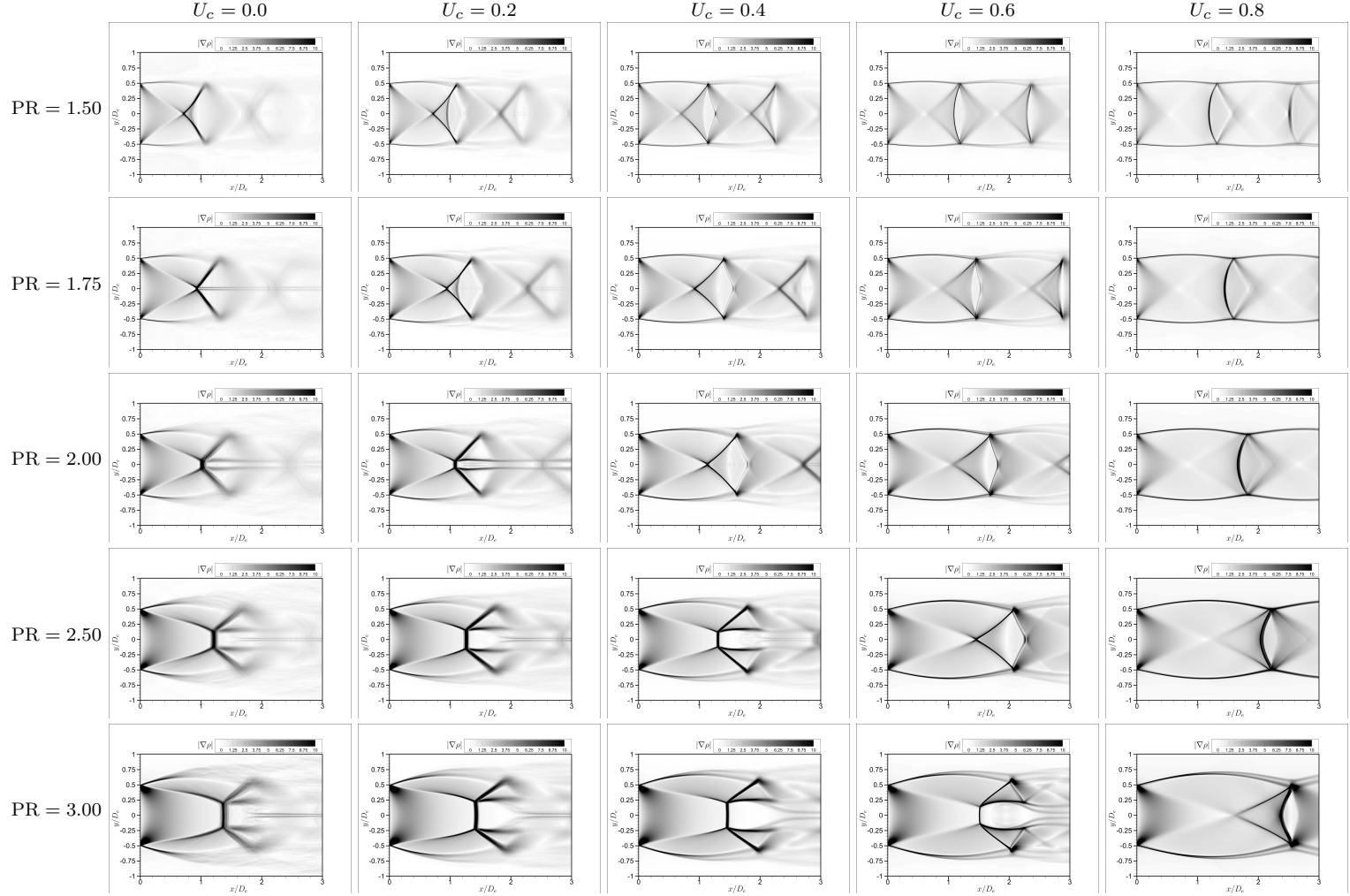
\captionof{figure}{Time-averaged density-gradient contours for $1.50 \leq PR \leq 3.00$ with increasing coflow velocity ratio $U_c$.}
\label{fig:app-densitygrad-sweep-pr150-pr300}

\end{minipage}
\end{landscape}

\begin{landscape}
\noindent
\begin{minipage}{\linewidth}
\centering
\scriptsize
\setlength{\tabcolsep}{1.5pt}
\renewcommand{\arraystretch}{0.65}

\begin{tabular}{c c c c c c}
    & $U_c=0.0$ & $U_c=0.2$ & $U_c=0.4$ & $U_c=0.6$ & $U_c=0.8$ \\

    \raisebox{6.0\height}{$\textrm{PR}=3.50$} &
    \JetSweepImg{0}{35} &
    \JetSweepImg{2}{35} &
    \JetSweepImg{4}{35} &
    \JetSweepImg{6}{35} &
    \JetSweepImg{8}{35} \\

    \raisebox{6.0\height}{$\textrm{PR}=4.00$} &
    \JetSweepImg{0}{40} &
    \JetSweepImg{2}{40} &
    \JetSweepImg{4}{40} &
    \JetSweepImg{6}{40} &
    \JetSweepImg{8}{40} \\

    \raisebox{6.0\height}{$\textrm{PR}=4.50$} &
    \JetSweepImg{0}{45} &
    \JetSweepImg{2}{45} &
    \JetSweepImg{4}{45} &
    \JetSweepImg{6}{45} &
    \JetSweepImg{8}{45} \\

    \raisebox{6.0\height}{$\textrm{PR}=5.00$} &
    \JetSweepImg{0}{50} &
    \JetSweepImg{2}{50} &
    \JetSweepImg{4}{50} &
    \JetSweepImg{6}{50} &
    \JetSweepImg{8}{50} \\
\end{tabular}

\vspace{0.15cm}


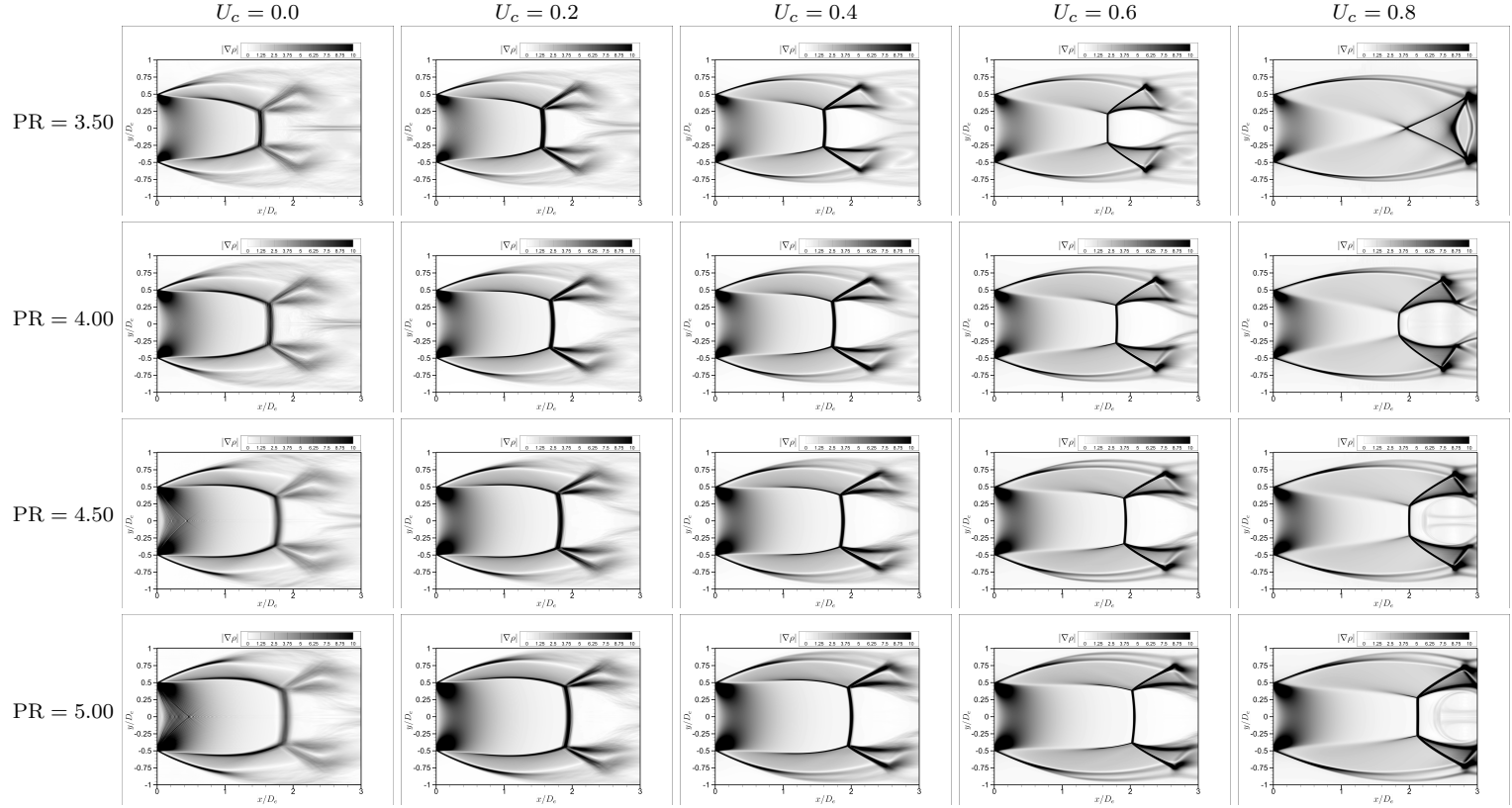
\captionof{figure}{Time-averaged density-gradient contours for $3.50 \leq PR \leq 5.00$ with increasing coflow velocity ratio $U_c$.}
\label{fig:app-densitygrad-sweep-pr350-pr500}

\end{minipage}
\end{landscape}

\section{Comparison of the method-of-characteristics and Navier--Stokes solutions}
\label{app:moc-validity}

\begin{figure}[htbp]
    \centering
    \includegraphics[width=\textwidth]{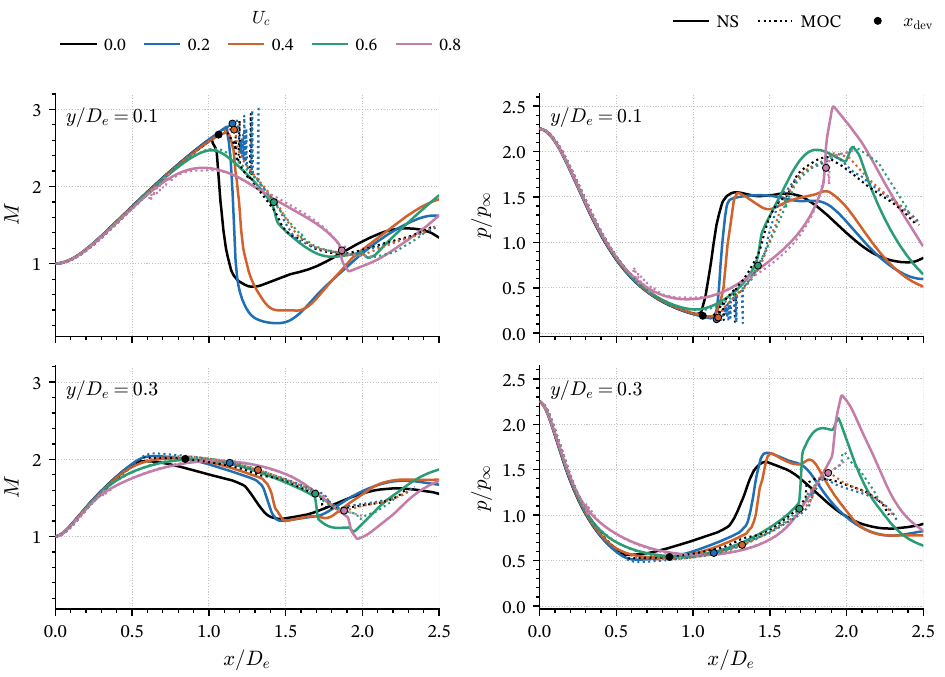}
    \caption{Comparison of the method-of-characteristics (MOC) and Navier--Stokes (NS) solutions at radial stations $y/D_e = 0.1$ (top) and $y/D_e = 0.3$ (bottom), for Mach number (left) and static pressure (right), at coflow velocity ratios $U_c = 0.0$ to $0.8$. Solid lines denote the NS solution and dotted lines the MOC solution, with the circles marking the location $x_{\mathrm{dev}}$ at which the two static pressures first differ by more than $0.1\,p_\infty$.}
    \label{fig:moc-vs-ns-stations}
\end{figure}

\begin{figure}[htbp]
    \centering
    \includegraphics[width=0.5\textwidth]{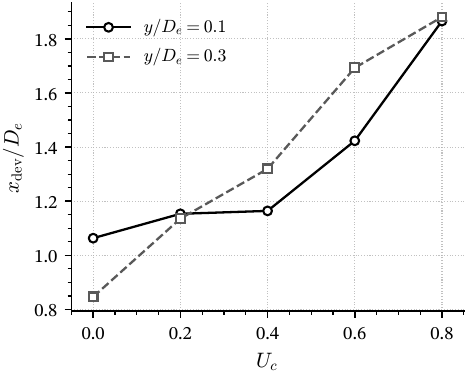}
    \caption{Axial location $x_{\mathrm{dev}}$ at which the method-of-characteristics and Navier--Stokes solutions begin to differ, plotted against the coflow velocity ratio $U_c$ for the two radial stations.}
    \label{fig:departure-vs-coflow}
\end{figure}

The method-of-characteristics (MOC) analysis of Section~\ref{sec:method-of-characteristics} assumes isentropic, irrotational flow, and accounts for the initial expansion fan, its reflection from the centerline, and the resulting compression waves reflected from the jet boundary. It does not perform the subsequent reflection of these compression waves from the centerline, nor does it capture the shock waves directly. Fig.~\ref{fig:moc-vs-ns-stations} compares the Mach number and static pressure obtained from the MOC analysis and the NS simulation at two radial stations, and Fig.~\ref{fig:departure-vs-coflow} shows the axial location at which they begin to differ, taken as the first point at which the static pressures depart by more than $0.1\,p_\infty$.

Over the initial portion of the jet the two solutions follow one another closely at all coflow ratios, reproducing the same expansion from the nozzle lip and the onset of the subsequent compression. The extent of this agreement varies with coflow. At  higher coflows, where the centerline reflection is regular, the two solutions remain in close agreement over most of the shown region. At lower coflows, where a Mach reflection and the associated shock system are present in the Navier--Stokes solution, the agreement extends over a shorter axial distance, the departure occurring where the flow passes through shock structures that the isentropic analysis does not represent. At the inner station $(y/D_e = 0.1)$ the MOC profiles for these cases also show non-physical oscillations in the same region, arising where the like characteristics begin to fold over one another.

The comparison is presented here to indicate the axial extent over which the two solutions exhibit the same behaviour. The conclusions drawn from the MOC analysis concern the compression waves reflected from the jet boundary and the onset of their coalescence, which occur within this extent.

\section*{Funding Sources}
This material is based upon work supported in part by the National Science Foundation (NSF) DMS Award 2513865 and National Aeronautics and Space Administration (NASA) under Grant NASA-80NSSC22M0050 issued through the NASA EPSCoR program. 

\section*{Acknowledgments}
The computational resources were provided by the Advanced Cyberinfrastructure Coordination Ecosystem: Services \& Support (ACCESS) grants PHY210037 and PHY240185 and the Auburn University Easley Cluster.

\section*{Data Availability}
The data that support the findings of this article are not publicly available. The data are available from the authors upon reasonable request.


\bibliography{apssamp}

\begin{thebibliography}{51}%
\makeatletter
\providecommand \@ifxundefined [1]{%
 \@ifx{#1\undefined}
}%
\providecommand \@ifnum [1]{%
 \ifnum #1\expandafter \@firstoftwo
 \else \expandafter \@secondoftwo
 \fi
}%
\providecommand \@ifx [1]{%
 \ifx #1\expandafter \@firstoftwo
 \else \expandafter \@secondoftwo
 \fi
}%
\providecommand \natexlab [1]{#1}%
\providecommand \enquote  [1]{``#1''}%
\providecommand \bibnamefont  [1]{#1}%
\providecommand \bibfnamefont [1]{#1}%
\providecommand \citenamefont [1]{#1}%
\providecommand \href@noop [0]{\@secondoftwo}%
\providecommand \href [0]{\begingroup \@sanitize@url \@href}%
\providecommand \@href[1]{\@@startlink{#1}\@@href}%
\providecommand \@@href[1]{\endgroup#1\@@endlink}%
\providecommand \@sanitize@url [0]{\catcode `\\12\catcode `\$12\catcode `\&12\catcode `\#12\catcode `\^12\catcode `\_12\catcode `\%12\relax}%
\providecommand \@@startlink[1]{}%
\providecommand \@@endlink[0]{}%
\providecommand \url  [0]{\begingroup\@sanitize@url \@url }%
\providecommand \@url [1]{\endgroup\@href {#1}{\urlprefix }}%
\providecommand \urlprefix  [0]{URL }%
\providecommand \Eprint [0]{\href }%
\providecommand \doibase [0]{https://doi.org/}%
\providecommand \selectlanguage [0]{\@gobble}%
\providecommand \bibinfo  [0]{\@secondoftwo}%
\providecommand \bibfield  [0]{\@secondoftwo}%
\providecommand \translation [1]{[#1]}%
\providecommand \BibitemOpen [0]{}%
\providecommand \bibitemStop [0]{}%
\providecommand \bibitemNoStop [0]{.\EOS\space}%
\providecommand \EOS [0]{\spacefactor3000\relax}%
\providecommand \BibitemShut  [1]{\csname bibitem#1\endcsname}%
\let\auto@bib@innerbib\@empty
\bibitem [{\citenamefont {Franquet}\ \emph {et~al.}(2015)\citenamefont {Franquet}, \citenamefont {Perrier}, \citenamefont {Gibout},\ and\ \citenamefont {Bruel}}]{franquet2015free}%
  \BibitemOpen
  \bibfield  {author} {\bibinfo {author} {\bibfnamefont {E.}~\bibnamefont {Franquet}}, \bibinfo {author} {\bibfnamefont {V.}~\bibnamefont {Perrier}}, \bibinfo {author} {\bibfnamefont {S.}~\bibnamefont {Gibout}},\ and\ \bibinfo {author} {\bibfnamefont {P.}~\bibnamefont {Bruel}},\ }\bibfield  {title} {\bibinfo {title} {Free underexpanded jets in a quiescent medium: A review},\ }\href@noop {} {\bibfield  {journal} {\bibinfo  {journal} {Progress in Aerospace Sciences}\ }\textbf {\bibinfo {volume} {77}},\ \bibinfo {pages} {25} (\bibinfo {year} {2015})}\BibitemShut {NoStop}%
\bibitem [{\citenamefont {Chang}(1973)}]{chang1973mach}%
  \BibitemOpen
  \bibfield  {author} {\bibinfo {author} {\bibfnamefont {I.~S.}\ \bibnamefont {Chang}},\ }\href@noop {} {\emph {\bibinfo {title} {Mach reflection, Mach disc, and the associated nozzle free-jet flows.}}}\ (\bibinfo  {publisher} {University of Illinois at Urbana-Champaign},\ \bibinfo {year} {1973})\BibitemShut {NoStop}%
\bibitem [{\citenamefont {Crist}\ \emph {et~al.}(1966)\citenamefont {Crist}, \citenamefont {Glass},\ and\ \citenamefont {Sherman}}]{crist1966study}%
  \BibitemOpen
  \bibfield  {author} {\bibinfo {author} {\bibfnamefont {S.}~\bibnamefont {Crist}}, \bibinfo {author} {\bibfnamefont {D.}~\bibnamefont {Glass}},\ and\ \bibinfo {author} {\bibfnamefont {P.}~\bibnamefont {Sherman}},\ }\bibfield  {title} {\bibinfo {title} {Study of the highly underexpanded sonic jet.},\ }\href@noop {} {\bibfield  {journal} {\bibinfo  {journal} {AIAA journal}\ }\textbf {\bibinfo {volume} {4}},\ \bibinfo {pages} {68} (\bibinfo {year} {1966})}\BibitemShut {NoStop}%
\bibitem [{\citenamefont {Ashkenas}\ and\ \citenamefont {Sherman}(1966)}]{ashkenas1966structure}%
  \BibitemOpen
  \bibfield  {author} {\bibinfo {author} {\bibfnamefont {H.}~\bibnamefont {Ashkenas}}\ and\ \bibinfo {author} {\bibfnamefont {F.~S.}\ \bibnamefont {Sherman}},\ }\href@noop {} {\emph {\bibinfo {title} {The Structure and Utilization of Supersonic Free Jets in Low Density Wind Tunnels}}},\ \bibinfo {type} {Tech. Rep.}\ \bibinfo {number} {NASA-CR-60423}\ (\bibinfo  {institution} {Jet Propulsion Laboratory, California Institute of Technology},\ \bibinfo {year} {1966})\BibitemShut {NoStop}%
\bibitem [{\citenamefont {Gibbings}\ \emph {et~al.}(1972)\citenamefont {Gibbings}, \citenamefont {Ingham},\ and\ \citenamefont {Johnson}}]{gibbings1972flow}%
  \BibitemOpen
  \bibfield  {author} {\bibinfo {author} {\bibfnamefont {J.~C.}\ \bibnamefont {Gibbings}}, \bibinfo {author} {\bibfnamefont {J.}~\bibnamefont {Ingham}},\ and\ \bibinfo {author} {\bibfnamefont {D.}~\bibnamefont {Johnson}},\ }\href@noop {} {\emph {\bibinfo {title} {Flow in a Supersonic Jet Expanding from a Convergent Nozzle}}},\ \bibinfo {type} {Current Papers}\ \bibinfo {number} {C.P. No. 1197}\ (\bibinfo  {institution} {Aeronautical Research Council},\ \bibinfo {address} {London},\ \bibinfo {year} {1972})\BibitemShut {NoStop}%
\bibitem [{\citenamefont {Addy}(1981)}]{addy1981effects}%
  \BibitemOpen
  \bibfield  {author} {\bibinfo {author} {\bibfnamefont {A.}~\bibnamefont {Addy}},\ }\bibfield  {title} {\bibinfo {title} {Effects of axisymmetric sonic nozzle geometry on mach disk characteristics},\ }\href@noop {} {\bibfield  {journal} {\bibinfo  {journal} {Aiaa Journal}\ }\textbf {\bibinfo {volume} {19}},\ \bibinfo {pages} {121} (\bibinfo {year} {1981})}\BibitemShut {NoStop}%
\bibitem [{\citenamefont {Antsupov}(1974)}]{antsupov1974properties}%
  \BibitemOpen
  \bibfield  {author} {\bibinfo {author} {\bibfnamefont {A.}~\bibnamefont {Antsupov}},\ }\bibfield  {title} {\bibinfo {title} {Properties of underexpanded and overexpanded supersonic gas jets},\ }\href@noop {} {\bibfield  {journal} {\bibinfo  {journal} {Soviet Physics Technical Physics}\ }\textbf {\bibinfo {volume} {19}},\ \bibinfo {pages} {234} (\bibinfo {year} {1974})}\BibitemShut {NoStop}%
\bibitem [{\citenamefont {Carlson}\ and\ \citenamefont {Lewis}(1964)}]{carlson1964normal}%
  \BibitemOpen
  \bibfield  {author} {\bibinfo {author} {\bibfnamefont {D.~J.}\ \bibnamefont {Carlson}}\ and\ \bibinfo {author} {\bibfnamefont {C.~H.}\ \bibnamefont {Lewis}},\ }\bibfield  {title} {\bibinfo {title} {Normal shock location in underexpanded gas and gas-particle jets},\ }\href@noop {} {\bibfield  {journal} {\bibinfo  {journal} {AIAA Journal}\ }\textbf {\bibinfo {volume} {2}},\ \bibinfo {pages} {776} (\bibinfo {year} {1964})}\BibitemShut {NoStop}%
\bibitem [{\citenamefont {D'Attorre}\ and\ \citenamefont {Harshbarger}(1965)}]{dattorre1965parameters}%
  \BibitemOpen
  \bibfield  {author} {\bibinfo {author} {\bibfnamefont {L.}~\bibnamefont {D'Attorre}}\ and\ \bibinfo {author} {\bibfnamefont {F.~C.}\ \bibnamefont {Harshbarger}},\ }\bibfield  {title} {\bibinfo {title} {Parameters affecting the normal shock location in underexpanded gas jets},\ }\href@noop {} {\bibfield  {journal} {\bibinfo  {journal} {AIAA Journal}\ }\textbf {\bibinfo {volume} {3}},\ \bibinfo {pages} {530} (\bibinfo {year} {1965})}\BibitemShut {NoStop}%
\bibitem [{\citenamefont {Werle}\ \emph {et~al.}(1970)\citenamefont {Werle}, \citenamefont {Shaffer},\ and\ \citenamefont {Driftmyer}}]{werle1970freejet}%
  \BibitemOpen
  \bibfield  {author} {\bibinfo {author} {\bibfnamefont {M.~J.}\ \bibnamefont {Werle}}, \bibinfo {author} {\bibfnamefont {D.~G.}\ \bibnamefont {Shaffer}},\ and\ \bibinfo {author} {\bibfnamefont {R.~T.}\ \bibnamefont {Driftmyer}},\ }\bibfield  {title} {\bibinfo {title} {On freejet terminal shocks},\ }\href@noop {} {\bibfield  {journal} {\bibinfo  {journal} {AIAA Journal}\ }\textbf {\bibinfo {volume} {8}},\ \bibinfo {pages} {2295} (\bibinfo {year} {1970})}\BibitemShut {NoStop}%
\bibitem [{\citenamefont {Driftmyer}(1972)}]{driftmyer1972correlation}%
  \BibitemOpen
  \bibfield  {author} {\bibinfo {author} {\bibfnamefont {R.~T.}\ \bibnamefont {Driftmyer}},\ }\bibfield  {title} {\bibinfo {title} {A correlation of freejet data},\ }\href@noop {} {\bibfield  {journal} {\bibinfo  {journal} {AIAA Journal}\ }\textbf {\bibinfo {volume} {10}},\ \bibinfo {pages} {1093} (\bibinfo {year} {1972})}\BibitemShut {NoStop}%
\bibitem [{\citenamefont {Ewan}\ and\ \citenamefont {Moodie}(1986)}]{ewan1986structure}%
  \BibitemOpen
  \bibfield  {author} {\bibinfo {author} {\bibfnamefont {B.~C.~R.}\ \bibnamefont {Ewan}}\ and\ \bibinfo {author} {\bibfnamefont {K.}~\bibnamefont {Moodie}},\ }\bibfield  {title} {\bibinfo {title} {Structure and velocity measurements in underexpanded jets},\ }\href@noop {} {\bibfield  {journal} {\bibinfo  {journal} {Combustion Science and Technology}\ }\textbf {\bibinfo {volume} {45}},\ \bibinfo {pages} {275} (\bibinfo {year} {1986})}\BibitemShut {NoStop}%
\bibitem [{\citenamefont {Love}\ and\ \citenamefont {Grigsby}(1955)}]{love1955some}%
  \BibitemOpen
  \bibfield  {author} {\bibinfo {author} {\bibfnamefont {E.~S.}\ \bibnamefont {Love}}\ and\ \bibinfo {author} {\bibfnamefont {C.~E.}\ \bibnamefont {Grigsby}},\ }\href@noop {} {\emph {\bibinfo {title} {Some studies of axisymmetric free jets exhausting from sonic and supersonic nozzles into still air and into supersonic streams}}},\ \bibinfo {type} {Tech. Rep.}\ (\bibinfo {year} {1955})\BibitemShut {NoStop}%
\bibitem [{\citenamefont {Murzinov}(1971)}]{murzinov1971similarity}%
  \BibitemOpen
  \bibfield  {author} {\bibinfo {author} {\bibfnamefont {I.~N.}\ \bibnamefont {Murzinov}},\ }\bibfield  {title} {\bibinfo {title} {Similarity parameters for the escape of a strongly underexpanded jet into a flooded space},\ }\href@noop {} {\bibfield  {journal} {\bibinfo  {journal} {Fluid Dynamics}\ }\textbf {\bibinfo {volume} {6}},\ \bibinfo {pages} {675} (\bibinfo {year} {1971})}\BibitemShut {NoStop}%
\bibitem [{\citenamefont {Jiang}\ \emph {et~al.}(2022)\citenamefont {Jiang}, \citenamefont {Han}, \citenamefont {Hu}, \citenamefont {Gao},\ and\ \citenamefont {Lee}}]{jiang2022theoretical}%
  \BibitemOpen
  \bibfield  {author} {\bibinfo {author} {\bibfnamefont {C.}~\bibnamefont {Jiang}}, \bibinfo {author} {\bibfnamefont {T.}~\bibnamefont {Han}}, \bibinfo {author} {\bibfnamefont {S.}~\bibnamefont {Hu}}, \bibinfo {author} {\bibfnamefont {Z.}~\bibnamefont {Gao}},\ and\ \bibinfo {author} {\bibfnamefont {C.-H.}\ \bibnamefont {Lee}},\ }\bibfield  {title} {\bibinfo {title} {Theoretical prediction for the mach-disk height in two-dimensional supersonic underexpanded jets},\ }\href@noop {} {\bibfield  {journal} {\bibinfo  {journal} {AIAA Journal}\ }\textbf {\bibinfo {volume} {60}},\ \bibinfo {pages} {2115} (\bibinfo {year} {2022})}\BibitemShut {NoStop}%
\bibitem [{\citenamefont {D'Ambrosio}\ \emph {et~al.}(1999)\citenamefont {D'Ambrosio}, \citenamefont {Socio},\ and\ \citenamefont {Gaffuri}}]{dambrosio1999physical}%
  \BibitemOpen
  \bibfield  {author} {\bibinfo {author} {\bibfnamefont {D.}~\bibnamefont {D'Ambrosio}}, \bibinfo {author} {\bibfnamefont {L.~M.~D.}\ \bibnamefont {Socio}},\ and\ \bibinfo {author} {\bibfnamefont {G.}~\bibnamefont {Gaffuri}},\ }\bibfield  {title} {\bibinfo {title} {Physical and numerical experiments on an under-expanded jet},\ }\href@noop {} {\bibfield  {journal} {\bibinfo  {journal} {Meccanica}\ }\textbf {\bibinfo {volume} {34}},\ \bibinfo {pages} {267} (\bibinfo {year} {1999})}\BibitemShut {NoStop}%
\bibitem [{\citenamefont {Sommerfeld}(1994)}]{sommerfeld1994structure}%
  \BibitemOpen
  \bibfield  {author} {\bibinfo {author} {\bibfnamefont {M.}~\bibnamefont {Sommerfeld}},\ }\bibfield  {title} {\bibinfo {title} {The structure of particle-laden, underexpanded free jets},\ }\href@noop {} {\bibfield  {journal} {\bibinfo  {journal} {Shock Waves}\ }\textbf {\bibinfo {volume} {3}},\ \bibinfo {pages} {299} (\bibinfo {year} {1994})}\BibitemShut {NoStop}%
\bibitem [{\citenamefont {Lee}(2004)}]{lee2004supersonic}%
  \BibitemOpen
  \bibfield  {author} {\bibinfo {author} {\bibfnamefont {K.~H.}\ \bibnamefont {Lee}},\ }\emph {\bibinfo {title} {Study of Supersonic, Dual, Coaxial, Swirling Jet}},\ \href@noop {} {\bibinfo {type} {Doctor of engineering dissertation}},\ \bibinfo  {school} {Saga University}, \bibinfo {address} {Saga, Japan} (\bibinfo {year} {2004}),\ \bibinfo {note} {department of Energy and Materials Science, Graduate School of Science and Engineering}\BibitemShut {NoStop}%
\bibitem [{\citenamefont {Muraoka}\ and\ \citenamefont {Hiejima}(2022)}]{muraoka2022onset}%
  \BibitemOpen
  \bibfield  {author} {\bibinfo {author} {\bibfnamefont {R.}~\bibnamefont {Muraoka}}\ and\ \bibinfo {author} {\bibfnamefont {T.}~\bibnamefont {Hiejima}},\ }\bibfield  {title} {\bibinfo {title} {Onset conditions for mach disk formation in underexpanded jet flows},\ }\href@noop {} {\bibfield  {journal} {\bibinfo  {journal} {Physics of Fluids}\ }\textbf {\bibinfo {volume} {34}} (\bibinfo {year} {2022})}\BibitemShut {NoStop}%
\bibitem [{\citenamefont {Masuda}\ and\ \citenamefont {Moriyama}(1994)}]{masuda1994aerodynamic}%
  \BibitemOpen
  \bibfield  {author} {\bibinfo {author} {\bibfnamefont {W.}~\bibnamefont {Masuda}}\ and\ \bibinfo {author} {\bibfnamefont {E.}~\bibnamefont {Moriyama}},\ }\bibfield  {title} {\bibinfo {title} {Aerodynamic characteristics of underexpanded coaxial impinging jets},\ }\href@noop {} {\bibfield  {journal} {\bibinfo  {journal} {JSME International Journal Series B Fluids and Thermal Engineering}\ }\textbf {\bibinfo {volume} {37}},\ \bibinfo {pages} {769} (\bibinfo {year} {1994})}\BibitemShut {NoStop}%
\bibitem [{\citenamefont {Narayanan}\ and\ \citenamefont {Damodaran}(1993)}]{narayanan1993mach}%
  \BibitemOpen
  \bibfield  {author} {\bibinfo {author} {\bibfnamefont {A.~K.}\ \bibnamefont {Narayanan}}\ and\ \bibinfo {author} {\bibfnamefont {K.}~\bibnamefont {Damodaran}},\ }\bibfield  {title} {\bibinfo {title} {Mach disk of dual coaxial axisymmetric jets},\ }\href@noop {} {\bibfield  {journal} {\bibinfo  {journal} {AIAA journal}\ }\textbf {\bibinfo {volume} {31}},\ \bibinfo {pages} {1343} (\bibinfo {year} {1993})}\BibitemShut {NoStop}%
\bibitem [{\citenamefont {Rao}\ \emph {et~al.}(1996)\citenamefont {Rao}, \citenamefont {Kumar},\ and\ \citenamefont {Kurian}}]{rao1996near}%
  \BibitemOpen
  \bibfield  {author} {\bibinfo {author} {\bibfnamefont {T.}~\bibnamefont {Rao}}, \bibinfo {author} {\bibfnamefont {R.}~\bibnamefont {Kumar}},\ and\ \bibinfo {author} {\bibfnamefont {J.}~\bibnamefont {Kurian}},\ }\bibfield  {title} {\bibinfo {title} {Near field shock structure of dual co-axial jets},\ }\href@noop {} {\bibfield  {journal} {\bibinfo  {journal} {Shock Waves}\ }\textbf {\bibinfo {volume} {6}},\ \bibinfo {pages} {361} (\bibinfo {year} {1996})}\BibitemShut {NoStop}%
\bibitem [{\citenamefont {Srinivasarao}\ \emph {et~al.}(2012)\citenamefont {Srinivasarao}, \citenamefont {Lovaraju},\ and\ \citenamefont {Rathakrishnan}}]{srinivasarao2012effect}%
  \BibitemOpen
  \bibfield  {author} {\bibinfo {author} {\bibfnamefont {T.}~\bibnamefont {Srinivasarao}}, \bibinfo {author} {\bibfnamefont {P.}~\bibnamefont {Lovaraju}},\ and\ \bibinfo {author} {\bibfnamefont {E.}~\bibnamefont {Rathakrishnan}},\ }\bibfield  {title} {\bibinfo {title} {Effect of co-flow on near field shock structure},\ }\href {https://doi.org/10.1115/1.4006911} {\bibfield  {journal} {\bibinfo  {journal} {Journal of Fluids Engineering}\ }\textbf {\bibinfo {volume} {134}},\ \bibinfo {pages} {074501} (\bibinfo {year} {2012})}\BibitemShut {NoStop}%
\bibitem [{\citenamefont {Ahmad}\ \emph {et~al.}(2022)\citenamefont {Ahmad}, \citenamefont {Hasan},\ and\ \citenamefont {Sanghi}}]{ahmad2022influence}%
  \BibitemOpen
  \bibfield  {author} {\bibinfo {author} {\bibfnamefont {H.}~\bibnamefont {Ahmad}}, \bibinfo {author} {\bibfnamefont {N.}~\bibnamefont {Hasan}},\ and\ \bibinfo {author} {\bibfnamefont {S.}~\bibnamefont {Sanghi}},\ }\bibfield  {title} {\bibinfo {title} {On the influence of co-flow on the shocks and vortex rings in the starting phases of under-expanded jets},\ }\href@noop {} {\bibfield  {journal} {\bibinfo  {journal} {Physics of Fluids}\ }\textbf {\bibinfo {volume} {34}} (\bibinfo {year} {2022})}\BibitemShut {NoStop}%
\bibitem [{\citenamefont {Avduevskii}\ \emph {et~al.}(1970)\citenamefont {Avduevskii}, \citenamefont {Ivanov}, \citenamefont {Karpman}, \citenamefont {Traskovskii},\ and\ \citenamefont {Yudelovich}}]{avduevskii1970flow}%
  \BibitemOpen
  \bibfield  {author} {\bibinfo {author} {\bibfnamefont {V.~S.}\ \bibnamefont {Avduevskii}}, \bibinfo {author} {\bibfnamefont {A.~V.}\ \bibnamefont {Ivanov}}, \bibinfo {author} {\bibfnamefont {I.~M.}\ \bibnamefont {Karpman}}, \bibinfo {author} {\bibfnamefont {V.~D.}\ \bibnamefont {Traskovskii}},\ and\ \bibinfo {author} {\bibfnamefont {M.~Y.}\ \bibnamefont {Yudelovich}},\ }\bibfield  {title} {\bibinfo {title} {Flow in supersonic viscous under expanded jet},\ }\href@noop {} {\bibfield  {journal} {\bibinfo  {journal} {Fluid Dynamics}\ }\textbf {\bibinfo {volume} {5}},\ \bibinfo {pages} {409} (\bibinfo {year} {1970})}\BibitemShut {NoStop}%
\bibitem [{\citenamefont {Norum}\ and\ \citenamefont {Shearin}(1984)}]{norum1984effects}%
  \BibitemOpen
  \bibfield  {author} {\bibinfo {author} {\bibfnamefont {T.~D.}\ \bibnamefont {Norum}}\ and\ \bibinfo {author} {\bibfnamefont {J.~G.}\ \bibnamefont {Shearin}},\ }\href@noop {} {\emph {\bibinfo {title} {Effects of Simulated Flight on the Structure and Noise of Underexpanded Jets}}},\ \bibinfo {type} {Tech. Rep.}\ \bibinfo {number} {NASA TP-2308}\ (\bibinfo  {institution} {NASA Langley Research Center},\ \bibinfo {address} {Hampton, VA},\ \bibinfo {year} {1984})\BibitemShut {NoStop}%
\bibitem [{\citenamefont {Norum}\ and\ \citenamefont {Shearin}(1988)}]{norum1988shock}%
  \BibitemOpen
  \bibfield  {author} {\bibinfo {author} {\bibfnamefont {T.~D.}\ \bibnamefont {Norum}}\ and\ \bibinfo {author} {\bibfnamefont {J.~G.}\ \bibnamefont {Shearin}},\ }\href@noop {} {\emph {\bibinfo {title} {Shock Structure and Noise of Supersonic Jets in Simulated Flight to {Mach} 0.4}}},\ \bibinfo {type} {Tech. Rep.}\ \bibinfo {number} {NASA TP-2785}\ (\bibinfo  {institution} {NASA Langley Research Center},\ \bibinfo {address} {Hampton, VA},\ \bibinfo {year} {1988})\BibitemShut {NoStop}%
\bibitem [{\citenamefont {Norum}\ and\ \citenamefont {Brown}(1993)}]{norum1993simulated}%
  \BibitemOpen
  \bibfield  {author} {\bibinfo {author} {\bibfnamefont {T.~D.}\ \bibnamefont {Norum}}\ and\ \bibinfo {author} {\bibfnamefont {M.~C.}\ \bibnamefont {Brown}},\ }\bibfield  {title} {\bibinfo {title} {Simulated high speed flight effects on supersonic jet noise},\ }in\ \href {https://doi.org/10.2514/6.1993-4388} {\emph {\bibinfo {booktitle} {15th AIAA Aeroacoustics Conference}}}\ (\bibinfo {address} {Long Beach, CA},\ \bibinfo {year} {1993})\ \bibinfo {note} {aIAA Paper 93-4388}\BibitemShut {NoStop}%
\bibitem [{\citenamefont {Morris}(1988)}]{morris1988note}%
  \BibitemOpen
  \bibfield  {author} {\bibinfo {author} {\bibfnamefont {P.~J.}\ \bibnamefont {Morris}},\ }\bibfield  {title} {\bibinfo {title} {A note on the effect of forward flight on shock spacing in circular jets},\ }\href {https://doi.org/10.1016/S0022-460X(88)80014-2} {\bibfield  {journal} {\bibinfo  {journal} {Journal of Sound and Vibration}\ }\textbf {\bibinfo {volume} {121}},\ \bibinfo {pages} {175} (\bibinfo {year} {1988})}\BibitemShut {NoStop}%
\bibitem [{\citenamefont {Tam}(1991)}]{tam1992broadband}%
  \BibitemOpen
  \bibfield  {author} {\bibinfo {author} {\bibfnamefont {C.~K.~W.}\ \bibnamefont {Tam}},\ }\bibfield  {title} {\bibinfo {title} {Broadband shock-associated noise from supersonic jets in flight},\ }\href {https://doi.org/10.1016/0022-460X(91)90656-5} {\bibfield  {journal} {\bibinfo  {journal} {Journal of Sound and Vibration}\ }\textbf {\bibinfo {volume} {151}},\ \bibinfo {pages} {131} (\bibinfo {year} {1991})}\BibitemShut {NoStop}%
\bibitem [{\citenamefont {Andr{\'e}}\ \emph {et~al.}(2017)\citenamefont {Andr{\'e}}, \citenamefont {Castelain},\ and\ \citenamefont {Bailly}}]{andre2016flighteffects}%
  \BibitemOpen
  \bibfield  {author} {\bibinfo {author} {\bibfnamefont {B.}~\bibnamefont {Andr{\'e}}}, \bibinfo {author} {\bibfnamefont {T.}~\bibnamefont {Castelain}},\ and\ \bibinfo {author} {\bibfnamefont {C.}~\bibnamefont {Bailly}},\ }\bibfield  {title} {\bibinfo {title} {Experimental study of flight effects on slightly underexpanded supersonic jets},\ }\href {https://doi.org/10.2514/1.J054797} {\bibfield  {journal} {\bibinfo  {journal} {AIAA Journal}\ }\textbf {\bibinfo {volume} {55}},\ \bibinfo {pages} {57} (\bibinfo {year} {2017})}\BibitemShut {NoStop}%
\bibitem [{\citenamefont {Hu}\ \emph {et~al.}(2010)\citenamefont {Hu}, \citenamefont {Wang},\ and\ \citenamefont {Adams}}]{hu2010adaptive}%
  \BibitemOpen
  \bibfield  {author} {\bibinfo {author} {\bibfnamefont {X.}~\bibnamefont {Hu}}, \bibinfo {author} {\bibfnamefont {Q.}~\bibnamefont {Wang}},\ and\ \bibinfo {author} {\bibfnamefont {N.~A.}\ \bibnamefont {Adams}},\ }\bibfield  {title} {\bibinfo {title} {An adaptive central-upwind weighted essentially non-oscillatory scheme},\ }\href@noop {} {\bibfield  {journal} {\bibinfo  {journal} {Journal of Computational Physics}\ }\textbf {\bibinfo {volume} {229}},\ \bibinfo {pages} {8952} (\bibinfo {year} {2010})}\BibitemShut {NoStop}%
\bibitem [{\citenamefont {Sv{\"a}rd}\ \emph {et~al.}(2007)\citenamefont {Sv{\"a}rd}, \citenamefont {Carpenter},\ and\ \citenamefont {Nordstr{\"o}m}}]{svard2007stable}%
  \BibitemOpen
  \bibfield  {author} {\bibinfo {author} {\bibfnamefont {M.}~\bibnamefont {Sv{\"a}rd}}, \bibinfo {author} {\bibfnamefont {M.~H.}\ \bibnamefont {Carpenter}},\ and\ \bibinfo {author} {\bibfnamefont {J.}~\bibnamefont {Nordstr{\"o}m}},\ }\bibfield  {title} {\bibinfo {title} {A stable high-order finite difference scheme for the compressible navier--stokes equations, far-field boundary conditions},\ }\href@noop {} {\bibfield  {journal} {\bibinfo  {journal} {Journal of Computational Physics}\ }\textbf {\bibinfo {volume} {225}},\ \bibinfo {pages} {1020} (\bibinfo {year} {2007})}\BibitemShut {NoStop}%
\bibitem [{\citenamefont {Sharan}\ \emph {et~al.}(2018)\citenamefont {Sharan}, \citenamefont {Pantano},\ and\ \citenamefont {Bodony}}]{sharan2018time}%
  \BibitemOpen
  \bibfield  {author} {\bibinfo {author} {\bibfnamefont {N.}~\bibnamefont {Sharan}}, \bibinfo {author} {\bibfnamefont {C.}~\bibnamefont {Pantano}},\ and\ \bibinfo {author} {\bibfnamefont {D.~J.}\ \bibnamefont {Bodony}},\ }\bibfield  {title} {\bibinfo {title} {Time-stable overset grid method for hyperbolic problems using summation-by-parts operators},\ }\href@noop {} {\bibfield  {journal} {\bibinfo  {journal} {Journal of Computational Physics}\ }\textbf {\bibinfo {volume} {361}},\ \bibinfo {pages} {199} (\bibinfo {year} {2018})}\BibitemShut {NoStop}%
\bibitem [{\citenamefont {Poinsot}\ and\ \citenamefont {Lele}(1992)}]{poinsot1992boundary}%
  \BibitemOpen
  \bibfield  {author} {\bibinfo {author} {\bibfnamefont {T.~J.}\ \bibnamefont {Poinsot}}\ and\ \bibinfo {author} {\bibfnamefont {S.~K.}\ \bibnamefont {Lele}},\ }\bibfield  {title} {\bibinfo {title} {Boundary conditions for direct simulations of compressible viscous flows},\ }\href@noop {} {\bibfield  {journal} {\bibinfo  {journal} {Journal of computational physics}\ }\textbf {\bibinfo {volume} {101}},\ \bibinfo {pages} {104} (\bibinfo {year} {1992})}\BibitemShut {NoStop}%
\bibitem [{\citenamefont {Bodony}(2006)}]{bodony2006analysis}%
  \BibitemOpen
  \bibfield  {author} {\bibinfo {author} {\bibfnamefont {D.~J.}\ \bibnamefont {Bodony}},\ }\bibfield  {title} {\bibinfo {title} {Analysis of sponge zones for computational fluid mechanics},\ }\href@noop {} {\bibfield  {journal} {\bibinfo  {journal} {Journal of Computational Physics}\ }\textbf {\bibinfo {volume} {212}},\ \bibinfo {pages} {681} (\bibinfo {year} {2006})}\BibitemShut {NoStop}%
\bibitem [{\citenamefont {Sharan}\ \emph {et~al.}(2022)\citenamefont {Sharan}, \citenamefont {Brady},\ and\ \citenamefont {Livescu}}]{sharan2022high}%
  \BibitemOpen
  \bibfield  {author} {\bibinfo {author} {\bibfnamefont {N.}~\bibnamefont {Sharan}}, \bibinfo {author} {\bibfnamefont {P.~T.}\ \bibnamefont {Brady}},\ and\ \bibinfo {author} {\bibfnamefont {D.}~\bibnamefont {Livescu}},\ }\bibfield  {title} {\bibinfo {title} {High-order dimensionally-split cartesian embedded boundary method for non-dissipative schemes},\ }\href@noop {} {\bibfield  {journal} {\bibinfo  {journal} {Journal of Computational Physics}\ }\textbf {\bibinfo {volume} {464}},\ \bibinfo {pages} {111341} (\bibinfo {year} {2022})}\BibitemShut {NoStop}%
\bibitem [{\citenamefont {Panda}\ and\ \citenamefont {Seasholtz}(1999)}]{panda1999measurement}%
  \BibitemOpen
  \bibfield  {author} {\bibinfo {author} {\bibfnamefont {J.}~\bibnamefont {Panda}}\ and\ \bibinfo {author} {\bibfnamefont {R.}~\bibnamefont {Seasholtz}},\ }\bibfield  {title} {\bibinfo {title} {Measurement of shock structure and shock--vortex interaction in underexpanded jets using rayleigh scattering},\ }\href@noop {} {\bibfield  {journal} {\bibinfo  {journal} {Physics of Fluids}\ }\textbf {\bibinfo {volume} {11}},\ \bibinfo {pages} {3761} (\bibinfo {year} {1999})}\BibitemShut {NoStop}%
\bibitem [{\citenamefont {Chatterjee}\ \emph {et~al.}(2009)\citenamefont {Chatterjee}, \citenamefont {Ghodake},\ and\ \citenamefont {Singh}}]{chatterjee2009screech}%
  \BibitemOpen
  \bibfield  {author} {\bibinfo {author} {\bibfnamefont {A.}~\bibnamefont {Chatterjee}}, \bibinfo {author} {\bibfnamefont {D.}~\bibnamefont {Ghodake}},\ and\ \bibinfo {author} {\bibfnamefont {A.}~\bibnamefont {Singh}},\ }\bibfield  {title} {\bibinfo {title} {Screech frequency prediction in underexpanded axisymmetric screeching jets},\ }\href@noop {} {\bibfield  {journal} {\bibinfo  {journal} {International Journal of Aeroacoustics}\ }\textbf {\bibinfo {volume} {8}},\ \bibinfo {pages} {499} (\bibinfo {year} {2009})}\BibitemShut {NoStop}%
\bibitem [{\citenamefont {Sugawara}\ \emph {et~al.}(2020)\citenamefont {Sugawara}, \citenamefont {Nakao}, \citenamefont {Miyazato}, \citenamefont {Ishino},\ and\ \citenamefont {Miki}}]{sugawara2020three}%
  \BibitemOpen
  \bibfield  {author} {\bibinfo {author} {\bibfnamefont {S.}~\bibnamefont {Sugawara}}, \bibinfo {author} {\bibfnamefont {S.}~\bibnamefont {Nakao}}, \bibinfo {author} {\bibfnamefont {Y.}~\bibnamefont {Miyazato}}, \bibinfo {author} {\bibfnamefont {Y.}~\bibnamefont {Ishino}},\ and\ \bibinfo {author} {\bibfnamefont {K.}~\bibnamefont {Miki}},\ }\bibfield  {title} {\bibinfo {title} {Three-dimensional reconstruction of a microjet with a mach disk by mach--zehnder interferometers},\ }\href@noop {} {\bibfield  {journal} {\bibinfo  {journal} {Journal of Fluid Mechanics}\ }\textbf {\bibinfo {volume} {893}},\ \bibinfo {pages} {A25} (\bibinfo {year} {2020})}\BibitemShut {NoStop}%
\bibitem [{\citenamefont {Singh}\ and\ \citenamefont {Chatterjee}(2007)}]{singh2007numerical}%
  \BibitemOpen
  \bibfield  {author} {\bibinfo {author} {\bibfnamefont {A.}~\bibnamefont {Singh}}\ and\ \bibinfo {author} {\bibfnamefont {A.}~\bibnamefont {Chatterjee}},\ }\bibfield  {title} {\bibinfo {title} {Numerical prediction of supersonic jet screech frequency},\ }\href@noop {} {\bibfield  {journal} {\bibinfo  {journal} {Shock Waves}\ }\textbf {\bibinfo {volume} {17}},\ \bibinfo {pages} {263} (\bibinfo {year} {2007})}\BibitemShut {NoStop}%
\bibitem [{\citenamefont {Panda}(1998)}]{panda1998shock}%
  \BibitemOpen
  \bibfield  {author} {\bibinfo {author} {\bibfnamefont {J.}~\bibnamefont {Panda}},\ }\bibfield  {title} {\bibinfo {title} {Shock oscillation in underexpanded screeching jets},\ }\href@noop {} {\bibfield  {journal} {\bibinfo  {journal} {Journal of Fluid Mechanics}\ }\textbf {\bibinfo {volume} {363}},\ \bibinfo {pages} {173} (\bibinfo {year} {1998})}\BibitemShut {NoStop}%
\bibitem [{\citenamefont {Ferri}(1949)}]{ferri1949elements}%
  \BibitemOpen
  \bibfield  {author} {\bibinfo {author} {\bibfnamefont {A.}~\bibnamefont {Ferri}},\ }\bibfield  {title} {\bibinfo {title} {Elements of aerodynamics of supersonic flows},\ }\href@noop {} {\bibfield  {journal} {\bibinfo  {journal} {(No Title)}\ } (\bibinfo {year} {1949})}\BibitemShut {NoStop}%
\bibitem [{\citenamefont {Anderson}(1990)}]{anderson1990modern}%
  \BibitemOpen
  \bibfield  {author} {\bibinfo {author} {\bibfnamefont {J.~D.}\ \bibnamefont {Anderson}},\ }\bibfield  {title} {\bibinfo {title} {Modern compressible flow: with historical perspective},\ }\href@noop {} {\bibfield  {journal} {\bibinfo  {journal} {(No Title)}\ } (\bibinfo {year} {1990})}\BibitemShut {NoStop}%
\bibitem [{\citenamefont {Shapiro}(1953)}]{shapiro1953dynamics}%
  \BibitemOpen
  \bibfield  {author} {\bibinfo {author} {\bibfnamefont {A.~H.}\ \bibnamefont {Shapiro}},\ }\bibfield  {title} {\bibinfo {title} {The dynamics and thermodynamics of compressible fluid flow},\ }\href@noop {} {\bibfield  {journal} {\bibinfo  {journal} {New York: Ronald Press}\ } (\bibinfo {year} {1953})}\BibitemShut {NoStop}%
\bibitem [{\citenamefont {Prandtl}(1904)}]{prandtl1904stationaren}%
  \BibitemOpen
  \bibfield  {author} {\bibinfo {author} {\bibfnamefont {L.}~\bibnamefont {Prandtl}},\ }\bibfield  {title} {\bibinfo {title} {{\"U}ber die station{\"a}ren wellen in einem gasstrahl},\ }\href@noop {} {\bibfield  {journal} {\bibinfo  {journal} {Physikalische Zeitschrift}\ }\textbf {\bibinfo {volume} {5}},\ \bibinfo {pages} {599} (\bibinfo {year} {1904})}\BibitemShut {NoStop}%
\bibitem [{\citenamefont {Powell}(2010)}]{powell2010prandtl}%
  \BibitemOpen
  \bibfield  {author} {\bibinfo {author} {\bibfnamefont {A.}~\bibnamefont {Powell}},\ }\bibfield  {title} {\bibinfo {title} {On {Prandtl's} formulas for supersonic jet cell length},\ }\href@noop {} {\bibfield  {journal} {\bibinfo  {journal} {International Journal of Aeroacoustics}\ }\textbf {\bibinfo {volume} {9}},\ \bibinfo {pages} {207} (\bibinfo {year} {2010})}\BibitemShut {NoStop}%
\bibitem [{\citenamefont {Pack}(1950)}]{pack1950note}%
  \BibitemOpen
  \bibfield  {author} {\bibinfo {author} {\bibfnamefont {D.~C.}\ \bibnamefont {Pack}},\ }\bibfield  {title} {\bibinfo {title} {A note on {Prandtl's} formula for the wave-length of a supersonic gas jet},\ }\href@noop {} {\bibfield  {journal} {\bibinfo  {journal} {The Quarterly Journal of Mechanics and Applied Mathematics}\ }\textbf {\bibinfo {volume} {3}},\ \bibinfo {pages} {173} (\bibinfo {year} {1950})}\BibitemShut {NoStop}%
\bibitem [{\citenamefont {Brown}\ and\ \citenamefont {Roshko}(1974)}]{brown1974density}%
  \BibitemOpen
  \bibfield  {author} {\bibinfo {author} {\bibfnamefont {G.~L.}\ \bibnamefont {Brown}}\ and\ \bibinfo {author} {\bibfnamefont {A.}~\bibnamefont {Roshko}},\ }\bibfield  {title} {\bibinfo {title} {On density effects and large structure in turbulent mixing layers},\ }\href@noop {} {\bibfield  {journal} {\bibinfo  {journal} {Journal of Fluid Mechanics}\ }\textbf {\bibinfo {volume} {64}},\ \bibinfo {pages} {775} (\bibinfo {year} {1974})}\BibitemShut {NoStop}%
\bibitem [{\citenamefont {Abramovich}(1963)}]{abramovich1963theory}%
  \BibitemOpen
  \bibfield  {author} {\bibinfo {author} {\bibfnamefont {G.~N.}\ \bibnamefont {Abramovich}},\ }\href@noop {} {\emph {\bibinfo {title} {The Theory of Turbulent Jets}}}\ (\bibinfo  {publisher} {MIT Press},\ \bibinfo {address} {Cambridge, MA},\ \bibinfo {year} {1963})\BibitemShut {NoStop}%
\bibitem [{\citenamefont {Dimotakis}(1986)}]{dimotakis1986entrainment}%
  \BibitemOpen
  \bibfield  {author} {\bibinfo {author} {\bibfnamefont {P.~E.}\ \bibnamefont {Dimotakis}},\ }\bibfield  {title} {\bibinfo {title} {Two-dimensional shear-layer entrainment},\ }\href@noop {} {\bibfield  {journal} {\bibinfo  {journal} {AIAA Journal}\ }\textbf {\bibinfo {volume} {24}},\ \bibinfo {pages} {1791} (\bibinfo {year} {1986})}\BibitemShut {NoStop}%
\end{thebibliography}%

\end{document}